\documentclass[%
 aip,
 amsmath,amssymb, nofootinbib,superscriptaddress,
 reprint,%
]{revtex4-1}

\usepackage{graphicx}
\usepackage{dcolumn}
\usepackage{bm}

\usepackage[utf8]{inputenc}
\usepackage[T1]{fontenc}
\usepackage{mathptmx}
\usepackage{afterpage}
\usepackage{rotating}
\usepackage[flushleft]{threeparttable}
\usepackage[utf8]{inputenc}
\usepackage{subcaption}
\usepackage{caption}
\usepackage{mwe,tikz}\usepackage[percent]{overpic}
\usepackage{siunitx}
\DeclareSIUnit[number-unit-product = {\,}]
\cal{cal}
\DeclareSIUnit\kcal{\kilo\cal}
\DeclareSIUnit\kcal{\kilo\joule\per\mole}
\DeclareSIUnit\molar{\mole\per\cubic\deci\metre}
\DeclareSIUnit\Molar{\textsc{m}}
\usepackage{multirow}
\usepackage{chemfig}
\usepackage[version=4]{mhchem}
\usepackage[symbol*]{footmisc}
\renewcommand{\thefootnote}{\fnsymbol{footnote}}

\begin{document}

\preprint{AIP/123-QED}

\title[]{Can the roles of polar and non-polar moieties be reversed in non-polar solvents?}

\author{Cedrix J. Dongmo Foumthuim$^\parallel$}
\homepage{cedrix.dongmo@unive.it}
\affiliation{Dipartimento di Scienze Molecolari e Nanosistemi, 
Universit\`{a} Ca' Foscari di Venezia,Campus Scientifico, Edificio Alfa,
via Torino 155,30170 Venezia Mestre, Italy.}
\author{Manuel Carrer$^\parallel$}%
 \homepage{manuel.carrer@kjemi.uio.no}
\affiliation{ 
Department of Chemistry and Hylleraas Centre for Quantum Molecular Sciences, University of Oslo, PO Box 1033 Blindern, 0315 Oslo, Norway. 
}%
\author{Maurine Houvet}
\homepage{maurine.houvet@orange.fr}
\affiliation{%
Polytech Nantes -Engineering school of the University of Nantes,
    Rue Christian Pauc, 44306 Nantes cedex 3. 
}%
\author{Tatjana \v{S}krbi\'{c}}
\homepage{tskrbic@uoregon.edu} 
\affiliation{%
Dipartimento di Scienze Molecolari e Nanosistemi, 
Universit\`{a} Ca' Foscari di Venezia,Campus Scientifico, Edificio Alfa,
via Torino 155,30170 Venezia Mestre, Italy.}
\affiliation{Department of Physics and Institute for Fundamental Science, University of Oregon, 
Eugene, OR 97403, USA.}

\author{Giuseppe Graziano}
\homepage{graziano@unisannio.it}
\affiliation{
Department of Science and Technology, University of Sannio-Benevento,
      via Francesco de Sanctis, 82100 Benevento, Italy.
}

\author{Achille Giacometti$^*$}
\homepage{achille.giacometti@unive.it}
\affiliation{%
Dipartimento di Scienze Molecolari e Nanosistemi, 
Universit\`{a} Ca' Foscari di Venezia,Campus Scientifico, Edificio Alfa,
via Torino 155,30170 Venezia Mestre, Italy.}
\affiliation{European Centre for Living Technology (ECLT)
Ca’ Bottacin, Dorsoduro 3911, Calle Crosera
30123 Venice, Italy.}
\date{\today}

\begin{abstract}
Using thermodynamics integration, we study the solvation free energy of 18 amino acid side chain equivalents in solvents with different polarity, ranging from the most polar water to the most non-polar cyclohexane. The amino acid side chain equivalents are obtained from the 20 natural amino acids by replacing the backbone part with a hydrogen atom, and discarding proline and glycine that have special properties. A detailed analysis of the relative solvation free energies suggests how it is possible to achieve a robust and unambiguous hydrophobic scale for the amino acids. By discriminating the relative contributions of the entropic and enthalpic terms, we find strong negative correlations in water and ethanol, associated with the well-known entropy-enthalpy compensation, and a much reduced correlation in cyclohexane. This shows that in general the role of the polar and non-polar moieties cannot be reversed in a non-polar solvent. Our findings are compared with past experimental as well as numerical results, and may shed additional light on the unique role of water as biological solvent. 
\end{abstract}

\maketitle

\def\thefootnote{$^\parallel$}\footnotetext{These authors contributed equally to this work}
\def\thefootnote{$^*$}\footnotetext{Corresponding author : achille.giacometti@unive.it}

\section{Introduction}
\label{sec:introduction}
The hydrophobic effect refers to the tendency of non-polar moieties to avoid the contact with water molecules and form an aggregate \cite{Tanford80}. This is also commonly viewed as one of the main driving force underlying the folding of a protein
\cite{Cantor80,Finkelstein16}. As a polypeptide chain is formed by a sequence of amino acids taken from a 20 letter alphabet, 50\% of which are roughly hydrophobic (i.e. tend to avoid the contact with water) and 50\% are polar (i.e. are happy to stay in contact with water), in water the chain tends to fold so to bury as much as possible the hydrophobic amino acids inside the folded chain. A number of concurring effects \cite{Camilloni16} prevent a perfect outcome of this scenario, but this would be the optimal configuration in terms of the hydrophobic effects. Hence, water clearly plays an essential role for protein folding and protein functioning.

On the other hand, some experimental studies have pointed out that several enzymes are stable and fully active in anhydrous non-polar solvents \cite{Klibanov01}. Pace and collaborators \cite{Pace04} noticed that folded proteins become unstable in polar solvents such as ethanol \ce{EtOH}, but return to be very stable, albeit essentially insoluble, in non-polar solvents such as cyclohexane \ce{cC6H12}. The idea is that intra-chain hydrogen bonds are effectively stronger in cyclohexane and similar liquids, because there is no competition to form hydrogen bonds with the non-polar solvent molecules. The denaturing action of ethanol is not as simple to rationalize because polypeptide chains, in aqueous solution with high concentrations of ethanol, populate conformations having a high content of $\alpha$ helices \cite{Hirota98}. A similar scenario emerged from two recent numerical studies \cite{Hayashi17,Hayashi18} using an approximate, albeit accurate, method to compute the solvation free energy. It was observed that, although the native state of globular proteins is the most stable in water compared with any other competing folds having the same sequence, this is no longer the case in ethanol and in cyclohexane, where the most stable folds are those having a high content of $\alpha$ helices, in agreement with experimental studies.


This scenario prompts the following question: is there a liquid, different from water, in which polypeptide chains fold by burying the polar amino acids and still possessing the ordered secondary structure elements? Clearly, non-polar solvents, such as cyclohexane for instance, appear to be optimal candidates for this, thus suggesting the two processes to be mirror images of one another. {\color{black} To the best of our knowledge, this issue has never been addressed before. As we shall see below, the results of the present study indicate that the two processes have very different driving forces.  This is also supported by recent results related to the possibility of forming micelles in non-polar solvents by surfactants having  a hydrophobic (rather than polar) head, and a polar (rather than hydrophobic) tail, an issue that has been addressed in a parallel work by some of the present authors \cite{Carrer20}, and that may be relevant for the existence of unconventional life forms in other planets where water is not available \cite{Sandstroem20}. In that case too, the self-assembly processes of these polarity-inverted surfactants have different driving forces.}

In order to tackle this problem in a fully fledged study, it would be necessary to characterize the behavior of a complete polypeptide chain in different solvents with molecular details, a task that is beyond our current computational capabilities. On the other hand, small peptides and all isolated natural amino acids are within our present reach.

Motivated by this scenario, in this paper we study the solvation free energies of natural amino acids in water, in ethanol, and in cyclohexane, as well as the free energy differences of moving one amino acid from one solvent into another one.
As natural amino acids cannot be isolated by their surrounding environment, we will then replace them with their amino acid side chain equivalents that can be obtained by substituting the backbone group with a single hydrogen atom to make the molecule neutral. This can be done for all amino acids but proline and glycine, the former because it does not have a proper side chain, the latter because it does not have a side chain at all. 

The problem is not new and experimental data are available,  -- see in particular the important contributions from  Wolfenden’s lab  \cite{Radzicka88,Wolfenden07}, but experimental data for solvation in ethanol are rather scanty. There are also several computer simulation studies considering solvation  of amino acid side chain equivalents in water and cyclohexane  \cite{Villa02,Chang07}, and comparing results for different water force fields \cite{Shirts03}. Another study also addressed the inclusion of the amino acids backbones \cite{Schauperl16}.

Using thermodynamic integration \cite{Frenkel01}, we perform an extensive analysis of the solvation free energies of the 18 amino acid side chain equivalents, in water, cyclohexane and ethanol at different temperatures. This also allows the separation of the entropy and enthalpy contributions, thus providing an exhaustive study of the solvation thermodynamics at an unprecedented scale.  

{\color{black} In summary, the key new elements provided in our study are (a) a comprehensive treatment in three contrasting solvents; (b) an extensive analysis of the temperature dependence, allowing entropies to be extracted; (c) a detailed discussion on the chemical-physics consequences, with a special focus on the unique role of water as a solvent for biological molecules.}



The plan of the paper is as follows. Section \ref{sec:theory} provides the general background of our analysis; results are presented
in Section \ref{sec:results}, and Section \ref{sec:conclusions} will provide some summarizing take-home messages. Additional tables and figures can be found in Appendixes.
\section{Theory and Methods}
\label{sec:theory}
\subsection{Thermodynamic integration}
\label{subsec:thermodynamic}
The solvation free energy $\Delta G_{solv}$ can be defined as the difference between the free energy of a single analyte molecule in a specified solvent $G_{solvent}$ and in vacuum $G_{vacuum}$
\begin{eqnarray}
  \label{sec2:eq1}
     \Delta G_{solv} &=& G_{solvent} - G_{vacuum}
\end{eqnarray}
If $\Delta G_{solv}<0$ ($\Delta G_{solv}>0$) the solvent is stabilizing (destabilizing) the molecule with respect to vacuum. This concept can clearly be extended to the free energy transfer  $\Delta \Delta G (S_1 \to S_2)$ between two different solvents $S_1$ and $S_2$
\begin{eqnarray}
  \label{sec2:eq2}
 \Delta \Delta G \left(S_1 \to S_2\right) &=& \Delta G_{S_{2}} - \Delta G_{S_{1}}
\end{eqnarray}
where $\Delta G_{S_{1}}$ and $\Delta G_{S_{2}}$ are the solvation free energy for solvents $S_1$ and $S_2$, respectively.

From the numerical viewpoint, free energy differences can be conveniently computed by using the well-known expression \cite{Frenkel01}
\begin{eqnarray}
  \label{sec2:eq3}
  \Delta G_{AB} &=& \int_{\lambda_{A}}^{\lambda_{B}} d\lambda \left \langle \frac{\partial V\left(\mathbf{r};\lambda\right)}{\partial \lambda} \right \rangle_{\lambda}
\end{eqnarray}
where $V(\mathbf{r},\lambda)$ is the potential energy of the system as a function of the coordinate vector $\mathbf{r}$, and $\lambda$ is a switching-on parameter allowing to go from state A to state B by changing its value from $\lambda_{A}$ to $\lambda_{B}$. The average $\langle \ldots \rangle_{\lambda}$ in Eq.(\ref{sec2:eq3}) is the usual thermal average with potential $V(\mathbf{r},\lambda)$. The $\lambda$ interval $[\lambda_A,\lambda_B]$ is partitioned
into a grid of small intervals, molecular dynamics simulations are performed for each value of $\lambda$
belonging to each interval, and the results are then integrated over all values of $\lambda$ to obtain the final free energy
difference.


\begin{figure}[htpb]
\centering
 \captionsetup{justification=raggedright,width=\linewidth}
   \begin{subfigure}{8cm}
  \includegraphics[width=\linewidth,trim=0cm 0cm 0cm 0cm, angle=0]{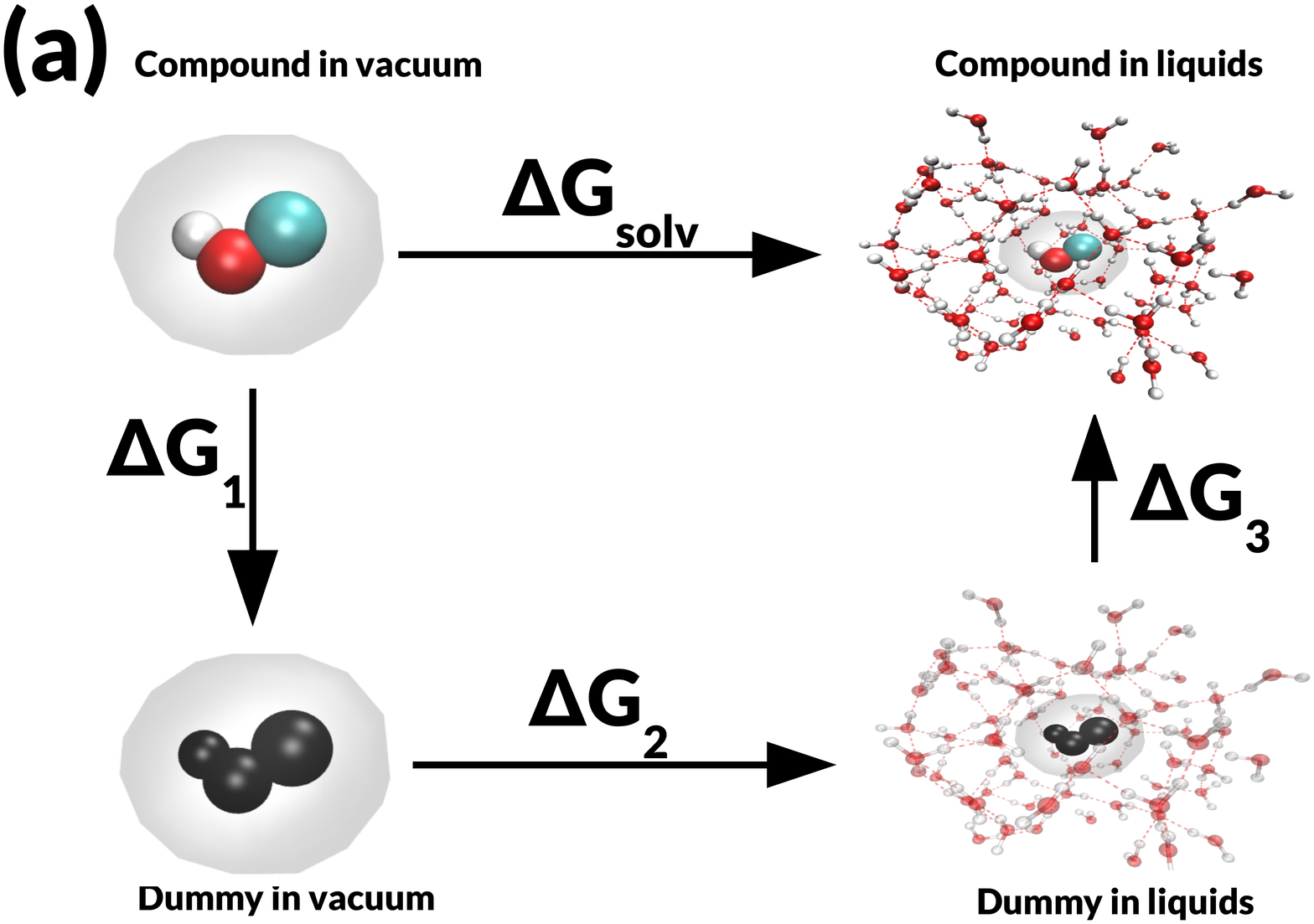}
  \caption{}\label{fig:fig1a}
  \end{subfigure}
   \begin{subfigure}{8cm}
  \includegraphics[width=\linewidth,trim=0cm 0cm 0cm 0cm, angle=0]{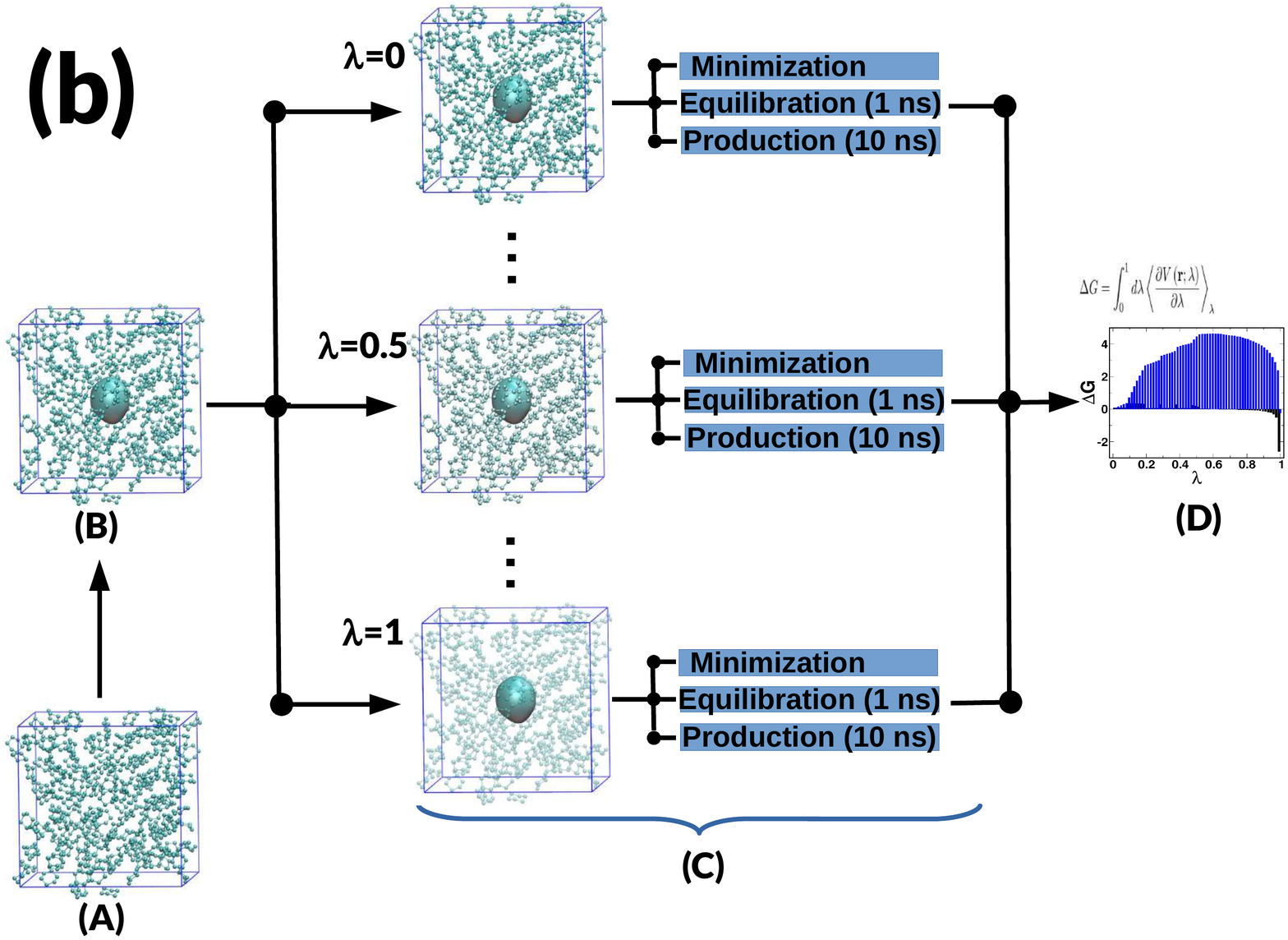}
  \caption{}\label{fig:fig1b}
  \end{subfigure}
  \caption{\textcolor{black}{\textbf{(a)} A thermodynamic cycle allowing the computation of the solvation free energy $\Delta G_{solv}$. \textbf{(b)}  Simulation workflow illustrated for the case of cyclohexane. \textbf{(A)} The simulation starts with a pre-equilibrated box of solvent; \textbf{(B)} The solute is then inserted into the equilibrated box;  \textbf{(C)} Parallel simulations are performed for each value of the coupling parameter $\lambda$. This includes an energy minimization (steepest descent + l-bfgs), an equilibration (NVT + NPT) and a production steps, as shown;  \textbf{(D)} The intermediate values of lambda are combined using the Bennet's acceptance ratio to obtain the solvation free energy. The value of lambda $\lambda =0$ represents the fully coupled (interacting) state while $\lambda=1$ is the fully uncoupled (non-interacting) state.}
  \label{fig:fig1}}
\end{figure}


\textcolor{black}{Best practices in free energies calculations \cite{Pohorille10,Klimovich15} suggest the use of alchemical transformation in the form of a thermodynamic cycle as defined in  Figure \ref{fig:fig1}(a) \cite{Ben-Naim80,Pohorille10,Klimovich15}. Firstly, all intramolecular non-bonded interactions in the solute compound are turned off to obtain the dummy compound in vacuum. Let $\Delta G_1$ be the free energy difference associated with this transition. Then, the dummy compound is transferred from vacuum to the solvent -- the liquid in Figure \ref{fig:fig1}(a).  As the free energy of non-interacting molecules does not depend on its environment, the corresponding free energy difference is effectively zero, so $\Delta G_2=0$. Finally, all the non-bonded interactions are turned on in the solvent with a free energy cost $\Delta G_3$ to achieve the final compound in the solvent (liquid). Then, clearly $\Delta G_{solv}=\Delta G_1+ \Delta G_2+\Delta G_3$, as summarized in Figure \ref{fig:fig1} (a). Note that in the presence of steric interactions only, $\Delta G_{solv}$ is purely of entropic nature and can be estimated using Scaled Particle Theory (SPT) (see Section \ref{sec:theory}). In practice, however, a direct calculation $\Delta G_{solv}$ can nowadays be achieved using an efficient application of thermodynamic integration \cite{Ruiter16} as schematically shown in Figure \ref{fig:fig1}(b). Here the solute is inserted into a pre-equilibrated solvent, and parallel simulations are involving energy minimization, NVT and NPT equilibration, and production runs are computed for several intermediate values of the coupling parameter $\lambda$, and then combined using Bennet's acceptance ratio \cite{Bennett76} to finally obtain the required solvation free energy. Here $\lambda=0$ refers to fully coupled case, whereas $\lambda=1$ to the fully uncoupled case. See Section \ref{subsec:numerical} for details.}

\begin{figure}[htpb]
\centering
 \captionsetup{justification=raggedright,width=\linewidth}
  \begin{subfigure}{8cm}
     \includegraphics[width=\linewidth]{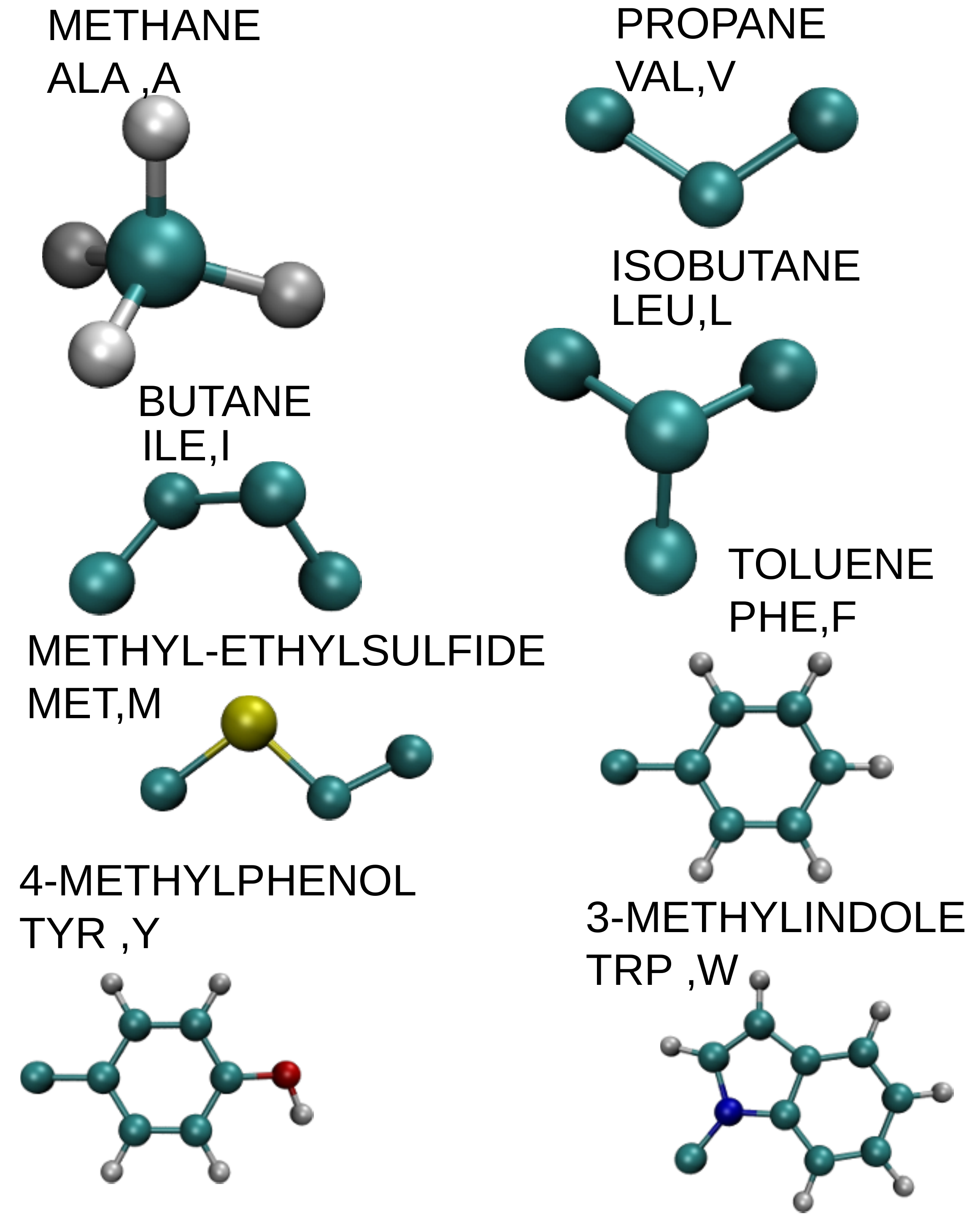}
    \caption{}\label{fig:fig2a}
  \end{subfigure}
  \begin{subfigure}{8cm}
     \includegraphics[width=\linewidth]{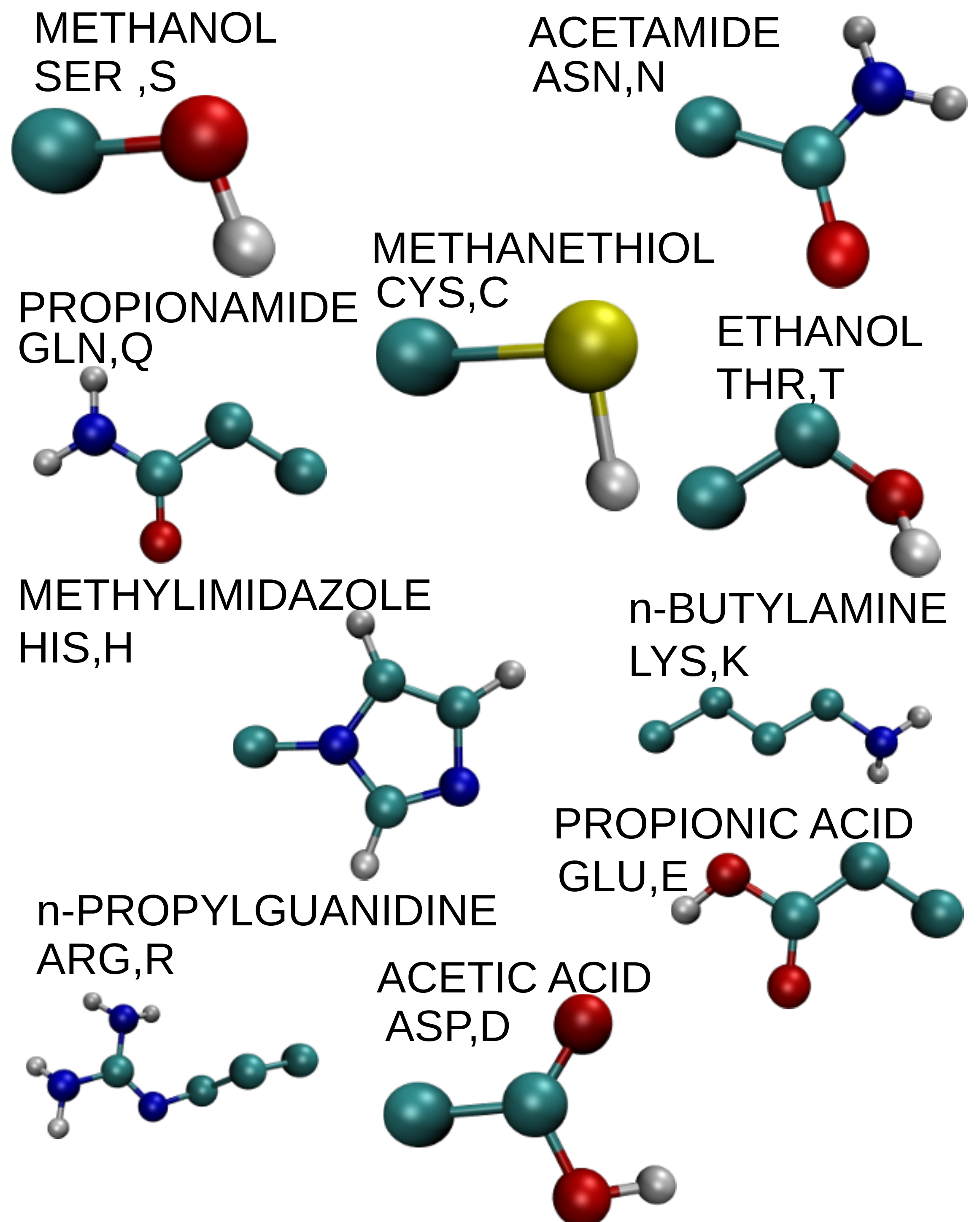}
    \caption{}\label{fig:fig2b}
  \end{subfigure}
  \caption{(a) Hydrophobic amino acid side chain equivalents. (b) Polar amino acid side chain equivalents. 
  \label{fig:fig2}}
\end{figure}

Fig. \ref{fig:aminoacids} depicts 18 of the 20 natural amino acids that are conventionally divided in hydrophobic (non-polar) Fig. \ref{fig:aminoacids_a}, and polar (hydrophilic) Fig. \ref{fig:aminoacids_b}. Although this division is accepted by general consensus, it relies purely on the chemical structure of the side chain.  As we will see below, computational as well as experimental results based on the above rigorous definition will provide further insights on these two classes. Two amino acids have special features and hence have not been included in Fig. \ref{fig:aminoacids}: proline because it does not have a proper side chain, glycine because essentially it has no side chains -- its side chain is a single hydrogen atom. In natural amino acids, side chains are attached to the backbone, as also visible in each of the 18 amino acids of Fig. \ref{fig:aminoacids}. 
Molecular equivalents of these 18 natural amino acids side chains can be obtained by capping them with a single hydrogen atom replacing the backbone part. This is presented in Figure \ref{fig:fig2} where each equivalent is
identified by the short hand notation of its natural side chain counterpart, as reported in  Table \ref{tab:equivalent}.

As for the amino acids, solvents too have their own hydrophobicity scale again relying essentially on indirect facts rather than on a robust thermodynamic relative measurement. One popular way is through the relative dielectric constant $\epsilon_{r}$ that is $78.5$ in water \ce{H2O}, $24.3$ in ethanol \ce{EtOH}, and $2.0$ in cyclohexane \ce{cC6H12} at $T= \SI{298}{\kelvin}$\cite{Lide04}. Accordingly, cyclohexane is much more hydrophobic than water and relatively more hydrophobic than ethanol. The rational beyond this choice of course stems for the fact that the dielectric constant is roughly proportional to the dipole strength that is providing the polarity of the solvent molecules, and dipole-dipole interactions are considerably stronger than any other interactions (quadrupole, van der Waals, etc.) appearing in the absence of a permanent dipole.
By computing the solvation free energy $\Delta G_{w}\equiv \Delta G_{\text{\ce{H2O}}}$ of each of these amino acid side chain equivalents in water, and then the solvation free energies $\Delta G_{c} \equiv \Delta G_{\text{\ce{cC6H12}}}$ and $\Delta G_{e}\equiv \Delta G_{\text{\ce{EtOH}}}$ in cyclohexane and ethanol, we can quantify their relatively polarity.
As hydrophobic molecules produce unfavourable interactions in water and favourable in cyclohexane, we expect $\Delta G_{w}>0$ and $\Delta G_{c}<0$ for hydrophobic amino acid side chain equivalents (ALA,VAL,ILE,LEU,MET,PHE,TYR,TRP),
and the opposite $\Delta G_{w}<0$ and $\Delta G_{c}>0$ for polar amino acid side chain equivalents (SER,ASN,GLN,CYS,THR,HIS,LYS,ARG,APS,GLU).
In addition, we can also quantify the free energy differences in the transfer water-cyclohexane $\Delta \Delta G_{w>c}=\Delta G_{c}-\Delta G_{w} \equiv \Delta G_{\text{\ce{cC6H12}}}- \Delta G_{\text{\ce{H2O}}}$, and water-ethanol $\Delta \Delta G_{w>e}=\Delta G_{e}-\Delta G_{w} \equiv \Delta G_{\text{\ce{EtOH}}}- \Delta G_{\text{\ce{H2O}}}$.
This difference provides a measure of the propensity for that particular amino acid side chain equivalent
to be solvated in one or the other solvent, and hence a robust scale of relative hydrophobicity with respect to water, as already suggested by Tanford many years ago \cite{Tanford80}.

Therefore we will label a particular amino acid side chain equivalent as hydrophobic (with respect to water), if
$\Delta \Delta G_{w>c}<0$, polar if $\Delta \Delta G_{w>c}>0$. Likewise, we can have an intermediate hydrophobicity values by computing the free energy difference of transferring an amino acid side chain equivalent from water to ethanol $\Delta \Delta G_{w>e}$.

A final interesting point is whether the particular solvation process is entropically or enthalpically dominated. This can be understood by separating out the enthalpic and the entropic contributions as obtained from the evaluation of the free energy difference $\Delta G(T)$ at different temperatures $T$, and then the calculation of the entropy via a differentiation with respect to the temperature. To this aim, we assume the following temperature dependence for the free energy difference \cite{Hajari15}
\begin{eqnarray}
  \label{sec2:eq4}
  \Delta G\left(T\right) &=& a + b T + cT \ln T
\end{eqnarray}
so that
\begin{eqnarray}
  \label{sec2:eq5}
  \Delta S\left(T\right) &=& - \left(\frac{\partial \Delta G\left(T\right)}{\partial T} \right)_{P} = -b -c\left[1+ \ln T\right]
\end{eqnarray}
and then the enthalpy change can be obtained from
\begin{eqnarray}
  \label{sec2:eq6}
  \Delta H\left(T\right) &=& \Delta G\left(T\right) + T \Delta S\left(T\right)
\end{eqnarray}
A numerical fit of the parameters $a$, $b$, and $c$ appearing in Eq.(\ref{sec2:eq4}) based on the results of simulations at different temperatures, will provide the required expressions for the entropy (Eq.(\ref{sec2:eq5})) and for the enthalpy (Eq.(\ref{sec2:eq6})). Standard deviation was evaluated using error block analysis \cite{Hajari15}.

{\color{black}A word of caution is in order here. As discussed in Ref. \cite{Hajari15}, the functional form given in Eq.(\ref{sec2:eq4}) is valid provided that the heat capacity change in solvating the amino acid side chain equivalents is approximately constant in the considered temperature range, $270-330\SI{}{\kelvin}$ in the present study.  This is comparable with $278-338\SI{}{\kelvin}$  considered in Ref. \cite{Hajari15} where only water was investigated. Moreover, temperature $\SI{270}{\kelvin}$ is below the freezing points of both water and cyclohexane, and will be used here as an extrapolated value from the liquid phase. That said, we will use the same functional form even for the two other solvents considered in the present study, knowing that this is an unwarranted approximation possibly invalid in some cases. Furthermore, $270\SI{}{\kelvin}$ does not appear to be a particular outlier to the fitted curves, see Fig .\ref{fig:water} - Fig.\ref{fig:etoh}. 
} 
\subsection{Scaled Particle Theory (SPT)}
\label{subsec:scaled}
According to a general theory of solvation \cite{Graziano19}, $\Delta G_{solv}$ can be calculated as the sum of the reversible work spent to create a cavity suitable to host the solute molecule, $\Delta G_{0}$, and of the reversible work to turn on the attractive solute-solvent interactions, $E_{a}$, usually assumed to be a purely energetic term. Reliable estimates of $\Delta G_{0}$ in any liquid can be calculated by means of the analytical relationships provided by classic Scaled Particle Theory (SPT) \cite{Pierotti76}. It is only necessary to assign an effective hard sphere diameter to solvent and solute molecules and to use the experimental solvent density at the temperature and pressure of interest. The use of experimental density is an indirect way to take into account the true interactions existing among solvent molecules in the pure liquid. Reliable estimates of $E_{a}$ can be calculated by means of simple expressions in the case of purely van der Waals attractions, whereas numerical calculations are in general necessary in the case of hydrogen bonds. Additional details on the theoretical aspects can be found in the original paper \cite{Graziano19}. 
\subsection{Numerical protocols}
\label{subsec:numerical}
The amino acid side chain equivalents used in this work are organic chemical moieties derived by truncating the natural amino acid side chains at position CB and capping the tail with a hydrogen atom. In particular, the initial structures for these latter compounds along with their building topology were retrieved from the Automated Topology Builder (ATB2.0) \cite{Koziara14}. While a united atom representation was used in setting up the systems, an in house modification of GROMOS96 (54a7) force field \cite{Schmid11} was required to account for non-natively parameterized molecules. The choice of this force field is in line with past work \cite{Villa02} where it was shown to provide good description of the solvent properties. Here we have used the latest 54a7 version of Gromos force field, while Villa et al \cite{Villa02} employed the 43a2 version. A good alternative for cyclohexane would have been the most recent version of the optimized OPLS (L-OPLS) \cite{Siu12} which yields improved values of hydrocarbon diffusion coefficients, viscosities, and gauche-trans ratios. Selection of the latter would have more faithfully compared with results of earlier simulations by Chang et al. \cite{Chang07} who used an older OPLS-AA force field. In view of the highly computational requirements of the present holistic analysis, we have made the reasonable compromise of selecting GROMOS96 (54a7) which was explicitly tuned to best reproduce the experimental hydration enthalpies of the side-chain analogs as well as to better preserve the protein secondary structure. Other choices \cite{Shirts03}, \cite{Schauperl16} provide comparable performances.

The simulations were performed in three different solvents covering a broad range of polarity, from non-polar cyclohexane  \ce{cC6H12}, to highly polar water \ce{H2O}, through the intermediate polar ethanol \ce{EtOH}. The 18 amino acid side chain equivalents were then inserted into a cubic box of \SI{3}{\nm} in size incorporating about 165, 290 and 881 molecules of \ce{cC6H12}, \ce{EtOH}, and \ce{H2O}, respectively. The simulations were performed with Gromacs simulation package (versions 2018.3 and 2018.7) \cite{Abraham15} and all the solutes were modeled in their neutral uncharged states.
{\color{black} As detailed in Section \ref{sec:theory}, free energy differences as given by Eq.\ref{sec2:eq3} have been computed from the fully coupled ($\lambda=0$) to the fully uncoupled $\lambda=1$ system, by gradually switching off all non-steric interactions. A grid of $\Delta \lambda =0.05$ has been used in all cases, resulting into a 21 binning points. See Fig. \ref{fig:decoupling_process}.}
The initial systems were initially prepared by minimizing the solute's potential energy and relaxing the solvent around solute atoms before running free energy molecular dynamics (MD) simulations. Two cycles of minimization rounds were performed embedding $10^{5}$ steps of steepest descent minimization algorithm with a minimization step size of $5 \times 10^{-4} \SI{}{\nm}$ and a maximum convergence force of $\SI{100.0}{\kJ \mol^{-1}\nm^{-1}}$ followed by $5\times 10^{4}$ steps of l-bfgs quasi-Newtonian minimization algorithm with a minimization step size of $10^{-3}\SI{}{\nm}$ and a maximum convergence force of $\SI{100.0}{\kJ \mol^{-1}\nm^{-1}}$.  Thereafter, an equilibration round in the canonical NVT ensemble was performed for $\SI{500}{\ps}$ using the accurate leap-frog stochastic dynamics integrator with the simulation time step of $\SI{2}{\fs}$. Due to the large numerical fluctuations recorded, a shorter time step of $\SI{0.5}{\fs}$ or $\SI{1}{\fs}$  was used in some simulations. While long-range electrostatics interactions were accounted with the Particle Mesh Ewald summation \cite{Essmann95}, short-range electrostatics and van der Waals interactions were truncated with a single-range cutoff at $\SI{12}{\angstrom}$ with the pair list updated every 20 steps. The simulations were performed at seven different temperatures in the range $270-330\SI{}{\kelvin}$. Each temperature was kept around the reference value by coupling the system to an external bath using the Berendsen thermostat (for less stable systems) \cite{Berendsen84}, 
with a coupling constant of 1.0 ps. All simulations were replicated in a 3D bulk-like phase using the periodic boundaries conditions and all bonds involving hydrogen atoms were restrained using LINCS algorithms \cite{Hess97}. For water we used the simple point charge SPC water model \cite{Berendsen1981}, whereas the parameters (or topology) for ethanol and cyclohexane were manually implemented. Small quantitative differences could be expected \cite{Schauperl16} by choosing more refined force fields for water, at the expenses of a significant increase in the computational time.
The second equilibration round was then performed for additional $\SI{500}{\ps}$ in the isobaric-isothermal NPT ensemble using the same parameters as in NVT. The pressure was equilibrated to the reference value of 1 bar using the Parrinello-Rahman pressure coupling (for more stable systems) \cite{Parrinello81} and the weak-coupling Berendsen barostat (for less stable systems) \cite{Berendsen84}, with a coupling constant of $\SI{1.0}{\ps}$. The isothermal compressibility (in bar$^{-1}$) of $4.5 \times 10^{-5}$, $1.2 \times 10^{-4}$, and $4.5 \times 10^{-5}$ was used for water, ethanol and cyclohexane, respectively. The final production runs for free energy calculation were performed for $\SI{10}{\ns}$ using 21 equidistant lambda points with a step size of 0.05. In the case of cyclohexane only the dispersive interactions were decoupled, while for ethanol and water also the coulomb interactions were considered.    
{\color{black} In short, we have closely followed the numerical protocol by Villa \textit{et al.} \cite{Villa02}, but we have improved it with an updated forcefield and extended the simulation timescales by performing longer free energy samplings. Moreover, in many cases we have employed a smaller time-step where Villa \textit{et al} \cite{Villa02} used 2 fs. Our simulation workflow is shown in Fig. \ref{fig:fig1b}.}


\section{Results}
\label{sec:results}
\subsection{Solvation free energy $\Delta G_{solv}$}
\label{subsec:solvation}
\begin{figure}[htpb]
\centering
 \captionsetup{justification=raggedright,width=\linewidth}
  \begin{subfigure}{9cm}
     \includegraphics[width=\linewidth]{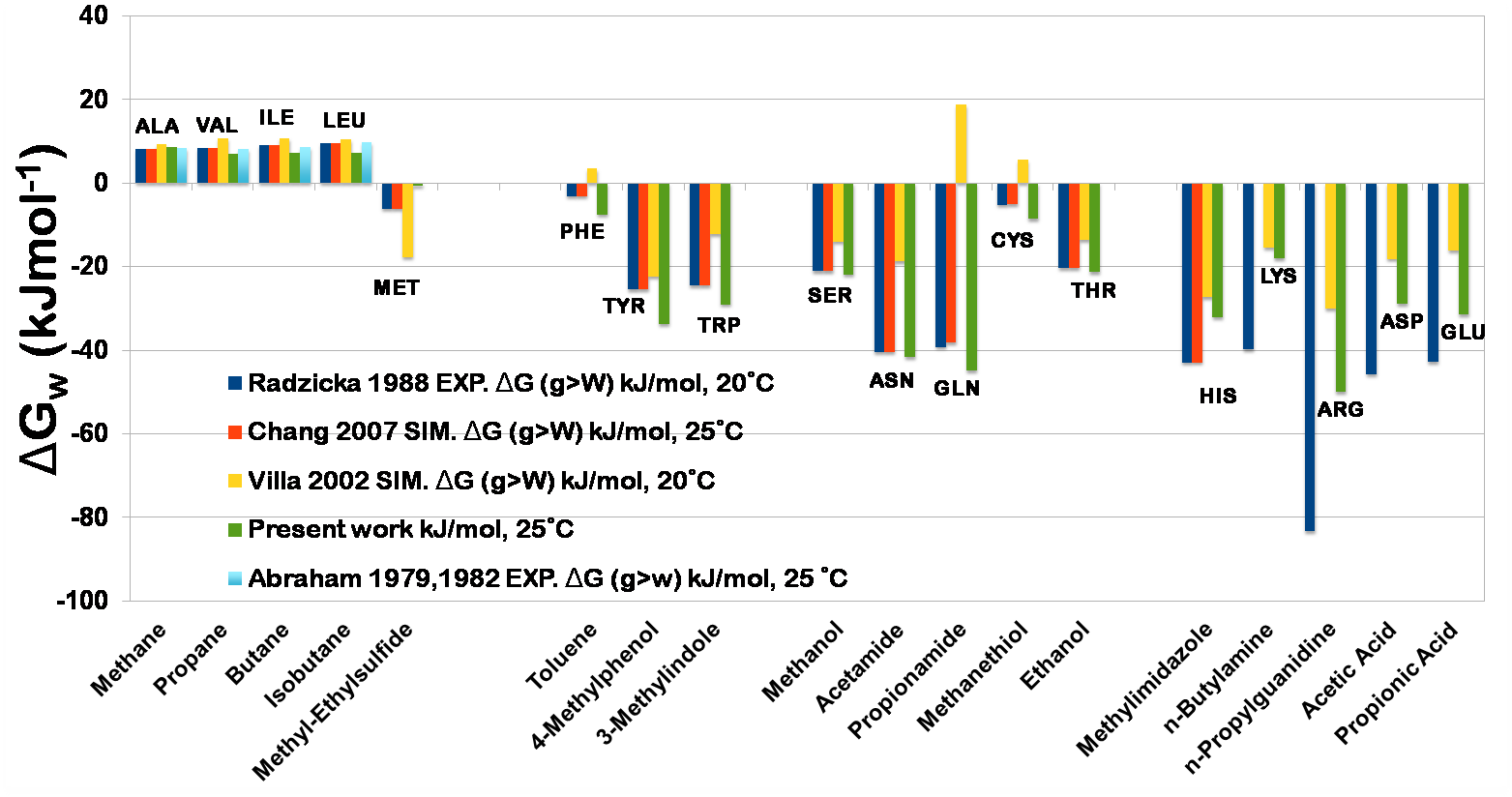}
    \caption{}\label{fig:fig3a}
  \end{subfigure}
  \begin{subfigure}{9cm}
     \includegraphics[width=\linewidth]{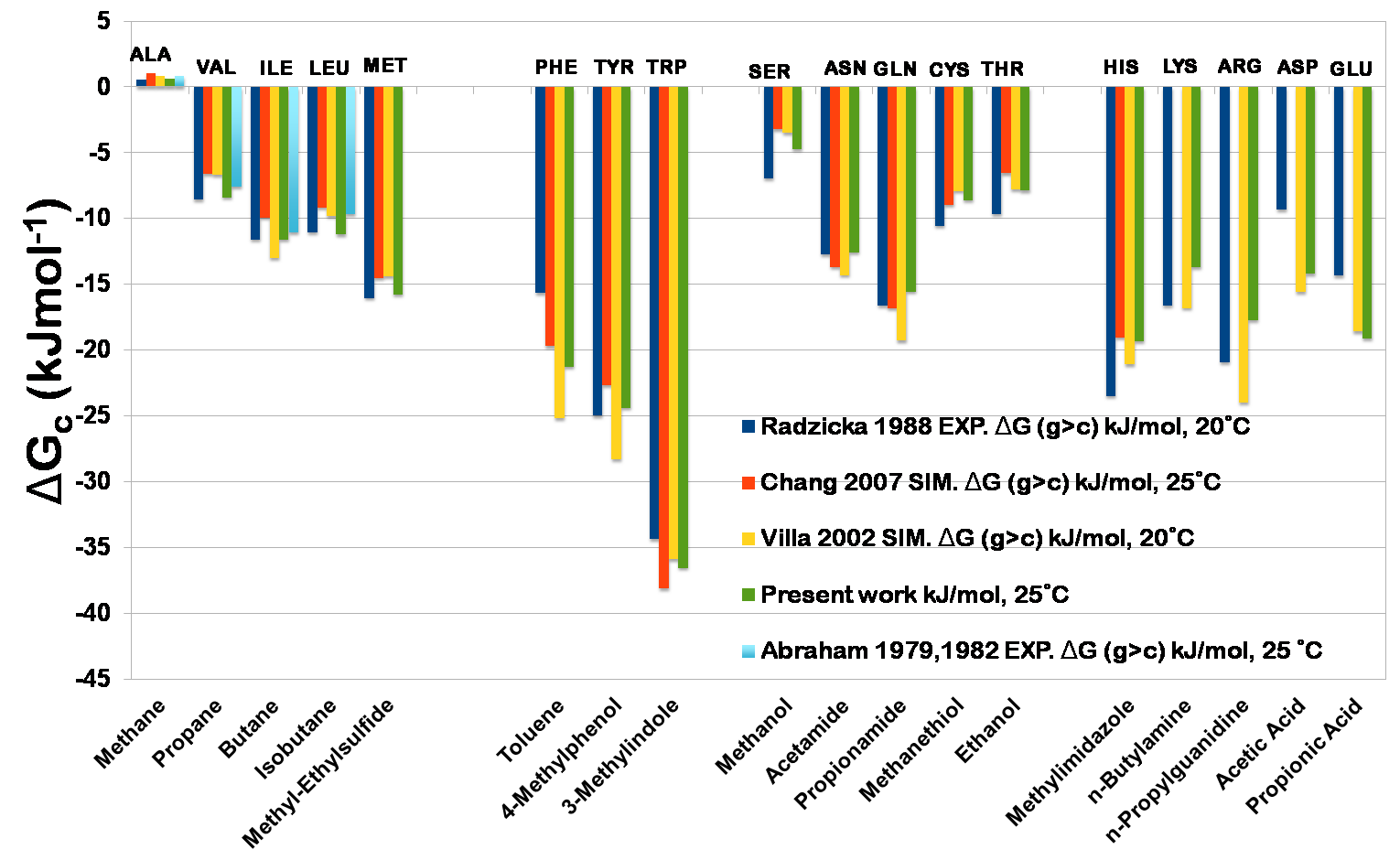}
    \caption{}\label{fig:fig3b}
  \end{subfigure}
  \begin{subfigure}{9cm}
     \includegraphics[width=\linewidth]{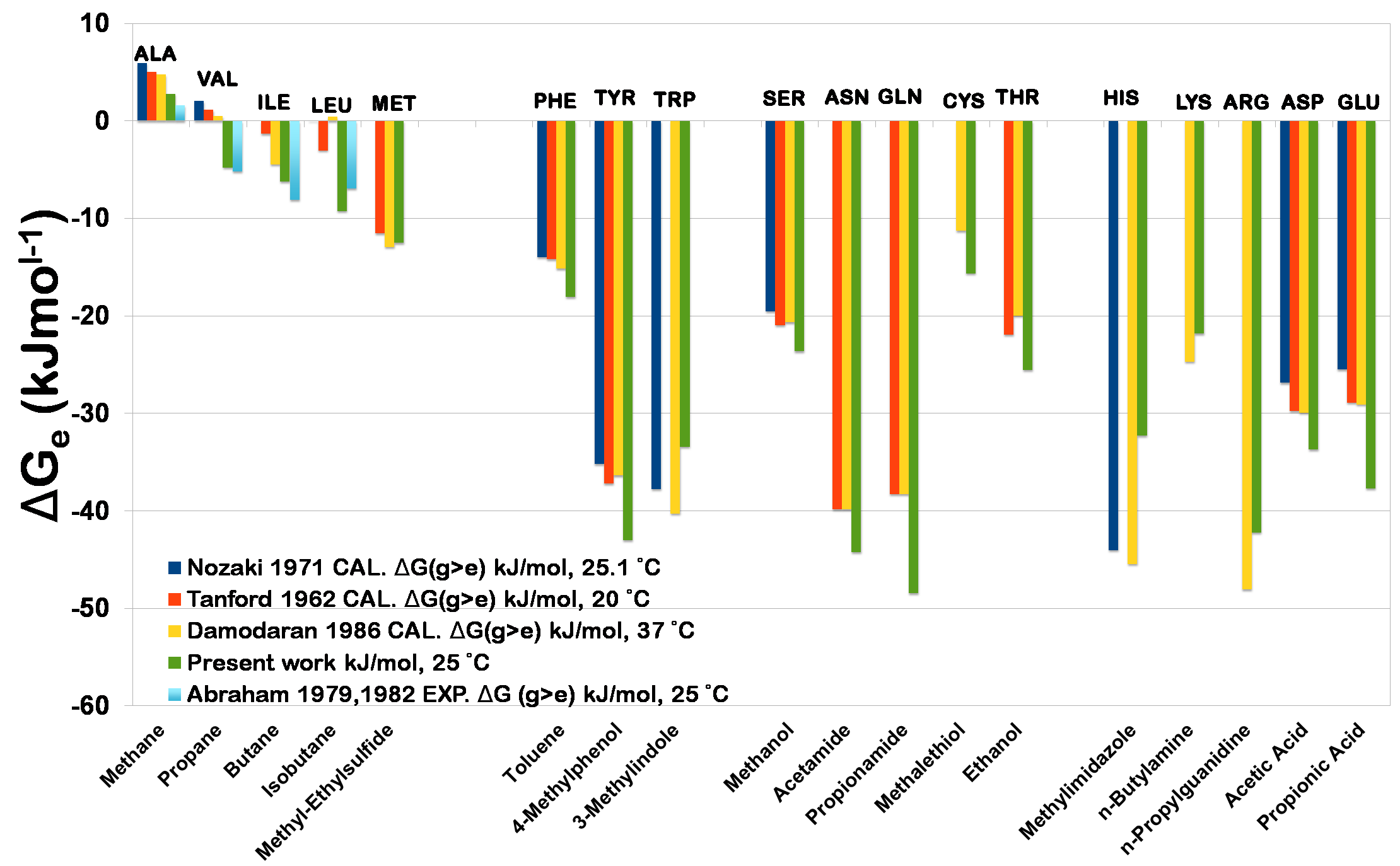}
    \caption{}\label{fig:fig3c}
  \end{subfigure}  
  \caption{(a) $\Delta G_{w} $ from vacuum to water \ce{H2O} {\color{black} at $25^{\circ}$C}; (b)  $\Delta G_{c}$ from vacuum to cyclohexane \ce{cC6H12}; (c) $\Delta G_{e}$ from vacuum to ethanol \ce{EtOH}. Results of the present work are also contrasted with past computational work of
    Refs. \cite{Villa02,Chang07}, as well as with experimental work of Ref. \cite{Radzicka88}. Each amino
    acid side chain equivalent is referred with the name ($x$-axis) and with the shorthand notations of the corresponding
    amino acid at the top of each value.
  \label{fig:fig3}}
\end{figure}

Figure \ref{fig:fig3} displays the solvation free energy for water \ce{H2O}
(Figure \ref{fig:fig3a}), cyclohexane  \ce{cC6H12} (Figure \ref{fig:fig3b}), and ethanol \ce{EtOH} (Figure \ref{fig:fig3c}),
and compares the results of the present work with both experimental and computational past works.
All corresponding values can be found in  \ref{tab:water}, \ref{tab:chex}, \ref{tab:etoh}.
Broadly speaking, the solvation free energies follow the general
division in hydrophobic and polar amino acids illustrated in Fig. \ref{fig:fig2}  -- note the sequence of the amino acids of Fig. \ref{fig:fig3} from left to right follow the same scheme hydrophobic $\to$ polar division of Fig. \ref{fig:fig2}, but there are exceptions. 
In water (Fig. \ref{fig:fig3a}), the $\Delta G_{w}$ values of polar amino acid side chain equivalents are negative, whereas they are positive for the hydrophobic methane (ALA), propane (VAL), butane (ILE) and isobutane (LEU), as largely expected. However, for MET, PHE, TYR and TRP we find  $\Delta G_{w}<0$. While odd at first sight, we note that this is in line with experimental data. For instance,  for toluene 
$\Delta G_{w} = -3.7$ \SI{}{\kilo\joule\per\mole} at $25^{\circ}$C and 1 atm, (see \ref{tab:spt}).
This means that aromatic side chains cannot be classified as purely hydrophobic, because they have favorable interactions with water molecules. 
This is a very interesting point. Indeed, it is known that benzene forms weak hydrogen bonds with two water molecules located over the two faces of the planar aromatic ring. In general, the partial positive charge of the hydrogen atom attached to very electronegative atoms (i.e. O or N) interact favorably with the delocalized $\pi$ electrons of the aromatic ring \cite{Suzuki92}. This is a specific example of a more general class of \ce{A-H . . . \phi}  hydrogen bonds, where $\phi$ represents an aromatic ring and A may be a N, O, or C atom \cite{Gierszal11}. In particular, this weak hydrogen bonds can form among water molecules and the aromatic side chains of PHE, TYR and TRP \cite{Gierszal11}. While the present forcefield was not devised to address this problem, it still proves instructive to test for this prediction in the case of the TRP amino acid side chain equivalent. This is reported in \ref{fig:hbonds} where for both water and ethanol we report the number of hydrogen bonds as a function of the simulation time at various stages of the decoupling process (i.e. different values of $\lambda$). 
\textcolor{black}{In the case of hydrophobic amino acid side chain equivalents our findings are in general good agreement with both experimental results \cite{Radzicka88}, as well as past computational results \cite{Villa02,Chang07}, with the exception of Methyl-ethylsulfide (MET) and Toluene (PHE). This could be ascribed to a more general difficulty in simulating the aromatic rings compared to the other acyclic compounds.}
By contrast, well grounded estimates were obtained for 3-methylindole (TRP) in agreement with past ones, thus supporting the reliability of the adopted force field.

In parallel to the case of water, our estimates of the solvation free energy in cyclohexane \ce{cC6H12} (Fig. \ref{fig:fig3b}) confirm the results that \textit{all} amino acid side chain equivalents have \textit{favourable} solvation free energy ($\Delta G_{c}<0$), with quantitative agreement with experimental data \cite{Radzicka88}  and previous computational investigations \cite{Villa02,Chang07}. As cyclohexane \ce{cC6H12} is a nonpolar liquid, unable to form hydrogen bonds, the negative $\Delta G_{c}$ values are due to the action of van der Waals attractions among the solute molecule and surrounding solvent molecules whose magnitude overcomes the free energy cost for creating the cavity. This is likely to be ascribed to the cyclohexane \ce{cC6H12} large molecular polarizability, a fundamental player of London dispersion interactions. This is confirmed by the finding that the largest $\vert \Delta G_{c} \vert$ value is found for 3-methylindole (TRP), that is the largest solute molecule in terms of surface area among those considered in this study. In general, the values $\Delta G_{c}$ obtained in the present work are closer to experimental data than previously calculated values (see Table \ref{tab:chex}).

The solvation free energy $\Delta G_{e}$ in ethanol \ce{EtOH} (see Fig. \ref{fig:fig3c}) is found negative for all amino acid side chain equivalents, but methane (ALA), paralleling the situation in cyclohexane \ce{cC6H12} (compare Figures  \ref{fig:fig3b} and \ref{fig:fig3c}), and in line with experimental data  (see Table \ref{tab:etoh}). We are not aware of any previous computational study providing estimates of $\Delta G_{e}$ for all the amino acid side chain equivalents considered here, so only limited comparisons can be carried out \cite{Nozaki71,Damodaran86,Tanford62}. Note that here the temperatures are also different. As a further remark, we stress that experimental $\Delta G_{e}$  by Nozaki and Tanford \cite{Nozaki71} were obtained by subtracting the $\Delta G_{e}$ value of GLY from those of the amino acids (i.e. including backbones) under an unwarranted additivity assumption. Indeed this assumption usually breaks down for very polar solutes, such as amino acids. Interestingly, even though ethanol \ce{EtOH} is a polar solvent able to form hydrogen bonds, its behavior resembles that of cyclohexane \ce{cC6H12}. This means that here too the attractive solute-solvent energetic interactions (accounting also for hydrogen bonds) are able to overcome the free energy cost of cavity creation. Not surprisingly, in fact, the largest  $\vert \Delta G_{e} \vert$ are associated with solutes, such as acetamide (ASN) and propionamide (GLN), able to be engaged in multiple hydrogen bonds with ethanol \ce{EtOH} molecules.

In this respect, it proves instructive to compare present findings with results that can be obtained from SPT and related theories, as discussed in Section \ref{subsec:scaled} \cite{Ashbaugh06,Graziano19}. Here solvation free energy can be estimated in all liquids as the sum of two contributions: (a) the reversible work $\Delta G_{0}$ to create a cavity in the liquid, suitable to host the solute molecule. This contribution is always positive so $\Delta G_{0}>0$ always; (b) the reversible work $E_{a}$ to turn on solute-solvent energetic attractions, both van der Waals interactions and hydrogen bonds. This second contribution can be considered purely energetic so that $E_{a}<0$ always. In other words, $E_{a}$  favors solvation, whereas $\Delta G_{0}$ opposes it. Reliable  estimates for $\Delta G_{0}$ of  simple geometric shapes are calculated by means of classic SPT. By assigning an effective hard sphere diameter to solvent molecules, and using the experimental density of the three liquids at 298 K and 1 atm, we find that  $\Delta G_{{0}_{w}} >  \Delta G_{{0}_{e}} > \Delta G_{{0}_{c}}$ for water \ce{H2O}, ethanol \ce{EtOH}, and cyclohexane \ce{cC6H12} in decreasing order, see \ref{fig:spt}. The ranking order can be easily rationalized by the fact that water molecules are the smallest (a cyclohexane molecule has a van der Waals volume roughly 5 times larger than that of a water molecule) and so liquid water is characterized by the largest number density, that in turn increases the entropy loss associated with cavity creation, due to the solvent-excluded volume effect. On the other hand, the energetic $E_a$ term consists of a van der Waals contribution, essentially of the same magnitude in the three liquids, and a hydrogen bond contribution, whose magnitude is large in water \ce{H2O}, slightly less in ethanol \ce{EtOH}, and zero in cyclohexane \ce{cC6H12}. Using this method sketched in Section \ref{subsec:scaled}, we estimated the solvation free energies for 
methane (ALA), propane (VAL), toluene (PHE) and methanol (SER) as $\Delta G_{{0}_{e}}+E_a$ and find them in agreement with experimental values in the three considered liquids, as shown in Table \ref{tab:spt}. Hence (a) in water \ce{H2O}, $\Delta G_{{0}_{w}} > \vert E_a \vert$ for aliphatic hydrocarbons, whereas the opposite holds true for aromatic hydrocarbons and polar molecules able to form hydrogen bonds with water molecules; (b) in cyclohexane \ce{cC6H12} and ethanol \ce{EtOH}, $\vert E_{a} \vert  > \Delta G_{0_{e,c}}$ for all the amino acids side chain equivalents but methane (ALA), because the free energy cost of cavity creation is not so large.

\subsection{Transfer free energies between solvents}
\label{subsec:transfer}
\begin{figure}[htpb]
\centering
 \captionsetup{justification=raggedright,width=\linewidth}
  \begin{subfigure}{9cm}
     \includegraphics[width=\linewidth]{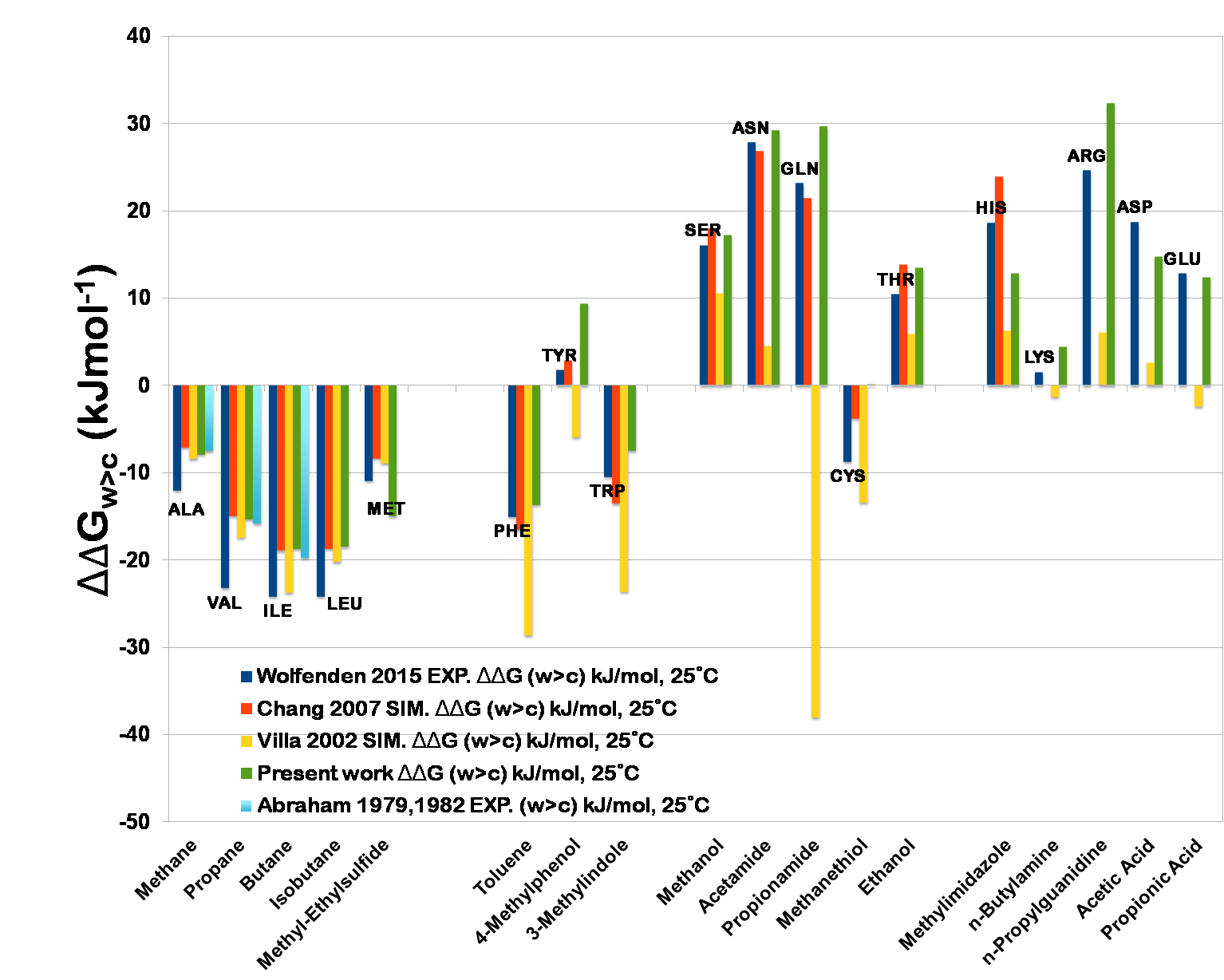}
    \caption{}\label{fig:fig4a}
  \end{subfigure}
  \begin{subfigure}{9cm}
     \includegraphics[width=\linewidth]{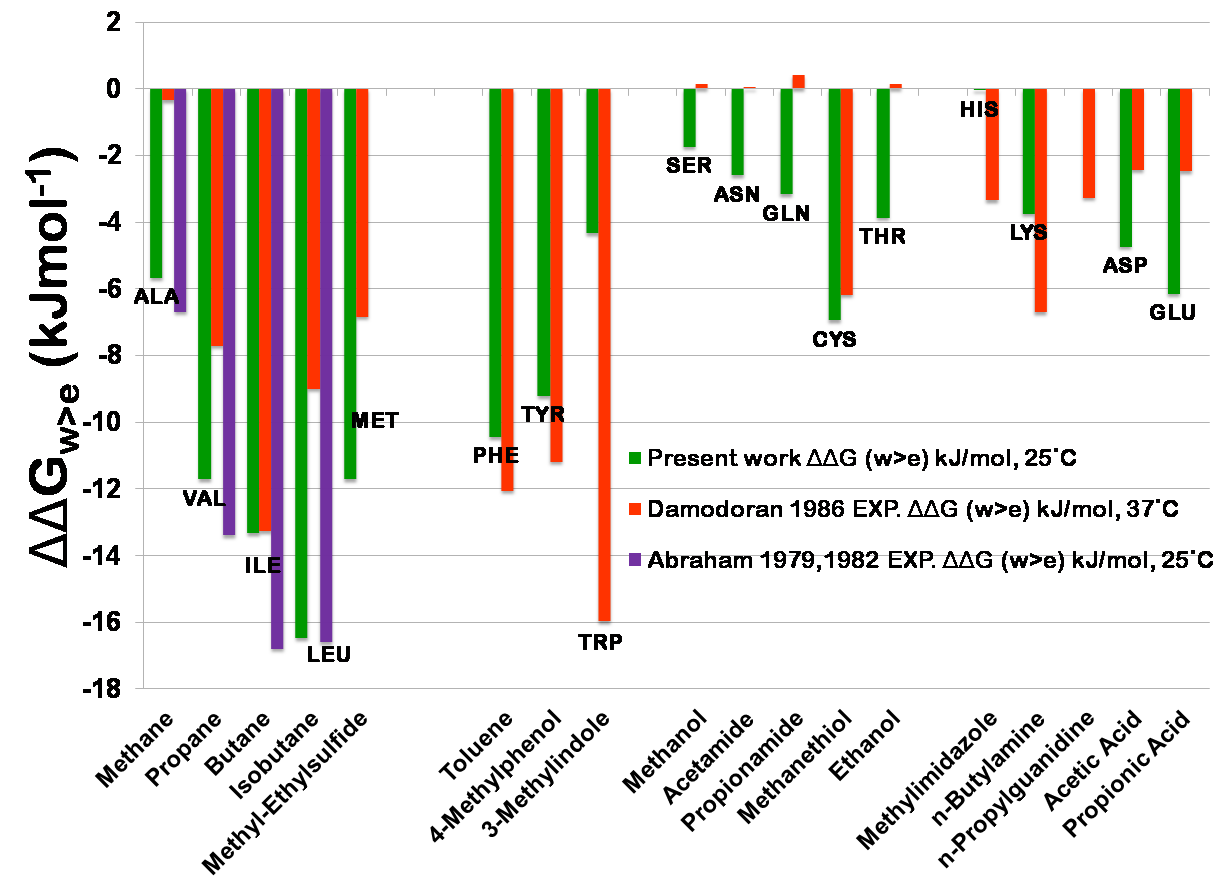}
    \caption{}\label{fig:fig4b}
  \end{subfigure}
  \caption{(a) $\Delta \Delta G_{w>c}$ from water \ce{H2O} to cyclohexane \ce{cC6H12} ; (b)  $\Delta \Delta G_{w>e}$ from water \ce{H2O} to ethanol \ce{EtOH} . Results of the present work are also contrasted with past computational work of
    Refs. \cite{Villa02,Chang07}, as well as with experimental works of Refs. \cite{Wolfenden15} (water \ce{H2O}  to cyclohexane \ce{cC6H12} ) and \cite{Damodaran86} (water \ce{H2O}  to ethanol \ce{EtOH}). Each amino
    acid side chain equivalent is referred with the name ($x$-axis) and with the shorthand notations of the corresponding
    amino acid at the top of each value.
  \label{fig:fig4}}
\end{figure}

Additional insights can be achieved by computing the change in the solvation free energy between different solvents. We shall refer to them as the transfer free energies in the following. This is shown in Fig. \ref{fig:fig4}. Figure \ref{fig:fig4a} depicts our results for the free energy transfer $\Delta \Delta G_{w>c} \equiv \Delta \Delta G (\text{\ce{H2O}} \to \text{\ce{cC6H12}}) $ from water \ce{H2O} to cyclohexane \ce{cC6H12} and contrast them with the past simulations \cite{Villa02,Chang07} and experiments \cite{Wolfenden15}. Rather evidently, here all hydrophobic amino acid side chain equivalents (except TYR) have
$\Delta \Delta G_{w>c}<0$ indicating their increased propensities in being solvated by a non-polar solvent such as cyclohexane \ce{cC6H12} rather than a polar solvent such water \ce{H2O}. Likewise, we find that $\Delta \Delta G_{w>c}>0$ for all polar amino acid side chain equivalents indicating their decreased propensities in being solvated by cyclohexane \ce{cC6H12} rather than water \ce{H2O}. 

The present values are quantitatively close to experimental data \cite{Wolfenden15} and perform better than previously calculated estimates \cite{Villa02,Chang07}. For instance, $\Delta \Delta G_{w>c}$ for propionamide (GLN) is positive and in line with the experimental value, whereas a previously numerical estimate was negative (see Fig. \ref{fig:fig4a}). Remarkably, $\Delta \Delta G_{w>c}$  almost perfectly divides polar amino acid side chain equivalents from hydrophobic ones, thus prompting the possibility of being used as a correct measure of hydrophobicity for amino acid side chains, as claimed a long time ago by Wolfenden \cite{Wolfenden81}. This finding is also an indication that cyclohexane \ce{cC6H12}, such as other non-polar organic liquids, does not act as a denaturant of the folded state of globular proteins, in agreement with experimental evidence \cite{Pace04}.

The values of the transfer free energy $\Delta \Delta G_{w>e} \equiv \Delta \Delta G (\text{\ce{H2O}} \to \text{\ce{EtOH}}) $  from water \ce{H2O} to ethanol \ce{EtOH} are shown in Fig. \ref{fig:fig4b}. They are negative for all amino acid side chain equivalents, regardless of their polarity.
This is in line with available experimental data \cite{Damodaran86}, even though the latter are very small positive for methanol (SER), acetamide (ASN) propionamide (GLN), and ethanol (THR). In particular, the calculated $\Delta \Delta G_{w>e}$ values for methane (ALA), propane (VAL), butane (ILE) and isobutane (LEU) are negative and fully consistent with experimental data (see Table \ref{tab:spt}).

Once more, it proves instructive to contrast the above findings with theoretical results stemming from the SPT analysis of Section \ref{subsec:scaled}. Here, we can build on the fact that the reversible work of cavity creation $\Delta G_{0_{w}}$ in water \ce{H2O} is larger than its counterpart $\Delta G_{0_{e}}$ in ethanol \ce{EtOH}, that is $\Delta G_{0_{w}} > \Delta G_{0_{e}}$
(see Table \ref{tab:spt}). By contrast, the reversible work of turning on solute-solvent attractions, in water $E_{a_{w}}$ \ce{H2O} is approximately equal to its counterpart $E_{a_{e}}$ in ethanol, ($E_{a_{w}} \approx E_{a_{e}}$), so that
$\Delta G_{0_{w}} + E_{a_{w}} > \Delta G_{0_{e}} +E_{a_{e}}$, thus predicting $\Delta \Delta G_{w>e} <0$ in agreement with the above numerical results. 
The fact that the $\Delta \Delta G_{w>e}$  is negative for almost all side chains indicates that: (a) it cannot be a correct measure of hydrophobicity; (b) ethanol has a denaturing action towards the folded state of globular proteins as confirmed by experimental studies \cite{Nozaki71}.
\subsection{Entropy-enthalpy compensation}
\label{subsec:compensation}
{\color{black} While solvation free energy is certainly the most insightful quantity for understanding a molecule's interaction with a solvent, a deeper understanding can be achieved by singling out its entropy and enthalpy components. In this case, experiments struggle to provide a detailed description and theoretical and numerical simulations prove to be very effective. In this framework, an useful approach is provided by the grid cell theory \cite{Gerogikas14} that is a refined version of partition function methods \cite{Irudayam10}. It is then of considerable interest to ask how the present study can contribute to this issue. }

\begin{figure}[htpb]
\centering
 \captionsetup{justification=raggedright,width=\linewidth}
  \begin{subfigure}{9cm}
     \includegraphics[width=\linewidth]{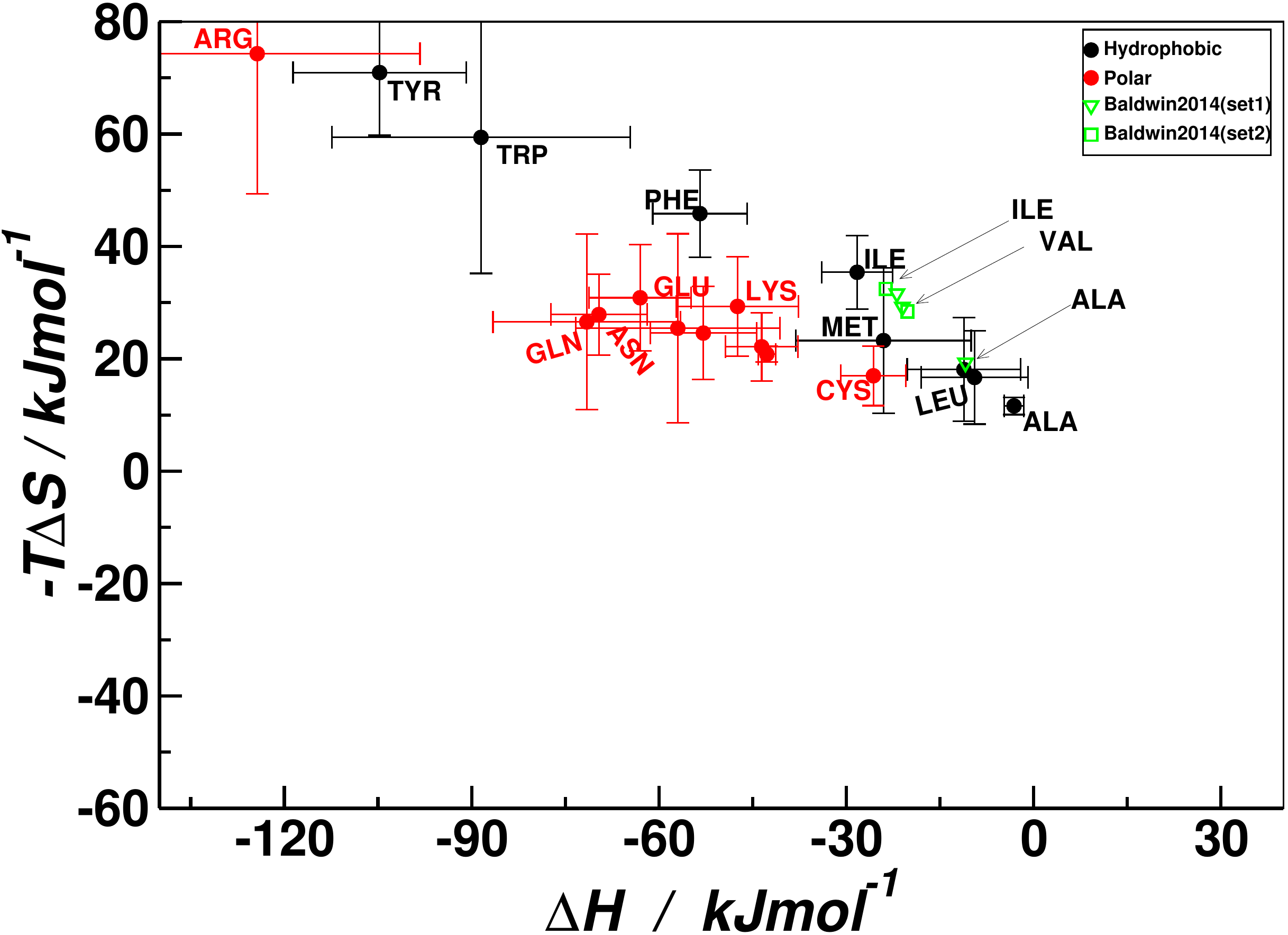}
    \caption{}\label{fig:fig5a}
  \end{subfigure}
  \begin{subfigure}{9cm}
     \includegraphics[width=\linewidth]{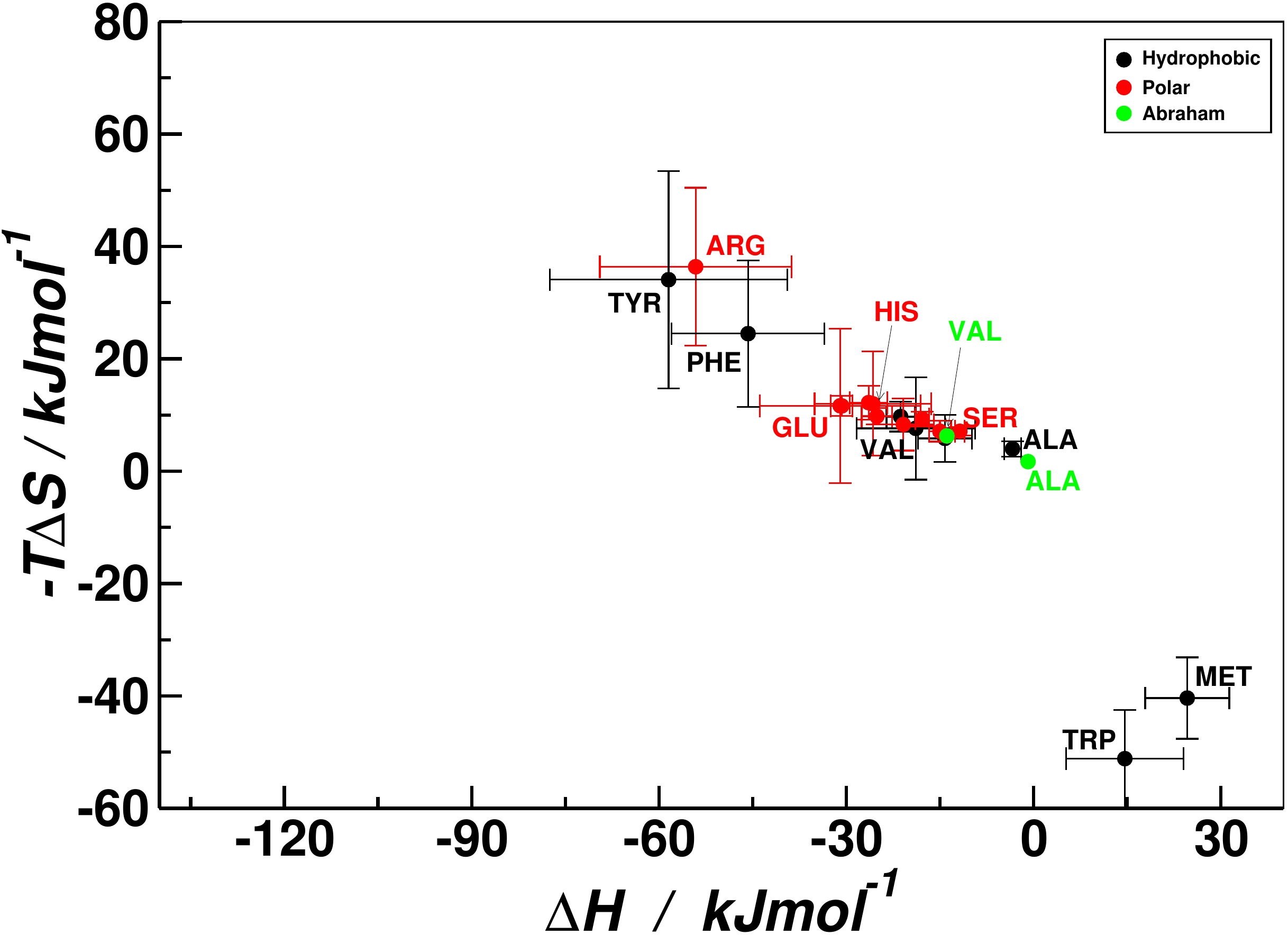}
    \caption{}\label{fig:fig5b}
  \end{subfigure}
  \begin{subfigure}{9cm}
     \includegraphics[width=\linewidth]{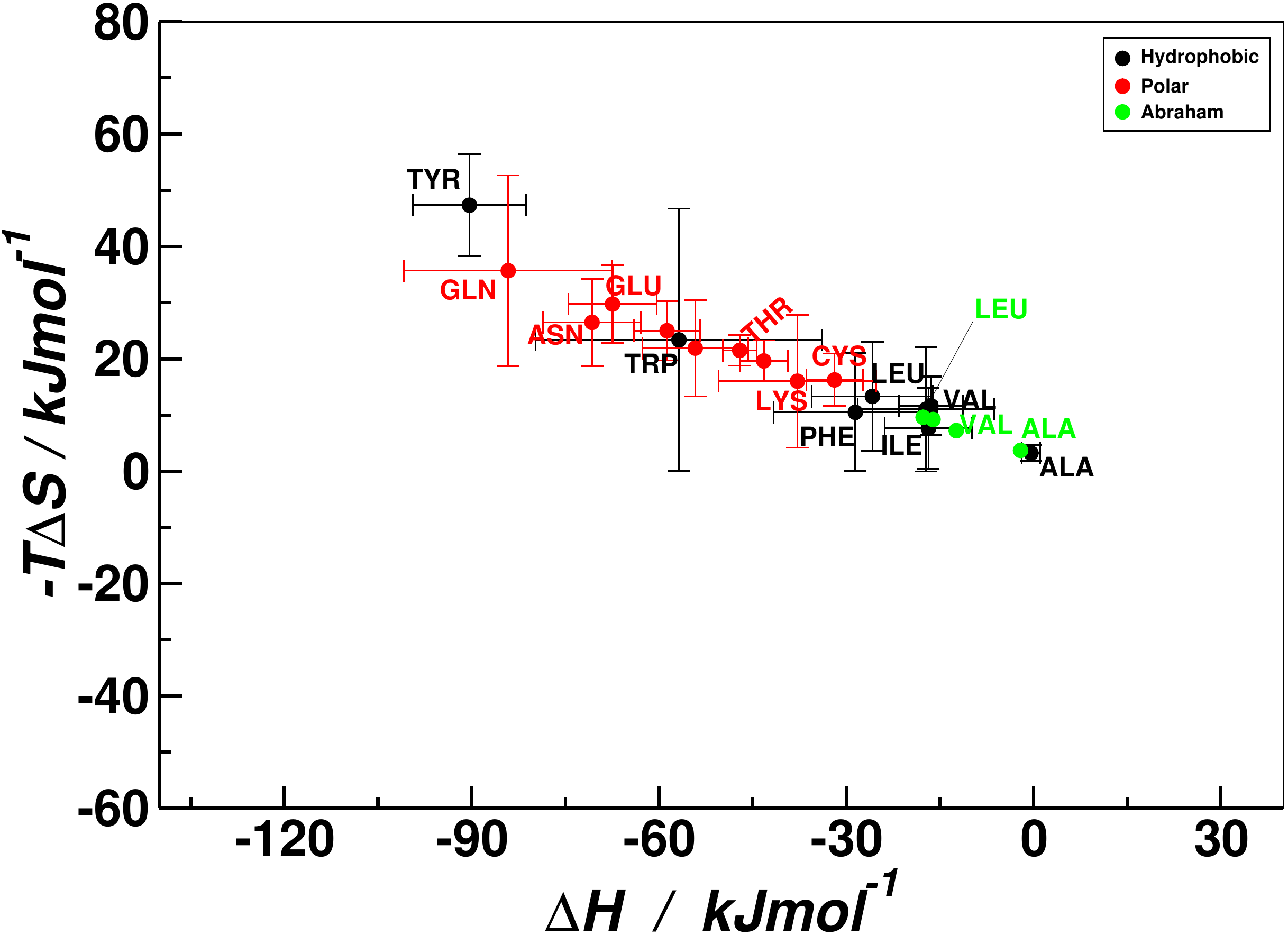}
    \caption{}\label{fig:fig5c}
  \end{subfigure}
  \caption{Entropic part $-T \Delta S$ of the solvation free energy $\Delta G$ as a function of the enthalpic part $\Delta H$ in the case of  (a) water \ce{H2O}; (b) cyclohexane \ce{cC6H12}; (c) ethanol \ce{EtOH}. In case of water \ce{H2O}, corresponding experimental results \cite{Baldwin14} are also included. Note that all plots are in the same scale.
  \label{fig:fig5}}
\end{figure}

As anticipated in Section \ref{subsec:thermodynamic} we follow the work of Schauperl et al.\cite{Schauperl16}, to separate the free energy of solvation in its enthalpic and entropic parts. 
To this aim, we compute the solvation free energy at seven different temperatures in the range $270-330\SI{}{\kelvin}$ and then used Eq.(\ref{sec2:eq4}) to fit the parameters $a$, $b$, and $c$ and hence obtain $\Delta S(T)$ from  Eq.(\ref{sec2:eq5}). All details on these calculations can be found in \ref{tab:fit_coefs}, \ref{fig:water}, \ref{fig:chex}, and \ref{fig:etoh}. Fig.\ref{fig:fig5} then reports the entropic term $-T\Delta S$ of the solvation free energy $\Delta G_{solv}$ as a function of the enthalpic term $\Delta H$ for different solvents: water \ce{H2O} (Fig. \ref{fig:fig5a}), cyclohexane \ce{cC6H12} (Fig. \ref{fig:fig5b}) and ethanol \ce{EtOH} (Fig. \ref{fig:fig5c}).

Visual inspection of the plots in Fig. \ref{fig:water}, \ref{fig:chex}, and \ref{fig:etoh} indicates that: (a) For water \ce{H2O} $\Delta G_{w} (T)$  is an increasing function of temperature for all amino acid side chain equivalents and so the calculated hydration entropy change is always negative, regardless of the solute polarity, in line with experimental data \cite{Graziano09} (this is a further support of the reliability of the adopted force fields and calculation procedure); (b) For cyclohexane \ce{cC6H12}  $\Delta G_{c} (T)$ does not have a temperature dependence common to all amino acid side chain equivalents; (c) For ethanol \ce{EtOH} $\Delta G_{e} (T)$  is an increasing function of temperature for almost all amino acid side chain equivalents, closely resembling the situation for water. A quantitative analysis performed using  Eqs.(\ref{sec2:eq4}) and Eqs.(\ref{sec2:eq5}) leads to the calculated solvation enthalpy and entropy values at $\SI{298.15}{\kelvin}$ listed in Table \ref{tab:water_solv},
Table \ref{tab:chex_solv}, and Table \ref{tab:etoh_solv}. These values have been used to build up plots of 
$-T \Delta S$ versus $\Delta H$ for all amino acid side chain equivalents in the three liquids, as displayed in Figure \ref{fig:fig5}.

The general expectation is that in water a large and negative $\Delta H$ term (i.e., strong solute-solvent energetic attractions) is associated with a large and positive $-T \Delta S$ term (i.e., a decrease of entropy). In other words, an enthalpy gain leads to an entropy loss, and a correlation with a negative slope emerges. This feature is commonly denoted as ' entropy-enthalpy compensation'.
This is indeed confirmed by our results reported in Fig.\ref{fig:fig5a}.
While qualitatively similar in the three solvents,  the entropy-enthalpy compensation is quantitatively much more  relevant in water \ce{H2O} as shown in Fig. \ref{fig:fig5a}. For instance, for toluene (PHE)  (a)  In water \ce{H2O} $ \Delta H = -53  \SI{}{\kilo\joule\per\mole}$  and $-T\Delta S = 46 \SI{}{\kilo\joule\per\mole}$; (b) In cyclohexane \ce{cC6H12} $\Delta H = -46  \SI{}{\kilo\joule\per\mole}$ and $-T\Delta S = 24 \SI{}{\kilo\joule\per\mole}$ ; (c) In ethanol \ce{EtOH}  $\Delta H = -29  \SI{}{\kilo\joule\per\mole}$  and $-T\Delta S = 10  \SI{}{\kilo\joule\per\mole}$. The large difference among the three liquids is mainly due to the structural reorganization of solvent molecules upon solute insertion, that should provide positive contributions to both the solvation enthalpy and entropy changes. This structural reorganization can be correlated to the isobaric thermal expansion coefficient $\alpha_{p}$ of the liquid. Indeed at room temperature $\SI{298}{\kelvin}$ and 1 atm, $\alpha_{p}$ is very small in water, but large in organic liquids ($\alpha_{p}  = 0.257\times 10^{-3} $ for water \ce{H2O}, $1.214 \times 10^{-3}$ for cyclohexane \ce{cC6H12}, and $1.089 \times 10^{-3} $ for ethanol \ce{EtOH} in $\SI{}{\kelvin} ^{-1}$) \cite{Lide04}, giving a simple explanation of the different magnitude of such structural reorganization in the three solvents. The net distinction between polar and hydrophobic amino acid side chain equivalents occurring in ethanol \ce{EtOH} is noteworthy and in striking contrast with the lack of such a separation in cyclohexane \ce{cC6H12} (compare Figs. \ref{fig:fig5b} and \ref{fig:fig5c}). Unfortunately, a detailed comparison with experimental or computational results is not possible due to a lack of such data for most of the amino acid side chain equivalents considered in the present study. 
As expected, in water \ce{H2O} (Fig. \ref{fig:fig5a})  polar amino acid side chain equivalents have large and negative $\Delta H$, mainly induced by the possibility of forming hydrogen bonds with water, compensated by an equally large and positive $-T\Delta S$ stemming from the solvent entropy reduction in the water cage \cite{Frank45} around a polar solute. Note that this trend is particularly emphasized in the case of n-propylguanidine (ARG) and much more reduced in Methanethiol (CYS) for which this enthalpy-entropy compensation mechanism is very limited. By contrast, hydrophobic amino acid side chain equivalents cannot form hydrogen bonds with water and hence do not trigger the re-orientation of the first solvation shell of water around them. As a consequence, they are nearly unaffected in terms of enthalpy change and show a small increase in $-T\Delta S$ originating by the entropy loss of forming the cavity for accommodating the solute (that is always present). The obtained results in the case of Methane (ALA), Propane (VAL), and Butane (ILE) are in reasonable agreement with calorimetric measurements \cite{Baldwin14}. Here too, 4-methylphenol (TRP) and toluene (PHE) appear to be outliers with large and negative $\Delta H$ and large and positive $-T\Delta S$. We remark that the results presented here are different from those appearing in the analogue plot of Schauperl et al.\cite{Schauperl16} (see their Figure 5) because their amino acids include the backbone that in our case is represented by a single hydrogen. Their analysis further show the sensibility of the obtained results with respect to a change of the water model, so care must be exercised in using them for quantitative comparisons. The slope of the negative correlation in water, however, appears to be $\approx 0.5$ consistent with them.

Fig. \ref{fig:fig5b} provides the same plot in cyclohexane \ce{cC6H12}. Here it is rather evident that we do not obtain the same results by inverting the role of polar and hydrophobic amino acid side chain equivalents. While a negative slope is overall visible, most of the amino acid side chain equivalents tend to cluster in a region of small enthalpic gain $\Delta H \approx $ \SI{-30}{\kilo\joule\per\mole} and neglegible entropic loss $-T\Delta S \approx 0$. Intriguingly, this appears to be \textit{independent} of the polar character of the amino acid side chain equivalents, as both polar and hydrophobic molecules belong to this cluster. There are however outliers in both senses. Hydrophobic 4-Methylphenol (TYR)  and polar n-Propylguanidine (ARG) show a much more marked enthalpy gain  with $\Delta H \approx $ \SI{-60}{\kilo\joule\per\mole} compensated by a significant entropy loss $ -T \Delta S \approx $ \SI{30}{\kilo\joule\per\mole}. At the opposite side, hydrophobic 3-Methylindole (TRP) and Methyl-ethylsulfide (MET)
present a significant entropic gain  $ -T \Delta S \approx $ \SI{-40}{\kilo\joule\per\mole} compensated by a corresponding enthalpy loss $\Delta H \approx $ \SI{20}{\kilo\joule\per\mole}.

Somewhat surprisingly, but in line with the discussion presented in previous Sections, the results in the case of ethanol \ce{EtOH} appear to be the cleanest ones, as reported in Fig. \ref{fig:fig5c}. Here a nearly perfect negative correlation is found, with all polar amino acid side chain equivalents gaining in enthalpy and losing in entropy upon being solvated.
This can be rationalized by recalling that in all cases a negative enthalpy of solvation, which provides a favourable contribution to the free energy is compensated by a positive solvation entropy.
This is due to the fact that for the insertion of a solute inside a solvent a cavity must be created and the solvent molecules have to rearrange themselves around it, independently of their polarity. Polar solutes however can form hydrogen bonds in water \ce{H2O} and ethanol \ce{EtOH} so the gain in enthalpy is sufficient to compensate this entropy loss.  On the other hand, the reverse is not true in cyclohexane \ce{cC6H12} where there is no general tendency of the hydrophobic amino acid side chain equivalent to display a favourable solvation in \ce{cC6H12} compared to the polar ones. 

Finally, even in this case it is useful to perform the same calculation by transfering the molecules from water to either cyclohexane or ethanol. This is reported in  Fig. \ref{fig:compensation_transfer}(a) for the case from water to cyclohexane and in  Fig. \ref{fig:compensation_transfer}(b) for the case from water to ethanol. In the case of water to cyclohexane (Fig. \ref{fig:compensation_transfer} (a)), experimental findings \cite{Wolfenden15} are also included for comparison. Irrespective of the polar nature of the amino acid side chain equivalent, Fig. \ref{fig:compensation_transfer}(b) generally shows a significant entropy gain with large variations from one amino acid side chain equivalent to another, and an equivalently large enthalpy loss especially for polar molecules, as expected. These results are also in reasonable agreement with experimental findings  \cite{Wolfenden15}.
A similar trend is also visible in case from water to ethanol, but in this case a splitting of the cluster of hydrophobic molecules from the cluster of polar ones is very appreciable, in line with the results of Fig. \ref{fig:fig5c}.


\section{Conclusions}
\label{sec:conclusions}
In this paper we have addressed the following issue. Imagine to have two molecules, one polar and one hydrophobic, and two solvents, one polar and one non-polar (hydrophobic). In water we expect a negative solvation free energy for the polar molecule, indicating the propensity of the molecule for being hydrated, as well as a weakly positive solvation free energy for the hydrophobic one, because in this case there is no gain in being hydrated. What happens in a non-polar (hydrophobic) solvent? Is the opposite true? To this aim we have performed detailed thermodynamics integration calculations to compute the solvation free energy of 18 amino acid side chain equivalents in water, cyclohexane and ethanol, the latter representing an intermediate case between a paradigmatic polar solvent such as water, and an equally  paradigmatic non-polar solvent such as cyclohexane. Our findings strongly suggest the answer to the above question to be negative, as we did not find any indication of a symmetry between the two cases.  We ascribed this to the different interactions, polar-polar in the case of polar amino acid side chain equivalents in water, van der Waals/quadrupolar in the case of hydropbobic amino acid side chain equivalents in cyclohexane. While these numerical simulations are notoriously difficult and very sensible to the details of the used force fields, we believe that our evidence is sufficient in view of the reasonable agreement with past available results, to make the above statement relatively sound. By repeating the calculations at different temperatures, we have been able to discriminate between the entropic and the enthalpic contributions. In water we found in all cases an entropy-enthalpy compensation, albeit with some unexpected and intriguing anomalies, in agreement with our expectations and past literature. No such compensation appears in the case of cyclohexane, thus supporting the above claim. Remarkably, a cleaner trend with no anomalies is found in the case of ethanol, with the hydrophobic and polar amino acid side chain equivalents arranging in two clearly separated clusters.

Our findings provide new insights on the biological role and the detailed mechanism of the hydrophobic effect, that is known to play a fundamental role in essentially all biological processes. In addition, they also suggest the possibility of defining a robust scheme to identify the relative polarities of the natural amino acids, thus rationalizing the zoo of different scales of hydrophobicity that have been proposed in the literature.

\section*{Conflicts of interest}
There are no conflicts to declare.

\section*{Acknowledgements}
We are indebted to Paul Dupire and Emanuele Petretto for their help at the initial stage of the project. The use of the SCSCF multiprocessor cluster at  the Universit\`{a} Ca' Foscari Venezia and of the high performance computer Talapas  at the University of Oregon is gratefully acknowledged. We also acknowledge the CINECA
projects HP10CYJPYK and HP10CGFUDT for the availability of high performance computing resources through the ISCRA initiative. The work was supported by MIUR PRIN-COFIN2017 \textit{Soft Adaptive Networks} grant 2017Z55KCW (A.G), Marie Sklodowska-Curie Fellowship No. 894784-EMPHABIOSIS and a Knight Chair to Prof. Jayanth Banavar at University of Oregon (T.S), and Erasmus mobility program (M.H). The authors would like to acknowledge networking support by the COST Action CA17139.


%


\providecommand*{\mcitethebibliography}{\thebibliography}
\csname @ifundefined\endcsname{endmcitethebibliography}
{\let\endmcitethebibliography\endthebibliography}{}

\onecolumngrid
\appendix

\section{Appendixes}

Free energy differences between initial and final states can be computed using Eq.\ref{TI_eq} below :
\begin{eqnarray}
  \label{TI_eq}
  \Delta G_{AB} &=& \int_{\lambda_{A}}^{\lambda_{B}} d\lambda \left \langle \frac{\partial V\left(\mathbf{r};\lambda\right)}{\partial \lambda} \right \rangle_{\lambda}
\end{eqnarray}
where V(\textbf{r}, $\lambda$) is the potential energy of the system as a function of
the coordinate vector \textbf{r}, and $\lambda$ is a switching-on parameter allowing to go from state A to state B by changing its value from $\lambda_A$ to $\lambda_B$.

The $\lambda$-dependence of the potential in bonded interaction is linear while non-bonded
interaction can be described with linear dependence or with Soft-core interaction. It should be noted that in our simulations we are analyzing only small molecules, so we are only interested in turning off the inter-molecular interactions such as Lennard-Jones and Coulomb potentials. We used the standard linear interpolation shown in Eq.\ref{eq2} 
\begin{eqnarray}
\begin{split}
  \label{eq2}
  V&=(1-\lambda)V^A+\lambda V^B\\
  \frac{\partial V}{\partial \lambda}& = V^B - V^A
  \end{split}
\end{eqnarray}
However, near off-states i.e. for values of $\lambda$ equal to 0 and 1 large numerical fluctuations are sometimes recorded leading to clashes between decoupling atoms, thereby preventing a smooth derivative of the potential in Eq.\ref{eq2}. A core sotftening  (Eq.\ref{eq3}) interacting potential was used to circumvent this issue 

\begin{eqnarray}
\begin{split}
  \label{eq3}
  V_{soft-core}(r)&=(1-\lambda)V^A(r_A)+\lambda V^B(r_B)\\
  r_A& = (\alpha R^6_A\lambda ^p +r^6)^{1/6}\\
  r_B& = (\alpha R^6_B(1-\lambda) ^p +r^6)^{1/6}
  \end{split}
\end{eqnarray}
where $\lambda$ and \textit{p} are respectively the soft-core and the soft-core power parameters, and \textit{R} is the interaction radius, which is equal to the ratio between the Lennard-Jones parameters $\sigma_{ij}$.

\section{Supplementary Tables}
\begin{table}[h]
\caption{The correspondence between the 20 amino acids and their neutral analog equivalent.}
  \label{tab:equivalent}
  \begin{center}
    \begin{tabular}{l r c c r}
      \hline
      \text{Character} & \text{Amino acid} & \text{Short name} & \text{Single letter} & \text{Equivalent } \\
      \hline
      \text{Hydrophobic} & \text{Alanine} & \text{ALA} & \text{A} & \text{Methane}\\
      \text{Hydrophobic} & \text{Valine} & \text{VAL} & \text{V} & \text{Propane}\\
      \text{Hydrophobic} & \text{Isoleucine} & \text{ILE} & \text{I} & \text{Butane}\\
      \text{Hydrophobic} & \text{Leucine} & \text{LEU} & \text{L} & \text{Isobutane}\\
      \text{Hydrophobic} & \text{Methionine} & \text{MET} & \text{M} & \text{Methyl-ethylsulfide}\\
      \text{Hydrophobic} & \text{Glycine} & \text{GLY} & \text{G} & \text{Hydrogen}\\
      \text{Hydrophobic} & \text{Phenylalanine} & \text{PHE} & \text{F} & \text{Toluene}\\
      \text{Hydrophobic} & \text{Tyrosine} & \text{TYR} & \text{Y} & \text{4-Methylphenol}\\
      \text{Hydrophobic} & \text{Tryptophan} & \text{TRP} & \text{W} & \text{3-Methylindole}\\
      \text{Polar} & \text{Serine} & \text{SER} & \text{S} & \text{Methanol}\\
      \text{Polar} & \text{Asparagine} & \text{ASN} & \text{N} & \text{Acetamide}\\
      \text{Polar} & \text{Glutamine} & \text{GLN} & \text{Q} & \text{Propionamide}\\
      \text{Polar} & \text{Cysteine} & \text{CYS} & \text{C} & \text{Methanethiol}\\
      \text{Polar} & \text{Threonine} & \text{THR} & \text{T} & \text{Ethanol}\\
      \text{Polar} & \text{Histidine} & \text{HIS} & \text{H} & \text{Methylimidazole}\\
      \text{Polar} & \text{Lysine} & \text{LYS} & \text{K} & \text{n-Butylamine}\\
      \text{Polar} & \text{Arginine} & \text{ARG} & \text{R} & \text{n-Propylguanidine}\\
      \text{Polar} & \text{Aspartic acid} & \text{ASP} & \text{D} & \text{Acetic Acid}\\
      \text{Polar} & \text{Glutamic acid} & \text{GLU} & \text{E} & \text{Propionic Acid}\\
      \text{-} & \text{Proline} & \text{PRO} & \text{P} & \text{-}\\
   \hline
   \end{tabular}
  \end{center}
\end{table}
\begin{table}[h]
  \centering
  \caption{Solvation free energies  (\SI{}{\kilo\joule\per\mole}) for hydrophobic and polar amino acid side chain analogs in water \ce{H2O}.}
  \label{tab:water}
   \begin{threeparttable}
    \begin{tabular}{l l r r r}
      \hline
      \text{Hydrophobic} & \text{This work, 25$^\circ$C} & Ref.\tnote{a} & Ref.\tnote{b} & Ref.\tnote{c} \\
      \hline
      Methane(Ala) & $8.47 \pm 0.12$ & $9.20$ & $8.12$ & $8.12$\\
      Propane(Val) & $6.93 \pm 0.50$ &$10.70$ & $8.33$ & $8.33$ \\
      Butane (Ile) & $7.11 \pm 1.83$ & $10.70$ & $9.00$ & $9.00$ \\
      Isobutane (Leu) & $7.24 \pm 1.34$ & $10.40$ & $9.54$ & $9.55$ \\
      Methyl-ethylsulfide (Met) & $-0.80 \pm 1.69$ & $-6.19$ & $-14.52$ & $-6.20$ \\ 
      3-methylindole (Trp) & $-29.09 \pm 2.34$ & $-12.30$ & $-24.60$ & $-24.62$ \\
      4-methylphenol (Tyr) & $-33.79\pm 3.04$ & $-22.40$ & $-25.56$ & $-25.58$ \\
      Toluene (Phe)        & $-7.62 \pm 1.12$ &  $-3.40$ & $-3.18$ & $-3.18$           \\
      \hline
      \text{Polar} & \text{This work, 25$^\circ$C} & Ref.\tnote{a} & Ref.\tnote{b} & Ref.\tnote{c}\\
      \hline
       Methanol (Ser) & $-21.92 \pm 0.21$ & $-14.10$ & $-21.17$ & $-21.19$ \\
       Ethanol (Thr) & $-21.41 \pm 0.35$ & $-13.70$ & $-20.42$ & $-20.43$   \\
       Acetamide (Asn) & $-41.75 \pm 0.95$ & $-18.80$ & $-40.50$ & $-40.53$     \\
       Propionamide (Gln) & $-44.97 \pm 1.41$ & $18.70$ & $-38.25$ & $-39.27$ \\
       Methanethiol (Cys) & $-8.70 \pm 2.88$ &  $5.50$ & $-5.19$ & $-5.28$  \\
       Methylimidazole (His) & $-32.16 \pm 1.74$ & $-27.40$ & $-42.97$ & $-43.00$ \\
       n-butylamine (Lys) & $-18.11 \pm 1.31$ & $-15.50$ & --- & $-39.86$ \\
       n-propylguanidine (Arg) & $-50.05 \pm 1.47$ & $-30.10$ & --- & $-83.40$ \\
       Acetic acid (Asp) & $-28.98 \pm 0.50$ & $-18.20$ & --- & $-45.85$ \\
      Propionic acid (Glu) & $-31.55 \pm 0.88$ & $-16.20$ & --- & $-42.87$ \\      
     \hline
    \end{tabular}
  	\begin{tablenotes}
     \item[a] Villa \& Mark (2002), 20$^\circ$C
     \item[b] Chang \textit{et al}. (2007), 25$^\circ$C
     \item[c] Radzicka \& Wolfenden (1988), 20$^\circ$C
     \end{tablenotes}
     \end{threeparttable}
\end{table}

\begin{table}[h]
\caption{Solvation free energies  (\SI{}{\kilo\joule\per\mole}) for hydrophobic and polar amino acid side chain analogs in cyclohexane \ce{cC6H12}.}
  \label{tab:chex}
  \begin{threeparttable}
    \begin{tabular}{l l r r r}
      \hline
      \text{Hydrophobic} & \text{This work, 25$^\circ$C} & Ref.\tnote{a} & Ref.\tnote{b} & Ref.\tnote{c} \\
      \hline
      Methane(Ala) & $0.60 \pm 0.11$ & $0.8 \pm 0.6$ & $1.05$ & $0.54$\\
      Propane(Val) & $-8.39 \pm 0.25$ &$-6.7\pm 0.9$ & $-6.61$ & $-8.58$ \\
      Butane (Ile) & $-11.60 \pm 0.89$ & $-13 \pm 1.5$ & $-9.91$ & $-11.59$ \\
      Isobutane (Leu) & $-11.23 \pm 0.86$ & $-9.8 \pm 1.5$ & $-9.16$ & $-11.05$ \\
      Methyl-ethylsulfide (Met) & $-15.77 \pm 0.61$ & $-14.4 \pm 1.3$ & $-14.52$ & $-16.02$ \\ 
      3-methylindole (Trp) & $-36.55 \pm 2.37$ & $-35.9 \pm 2.8$ & $-38.12$ & $-34.35$ \\
      4-methylphenol (Tyr) & $-24.38 \pm 0.77$ & $-28.3 \pm 1.2$ & $-22.68$ & $-24.98$ \\
      Toluene (Phe)        & $-21.27 \pm 1.65$ & $-25.2 \pm 1.0$ & $-19.71$ & $-15.65$ \\
      \hline
     \text{Polar} & \text{This work, 25$^\circ$C} & Ref.\tnote{a} & Ref.\tnote{b} & Ref.\tnote{c}\\
      \hline
       Methanol (Ser) & $-4.73 \pm 0.13$ & $-3.5 \pm 0.9$ & $-3.22$ & $-6.94$ \\
       Ethanol (Thr) & $-7.89 \pm 0.20$ & $-7.8 \pm 0.8$ & $-6.57$ & $-9.66$ \\
       Acetamide (Asn) & $-12.57 \pm 0.41$ & $-14.3 \pm 1.0$ & $-13.68$ & $-12.72$ \\
       Propionamide (Gln) & $-15.32 \pm 0.63$ & $-19.3 \pm 0.9$ & $-16.82$ & $-16.07$ \\
       Methanethiol (Cys) & $-8.55 \pm 0.18$ & $-7.9 \pm 0.8$ & $-8.95$ & $-10.54$ \\
       Methylimidazole (His) & $-19.34 \pm 1.03$ & $-21.1 \pm 1.0$ & $-19.04$ & $-23.47$ \\
       n-butylamine (Lys) & $-13.71 \pm 0.55$ & $-16.8 \pm 1.8$ & --- & $-16.61$ \\
       n-propylguanidine (Arg) & $-17.74 \pm 2.08$ & $-24 \pm 1.8$ & --- & $-20.92$ \\
       Acetic acid (Asp) & $-14.22 \pm 0.18$ & $-15.6 \pm 1.1$ & --- & $-9.33$ \\
      Propionic acid (Glu) & $-19.13 \pm 0.55$ & $-18.6 \pm 1.1$ & --- & $-14.35$ \\      
    \hline
    \end{tabular}
  		\begin{tablenotes}
       \item[a] Villa \& Mark (2002), 20$^\circ$C
       \item[b] Chang et al. (2007), 25$^\circ$C
       \item[c] Radzicka \& Wolfenden (1988), 20$^\circ$C
       \end{tablenotes}
       \end{threeparttable}
\end{table}
\begin{table}[h]
  \caption{Solvation free energies  (\SI{}{\kilo\joule\per\mole}) for hydrophobic and polar amino acid side chain analogs in ethanol \ce{EtOH}. Note that here and below, results from Damodaran \& Song (1986) are those from Nozaki \& Tanford (1971) estrapolated at higher temperatures, and are included here for completeness.}
     \label{tab:etoh}
     \begin{threeparttable}
    \begin{tabular}{l l r r r}
      \hline
      \text{Hydrophobic} & \text{This work, 25$^\circ$}C & Ref.\tnote{a} & Ref.\tnote{b} & Ref.\tnote{c} \\
      \hline
      Methane(Ala) & $2.79 \pm 0.17$ & $5.92$ & $4.75$ & $5.05$\\
      Propane(Val) & $-4.79 \pm 0.71$ &$2.03$ & $0.50$ & $1.13$ \\
      Butane (Ile) & $-6.23 \pm 0.76$ & --- & $-4.46$ & $-1.33$ \\
      Isobutane (Leu) & $-9.24 \pm 0.47$ & $0.01$ & $0.40$ & $3.03$ \\
      Methyl-ethylsulfide (Met) & $-12.52 \pm 1.01$ & --- & $-12.93$ & $-11.54$ \\ 
      3-methylindole (Trp) & $-33.43 \pm 1.43$ & $-33.77$ & $-40.27$ & --- \\
      4-methylphenol (Tyr) & $-43.01 \pm 0.95$& $35.20$ & $-36.38$ & $-37.21$ \\
      Toluene (Phe)        & $-18.07 \pm 0.68$ & $-13.98$ & $-15.15$ & $-14.19$ \\   
      \hline
      \text{Polar} & \text{This work, 25$^\circ$}C & Ref.\tnote{a} & Ref.\tnote{b} & Ref.\tnote{c}\\
      \hline
       Methanol (Ser) & $-23.56 \pm 0.37$ & $-19.54$ & $-20.67$ & $-20.97$ \\
       Ethanol (Thr) & $-25.56 \pm 0.20$ & --- & $-19.97$ & $-21.94$ \\
       Acetamide (Asn) & $-44.24 \pm 0.58$ & --- & $-39.86$ & $-39.86$ \\
       Propionamide (Gln) & $-48.46 \pm 0.42$ & --- & $-38.28$ & $-38.28$ \\
       Methanethiol (Cys) & $-15.64 \pm 0.41$ & --- & $-11.28$ & --- \\     
       Methylimidazole (His) & $-32.29 \pm 0.57$ & $-44.02$ & $-45.45$ & --- \\
       n-butylamine (Lys) & $-21.83 \pm 1.80$ & --- & $-24.70$ & --- \\
       n-propylguanidine (Arg) & $-42.24 \pm 1.18$ & --- & $-48.06$ & --- \\
       Acetic acid (Asp) & $-33.72 \pm 0.62$ & $-26.87$ & $-29.94$ & $-29.76$ \\
       Propionic acid (Glu) & $-37.71 \pm 0.42$ & $-25.51$ & $-29.08$ & $-28.90$ \\    
    \hline
    \end{tabular}
        \begin{tablenotes}
         \item[a] Nozaki \& Tanford (1971), 25.10$^\circ$C
         \item[b] Damodaran \& Song (1986), 37$^\circ$C
         \item[c] Tanford (1962), 20$^\circ$C
         \end{tablenotes}
         \end{threeparttable}
\end{table}


\begin{table}
  \begin{center}
  \caption{Classic SPT estimates of the Gibbs energy change, $ \Delta G_0$, associated with the creation in water \ce{H2O}, cyclohexane \ce{cC6H12} and ethanol \ce{EtOH} of a spherical cavity suitable to host methane, propane, toluene and methanol, at $28 ^{\circ}$ and 1 atm; estimates of the solute-solvent interaction energy, consisting of a van der Waals contribution (assumed to be solvent-independent) and a H-bond contribution. A comparison between the $\Delta G_0 + E_a$ values and the experimental $\Delta G$ is shown in the last two columns (no optimization has been performed). For each solute, the first line refers to water \ce{H2O}, the second to cyclohexane \ce{cC6H12}, and the third to ethanol \ce{EtOH}. Units are (\SI{}{\kilo\joule\per\mole}).}
  \label{tab:spt}
    \begin{tabular}{l c c c c}  

      \hline
       & $\Delta G_{0}$ & $E_{a}$ & $\Delta G_{0}+E_{a}$  & $\Delta G$ \\
      \hline
      Methane(ALA) $\sigma = 3.70$ \AA & $22.9$ & -$15.0$ & $7.9$ & $8.3$\\
                                       & $16.0$ & -$15.0$ & $1.0$ & $0.8$ \\
                                       & $17.7$ & -$15.0$ & $2.7$ & $1.6$ \\      
      Propane(VAL) $\sigma = 5.06$ \AA & $38.7$ & -$31.0$ & $7.7$ & $8.2$ \\
                                       & $25.7$ & -$31.0$ &-$5.3$ &-$7.6$ \\
                                       & $29.0$ & -$31.0$ &-$2.0$ &-$5.2$ \\   
      Toluene(PHE)$\sigma = 5.64$ \AA  & $46.7$ & -$50.0$ &-$3.3$ &-$3.7$ \\
                                       & $30.6$ & -$50.0$ &-$19.4$&-$18.7$ \\
                                       & $34.7$ & -$50.0$ &-$15.3$&-$14.2$ \\
     Methanol(SER)$\sigma = 3.83$ \AA  & $24.2$ & -$45.0$ &-$20.8$&-$21.4$ \\
                                       & $16.8$ & -$22.0$ &-$5.2$&-$5.3$ \\
                                       & $18.7$ & -$39.0$ &-$20.3$&-$21.0$ \\      
      \hline 
    \end{tabular}
  \end{center}
\end{table}

\begin{table}[h]
  \begin{center}
  \caption{Enthalpic and entropic contributions to the solvation free energies  (\SI{}{\kilo\joule\per\mole}) for hydrophobic and polar amino acid side chain analogs in water \ce{H2O}.}
  \label{tab:water_solv}
    \begin{tabular}{l| r r r| r r r}\cline{2-7}
      \hline
      &\multicolumn{3}{r|}{This work, 25$^\circ$C}&\multicolumn{3}{r}{Baldwin (2014), 25$^\circ$C}\\
     \cline{2-7}
     Hydrophobic &\text{$\Delta$G} &\text{$\Delta$H} & \text{$-\text{T}\Delta$S} &\text{$\Delta$G} & \text{$\Delta$H} & \text{$-\text{T}\Delta$S} \\
      \hline
      Methane(Ala) & $8.47 \pm 0.12$ & $-3.14 \pm 1.55$ & $11.61 \pm 1.55$  & $8.29$ & $-2.61$ & $4.61$\\
      Propane(Val) & $6.93 \pm 0.50$ &$-11.17 \pm 9.09$ & $18.10 \pm 9.25$ & $8.21 (8.21)$  &$ -5.02 (-4.83)$ & $ 6.98 (6.79)$ \\
      Butane (Ile) & $7.11 \pm 1.83$ & $-28.28 \pm 5.66$ & $35.39 \pm 6.52$ & $8.75$& $-5.66$ & $7.75$ \\
      Isobutane (Leu) & $7.24 \pm 1.34$ & $ -9.46\pm 8.56$ & $16.69 \pm 8.30$ & $9.71$ & $-5.23$ & $7.55$ \\
      Methyl-ethylsulfide (Met) & $-0.80 \pm 1.69$ & $-24.04 \pm 14.04$ & $23.23 \pm 12.94$ & $--$ & $--$ & $--$ \\ 
      3-methylindole (Trp) & $-29.09 \pm 2.34$ & $-88.50 \pm 23.89$ & $59.41 \pm 24.22$ & $--$ & $--$ & $--$ \\
      4-methylphenol (Tyr) &$-33.79 \pm 3.04$& $-104.73 \pm 13.86$& $70.93\pm 11.17$ & $--$ & $--$ & $--$ \\
      Toluene (Phe)        & $-7.62 \pm 1.12$ & $-53.44 \pm 7.56$ & $45.83 \pm 7.73$ & $--$ & $--$ & $--$ \\   
      \hline
      Polar & \text{$\Delta$G} &\text{$\Delta$H} & \text{$-\text{T}\Delta$S} &\text{$\Delta$G} & \text{$\Delta$H} & \text{$-\text{T}\Delta$S} \\
      \hline
      Methanol (Ser) & $-21.92 \pm 0.21$ & $-42.73 \pm 1.38$ & $20.82 \pm 1.43$ & $--$ & $--$ & $--$ \\
      Ethanol (Thr) & $-21.41 \pm 0.35$ & $-43.56 \pm 5.82$ & $22.14 \pm 6.08$ & $--$ & $--$ & $--$ \\
      Acetamide (Asn) & $-41.75 \pm 0.95$ & $-69.62 \pm 7.73$ & $27.87 \pm 7.19$ & $--$ & $--$ & $--$ \\
      Propionamide (Gln) & $-44.97 \pm 1.41$ & $-71.56 \pm 15.02$ & $ 26.57 \pm 15.64$ & $--$ & $--$ & $--$ \\
      Methanethiol (Cys) & $-8.70 \pm 2.88$ & $-25.67 \pm 5.21$ & $16.97 \pm 5.30$ & $--$ & $--$ & $--$ \\     
      Methylimidazole (His) & $-32.16 \pm 1.74$ & $-63.02 \pm 8.18$ & $30.86 \pm 9.45$ & $--$ & $--$ & $--$ \\
      n-butylamine (Lys) & $-18.11 \pm 1.31$ & $-47.43 \pm 9.80$ & $29.32 \pm 8.88$ & $--$ & $--$ & $--$ \\
      n-propylguanidine (Arg) & $-50.05 \pm 1.47$ & $-124.34 \pm 26.07$ & $74.29 \pm 24.94$ & $--$ & $--$ & $--$ \\
      Acetic acid (Asp) & $-28.98 \pm 0.50$ & $-52.89 \pm 8.53$ & $24.59 \pm 8.28$ & $--$ & $--$ & $--$ \\
      Propionic acid (Glu) & $-31.55 \pm  0.88$ & $-56.98 \pm 16.34$ & $25.43 \pm 16.84$ & $--$ & $--$ & $--$ \\    
    \hline
    \end{tabular}
  \end{center}
\end{table}

\begin{table}[h]
  \begin{center}
  \caption{Enthalpic and entropic contributions to the solvation free energies  (\SI{}{\kilo\joule\per\mole}) for hydrophobic and polar amino acid side chain analogs in cyclohexane \ce{cC6H12}.}
  \label{tab:chex_solv}
    \begin{tabular}{l| r r r| r r r}\cline{2-7}
      \hline
      &\multicolumn{3}{r|}{This work, 25$^\circ$C}&\multicolumn{3}{r}{Abraham (1979,1982), 25$^\circ$C}\\
     \cline{2-7}
     Hydrophobic & \text{$\Delta$G}&\text{$\Delta$H} & \text{$-\text{T}\Delta$S} & \text{$\Delta$G}&\text{$\Delta$H} & \text{$-\text{T}\Delta$S} \\
      \hline
      Methane(Ala) & $0.60\pm0.11$&$-3.38 \pm 1.37$ & $3.98 \pm 1.36$  & $0.8$ & $-0.9$ & $1.7$\\
      Propane(Val)& $-8.39\pm0.25$& $-14.22 \pm 4.31$ & $5.84 \pm 4.18$ & $-7.6$ & $-13.9$ & $ 6.29 $ \\
      Butane (Ile) & $-11.60\pm0.89$& $-21.31 \pm 2.25$ & $9.71 \pm 2.66$ & $-11.1$& $--$ & $--$ \\
      Isobutane (Leu) & $-11.23\pm0.86$& $ -18.88 \pm 9.48$ & $7.62 \pm 9.12$ & $-9.7$ & $--$ & $--$ \\
      Methyl-ethylsulfide (Met) & $-15.77\pm0.61$ & $24.59 \pm 6.73$ & $-40.37 \pm 7.24$ & $--$ & $--$ & $--$ \\ 
      3-methylindole (Trp) & $-36.55\pm2.37$ & $14.60 \pm 9.40$ & $-51.16 \pm 8.67$ & $--$ & $--$ & $--$ \\
      4-methylphenol (Tyr) & $-24.38\pm0.77$ & $-58.46 \pm 19.00$& $34.08\pm 19.35$ & $--$&$--$ & $--$ \\
      Toluene (Phe)        & $-21.27\pm1.65$ & $-45.74 \pm 12.23$ & $24.47 \pm 13.06$ & $--$ &$--$ & $--$ \\   
      \hline
      Polar &\text{$\Delta$G}&\text{$\Delta$H} & \text{$-\text{T}\Delta$S} & \text{$\Delta$G}&\text{$\Delta$H} & \text{$-T\Delta$S} \\
      \hline
      Methanol (Ser) & $-4.73\pm0.13$& $-11.83 \pm 0.74$ & $7.10 \pm 0.73$ & $--$ &$--$ & $--$ \\
      Ethanol (Thr) & $-7.89\pm0.20$&$-15.00 \pm 2.78$ & $7.11 \pm 2.86$ & &$--$ & $--$ \\
      Acetamide (Asn) & $-12.57\pm0.41$& $-20.88 \pm 2.18$ & $8.30 \pm 1.98$ & $--$ &$--$ & $--$ \\
      Propionamide (Gln) & $-15.32\pm0.63$& $-25.10 \pm 5.02$ & $ 9.78 \pm 4.93$ & $--$ &$--$ & $--$ \\
      Methanethiol (Cys) & $-8.55\pm0.18$& $-17.85 \pm 1.08$ & $9.30 \pm 1.30$ & $--$ & $--$ & $--$ \\     
      Methylimidazole (His) & $-19.34\pm1.03$& $-30.98 \pm 12.85$ & $11.64 \pm 13.74$ & $--$ & $--$ & $--$ \\
      n-butylamine (Lys) & $-13.71\pm0.55$ & $-25.71 \pm 9.33$ & $12.00 \pm 9.27$ & $--$ & $--$ & $--$ \\
      n-propylguanidine (Arg) & $-17.74\pm2.08$ &$-54.12 \pm 15.35$ & $36.38 \pm 14.06$ & $--$ &$--$ & $--$ \\
      Acetic acid (Asp) & $-14.22\pm0.18 $& $-26.42 \pm 3.00$ & $12.20 \pm 3.00$ &  $--$& $--$ & $--$ \\
      Propionic acid (Glu) & $-19.13\pm0.55$&$-30.74 \pm 3.26$ & $11.61 \pm 3.59$ & $--$ & $--$ & $--$ \\    
    \hline
    \end{tabular}
  \end{center}
\end{table}

\begin{table}[h]
  \begin{center}
  \caption{Enthalpic and entropic contributions to the solvation free energies  (\SI{}{\kilo\joule\per\mole}) for hydrophobic and polar amino acid side chain analogs in ethanol \ce{EtOH}.}
  \label{tab:etoh_solv}
    \begin{tabular}{l| r r r| r r r}\cline{2-7}
      \hline
      &\multicolumn{3}{r|}{This work, 25$^\circ$C}&\multicolumn{3}{r}{Abraham (1979,1982), 25$^\circ$C}\\
     \cline{2-7}
     Hydrophobic &\text{$\Delta$G}&\text{$\Delta$H} & \text{$-\text{T}\Delta$S} & \text{$\Delta$G}&\text{$\Delta$H} & \text{$-\text{T}\Delta$S} \\
      \hline
      Methane(Ala) & $2.79\pm0.17$ & $-0.43 \pm 1.50$ & $3.22 \pm 1.43$  &$1.6$& $-2.1$ & $3.7$\\
      Propane(Val) &$-4.79\pm0.71$ & $-16.44 \pm 5.15$ & $11.65 \pm 5.19$ &$-5.2$ & $ -12.4$ & $ 7.22 $ \\
      Butane (Ile) & $-6.23\pm0.76$ & $-17.25 \pm 10.92$ & $11.02 \pm 11.10$ & $-8.1$& $-17.7$ & $9.6$ \\
      Isobutane (Leu)& $-9.24\pm0.47$& $ -16.86 \pm 6.96$ & $7.62 \pm 7.16$ & $-6.9$& $-16.1$ & $9.21$ \\
      Methyl-ethylsulfide (Met) & $-12.52\pm1.01$& $-25.82 \pm 9.72$ & $13.31 \pm 9.65$ & $--$& $--$ & $--$ \\ 
      3-methylindole (Trp) & $-33.43\pm1.43$& $-56.80 \pm 22.95$ & $23.37 \pm 23.62$ & $--$& $--$ & $--$ \\
      4-methylphenol (Tyr) & $-43.01\pm0.95$& $-90.35 \pm 9.04$& $47.34\pm 9.08$ & $--$ & $--$ & $--$ \\
      Toluene (Phe)        & $-18.07\pm0.68$& $-28.56 \pm 13.07$ & $10.49 \pm 13.08$ & $--$ & $--$ & $--$ \\   
      \hline
      Polar & \text{$\Delta$G}&\text{$\Delta$H} & \text{$-\text{T}\Delta$S} & \text{$\Delta$G}& \text{$\Delta$H} & \text{$-\text{T}\Delta$S} \\
      \hline
      Methanol (Ser) & $-23.64\pm0.37$& $-43.22 \pm 3.87$ & $19.62 \pm 3.69$ & $--$ & $--$ & $--$ \\
      Ethanol (Thr) &$-25.56\pm0.20$ & $-47.06 \pm 2.70$ & $21.50 \pm 2.76$ & $--$ & $--$ & $--$ \\
      Acetamide (Asn) & $-44.24\pm0.58$& $-70.71 \pm 7.82$ & $26.47 \pm 7.78$ & $--$ & $--$ & $--$ \\
      Propionamide (Gln) & $-48.46\pm0.42$& $-84.15 \pm 16.67$ & $ 35.69 \pm 16.98$ & $--$ & $--$ & $--$ \\
      Methanethiol (Cys) & $-15.64\pm0.41$& $-31.89 \pm 4.51$ & $16.25 \pm 4.67$ & $--$ & $--$ & $--$ \\     
      Methylimidazole (His) &$-32.29\pm0.57$ & $-54.20 \pm 8.45$ & $21.91 \pm 8.57$ & $--$ & $--$ & $--$ \\
      n-butylamine (Lys) & $-21.83\pm1.80$& $-37.84 \pm 12.64$ & $16.01 \pm 11.80$ & $--$ & $--$ & $--$ \\
      n-propylguanidine (Arg) & $--\pm--$& $-- \pm-- $ & $-- \pm-- $ & $--$ & $--$ & $--$ \\
      Acetic acid (Asp) & $-33.72\pm0.62$& $-58.71 \pm 5.25$ & $24.99 \pm 5.24$ & $--$ & $--$ & $--$ \\
      Propionic acid (Glu) & $-37.70\pm0.42$& $-68.60 \pm 7.06$ & $30.90 \pm 6.95$ & $--$& $--$ & $--$ \\    
    \hline
    \end{tabular}
  \end{center}
\end{table}

\begin{table}[h]
  \begin{center}
  \caption{Enthalpic and entropic contributions to the transfer free energies (\SI{}{\kilo\joule\per\mole})  from water \ce{H2O} to cyclohexane \ce{cC6H12} for hydrophobic and polar amino acid side chain analogs.}
  \label{tab:transfer_water_chex}
    \begin{tabular}{l| r r r| r r r| r r r}\cline{2-10}
      \hline
      &\multicolumn{3}{r|}{This work, 25$^\circ$C} & \multicolumn{3}{r|}{Wolfenden (2015), 25$^\circ$C} &\multicolumn{3}{r}{Abraham (1979,1982), 25$^\circ$C}\\
     \cline{2-10}
     Hydrophobic & \text{$\Delta\Delta$G} &\text{$\Delta\Delta$H} & \text{$-\text{T}\Delta\Delta$S} & \text{$\Delta\Delta$G}&\text{$\Delta\Delta$H} & \text{$-\text{T}\Delta\Delta$S} & \text{$\Delta\Delta$G}&\text{$\Delta\Delta$H} & \text{$-\text{T}\Delta\Delta$S}\\
      \hline
      Methane(Ala) &$-7.87\pm0.23$ & $-0.24 \pm 2.92 $ & $-7.63 \pm 2.91$  &$-12.02$ & $10.68$ & $-22.69$ & $-7.50$& $10.00$ & $-17.5$ \\
      Propane(Val) &$-15.32\pm0.75$ & $-3.05 \pm 13.40$ & $-12.26 \pm 13.44 $ &$-23.28$ & $ 6.20$ & $ -29.48 $ &$-15.80$ & $7.10$ & $-22.29$\\
      Butane (Ile) & $-18.71\pm2.72$& $6.97 \pm 7.91$ & $-25.68 \pm 9.18$ &$-24.16$ & $4.31$ & $-28.47$ &$-19.80$ &$--$ & $--$ \\
      Isobutane (Leu) &$-18.47\pm2.20$ & $ -10.42 \pm 18.04$ & $-8.03 \pm 17.42$ &$-24.16$ &$2.51$ & $-26.67$ &$-19.40$ &$--$ & $--$ \\
      Methyl-ethylsulfide (Met) &$-14.97\pm2.30$ & $48.63 \pm 20.77$ & $-66.60 \pm 20.18$ & $-10.89 $ & $3.35$ & $-14.24$ & $--$&$--$ & $--$\\ 
      3-methylindole (Trp) & $-7.46\pm4.71$ & $103.1 \pm 33.29$ & $-110.57 \pm 32.89$ &$-10.42$ & $-0.54$ & $-9.88$ &$--$ & $--$ & $--$ \\
      4-methylphenol (Tyr) &$9.41\pm3.81$ & $46.27 \pm 32.86$& $-36.85\pm 30.52$ & $1.76$& $18.21$ & $-16.45$ & $--$& $--$ & $--$\\
      Toluene (Phe)       & $-13.65\pm2.77$& $7.70 \pm 19.79$ & $-21.36 \pm 20.79$ & $-15.04$&$-1.05$ & $-13.98$ &$--$ &$--$ & $--$ \\   
      \hline
      Polar &\text{$\Delta\Delta$G}&\text{$\Delta\Delta$H} & \text{$-\text{T}\Delta\Delta$S} & \text{$\Delta\Delta$G}&\text{$\Delta\Delta$H} & \text{$-\text{T}\Delta\Delta$S} & \text{$\Delta\Delta$G}&\text{$\Delta\Delta$H} & \text{$-\text{T}\Delta\Delta$S}\\
      \hline
      Methanol (Ser)& $17.19\pm0.34$& $30.91 \pm 2.12$ & $-13.71 \pm 2.16$ &$16.08$ &$26.00$ & $-9.92$ &$--$ &$--$ & $--$\\
      Ethanol (Thr) & $13.52\pm0.55$& $28.56 \pm 8.60$ & $-15.03 \pm 8.94$ & $10.42$&$28.14$ & $-17.71$ &$--$ &$--$ & $--$ \\
      Acetamide (Asn)&$29.18\pm1.36$ & $48.74 \pm 9.91$ & $-19.57 \pm 9.17$ &$27.80$ &$29.89$ & $-2.14$ &$--$ &$--$ & $--$ \\
      Propionamide (Gln) &$29.65\pm2.04$ & $46.46 \pm 20.04$ & $ -16.79 \pm 20.54$ &$23.19$ &$36.80$ & $-13.61$ & $--$&$--$ &$--$\\
      Methanethiol (Cys) & $0.15\pm3.06$ & $7.72 \pm 6.29$ & $-7.67 \pm 6.6$ & $-8.71$&$13.06$ & $-21.77$ & $--$&$--$ & $--$ \\     
      Methylimidazole (His)&$12.82\pm2.77$ & $32.04 \pm 21.03$ & $-19.22 \pm 23.19$ & $19.89$&$48.53$ & $-28.64$ &$--$ &$--$ & $--$ \\
      n-butylamine (Lys) & $4.40\pm1.86$ & $21.72 \pm 19.13$ & $-17.32 \pm 18.15$ & $1.55$&$22.23$ & $-20.68$ &$--$ &$--$ & $--$ \\
      n-propylguanidine (Arg) & $32.31\pm3.55$&$70.22 \pm 41.42$ & $ -37.91 \pm 39.00$ & $24.62$& $57.53$ & $-32.91$ & $--$&$--$ & $--$ \\
      Acetic acid (Asp) & $14.76\pm0.68$ &$26.47 \pm 11.53 $ & $ -12.39 \pm 11.28$ & $18.71$&$34.92$ & $-16.16$ & $--$&$--$ & $--$ \\
      Propionic acid (Glu)&$12.42\pm1.61$ & $26.24 \pm 19.59$ & $-13.82 \pm 20.41$ &$12.85$ & $43.43$ & $30.48$ & $--$&$--$ & $--$ \\    
    \hline
    \end{tabular}
  \end{center}
\end{table}

\begin{table}[h]
  \begin{center}
  \caption{Enthalpic and entropic contributions to the transfer free energies (\SI{}{\kilo\joule\per\mole}) from water \ce{H2O} to ethanol \ce{EtOH} for hydrophobic and polar amino acid side chain analogs.}
  \label{tab:transfer_water_etoh}
    \begin{tabular}{l| r r r| r r r}\cline{2-7}
      \hline
      &\multicolumn{3}{r|}{This work, 25$^\circ$C}&\multicolumn{3}{r}{Abraham (1979,1982), 25$^\circ$C}\\
     \cline{2-7}
     Hydrophobic &\text{$\Delta\Delta$G}&\text{$\Delta\Delta$H} & \text{$-\text{T}\Delta\Delta$S} & \text{$\Delta\Delta$G}&\text{$\Delta\Delta$H} & \text{$-\text{T}\Delta\Delta$S} \\
      \hline
      Methane(Ala) &$-5.68\pm0.29$ & $2.71 \pm 3.05$ & $-8.39 \pm 2.98 $  &$-6.70$ & $8.8$ & $-15.50$\\
      Propane(Val) & $-11.72\pm1.21$& $-5.27 \pm 14.24$ & $-6.45 \pm 14.44$ &$-13.40$ & $ 8.6$ & $ -21.97 $ \\
      Butane (Ile) &$-13.34\pm2.59$ & $11.03 \pm 16.58$ & $-24.37 \pm 17.62$ & $-16.80$& $5.9$ & $-22.69$ \\
      Isobutane (Leu)&$-16.48\pm1.81$ & $ -7.40 \pm 15.52$ & $-9.07 \pm 15.46$ &$-16.60$ & $5.8$ & $-22.39$ \\
      Methyl-ethylsulfide (Met) &$-11.72\pm2.70$ & $-1.78 \pm 23.76$ & $-9.92 \pm 22.59$ & $--$& $--$ & $--$ \\ 
      3-methylindole (Trp) & $-4.34\pm3.77$& $31.70 \pm 46.84$ & $-36.04 \pm 47.84$ &$--$ &$--$ & $--$ \\
      4-methylphenol (Tyr) &$-9.22\pm3.99$ & $14.38 \pm 22.90$& $-23.59\pm 20.25$ &$--$ &$--$ & $--$ \\
      Toluene (Phe)        &$-10.45\pm1.80$ & $24.88 \pm 20.63$ & $-35.34 \pm 20.81$ &$--$ &$--$ & $--$ \\   
      \hline
      Polar & \text{$\Delta\Delta$G}&\text{$\Delta\Delta$H} & \text{$-\text{T}\Delta\Delta$S} &\text{$\Delta\Delta$G} & \text{$\Delta\Delta$H} & \text{$-\text{T}\Delta\Delta$S} \\
      \hline
      Methanol (Ser) &$-1.75\pm0.58$ & $-0.31 \pm 5.25$ & $-1.46 \pm 5.12$ &$--$ &$--$ & $--$ \\
      Ethanol (Thr) &$-3.87\pm0.55$ &$-3.47 \pm 8.52$ & $-0.39 \pm 8.84$ &$--$ &$--$ & $--$ \\
      Acetamide (Asn) & $-2.60\pm1.53$&$-1.43 \pm 15.55$ & $-1.17 \pm 14.96$ & $--$&$--$ & $--$ \\
      Propionamide (Gln) &$-3.17\pm1.83$ & $-20.90 \pm 31.69$ & $ 17.75 \pm 35.62$ &$--$ &$--$ & $--$ \\
      Methanethiol (Cys) &$-6.95\pm3.29$ & $-4.96 \pm 9.72$ & $-1.99 \pm 9.97$ & $--$&$--$ & $--$ \\     
      Methylimidazole (His) &$-0.05\pm2.31$ & $8.24 \pm 16.63$ & $-8.26 \pm 18.02$ &$--$ &$--$ & $--$ \\
      n-butylamine (Lys) &$-3.75\pm3.11$ & $9.72 \pm 22.44$ & $-13.47 \pm 20.68$ & $--$&$--$ & $--$ \\
      n-propylguanidine (Arg) &$--\pm--$ & $-- \pm-- $ & $-- \pm -- $ & $--$&$--$ & $--$ \\
      Acetic acid (Asp) &$-4.74\pm1.12$ &$-5.93 \pm 13.78$ & $0.51 \pm 13.58$ & $--$&$--$ & $--$ \\
      Propionic acid (Glu) & $-6.16\pm1.30$& $-10.47 \pm 23.40$ & $4.31 \pm 23.79$ &$--$ &$--$ & $--$ \\    
    \hline
    \end{tabular}
  \end{center}
\end{table}
\newpage

\begin{turnpage}
\begin{table}[h!]
  \caption{Fitting coefficients for different amino acid side chain analogs used to compute the thermodynamics parameters along with their Pearson’s correlation coefficient $R^\textrm{2}$.}
   \label{tab:fit_coefs}
    \begin{tabular}{l| r r r r| r r r r | r r r r}\cline{2-13}
      \hline
      &\multicolumn{4}{r|}{\ce{cC6H12}}&\multicolumn{4}{r|}{EtOH}&\multicolumn{4}{r}\ce{H2O}\\
     \cline{2-13}
     Hydrophobic & \textit{a} & \textit{b} & \textit{c} & $R^\textrm{2}$ & \textit{a} & \textit{b} & \textit{c} & $R^\textrm{2}$ & \textit{a} & \textit{b} & \textit{c} & $R^\textrm{2}$\\
      \hline
      Methane(Ala) & $-29.3694$ & $0.5972$ & $-0.0872$  & $0.95818$ & $10.4955$ & $-0.2346$ & $0.0366$ & $0.99678$ & $-64.9942$ & $1.4308$ & $-0.2079$ & $0.99676$\\
      Propane(Val) & $8.9816$ &$-0.5017$ & $0.0778$ & $0.96322$  &$-7.3071$ & $-0.1660$ & $0.0306$ & $0.97862$& $-111.8008$ & $ 2.3413$& $ -0.3409$ & $0.90212$ \\
      Butane (Ile) & $-93.0027$ & $1.6432$ & $-0.2405$ & $0.97181$& $-104.5813$ & $1.9986$ & $-0.2929$ & $0.92449$ & $-577.1131$ & $12.5018$ & $-1.8502$ & $0.95994$ \\
      Isobutane (Leu) & $74.7048$ & $-2.0768$ & $0.3139$ & $0.94533$ & $252.7160$ & $-6.0302$ & $0.9042$ & $ 0.87209$ & $ -137.4634$ & $ 2.9061$ & $ -0.4250$ & $0.88111$ \\
      Methyl-ethylsulfide (Met) & $1117.2886$ & $-24.6814$ & $3.6649$ & $0.92475$ & $-182.5088$ & $3.5643$ & $ -0.5255$ & $0.98937$ & $ 139.2361 $ & $ -3.5924$ & $0.5486$ & $0.82625$ \\ 
      3-methylindole (Trp) & $2288.4173$ & $-51.2502$ & $7.6264$ & $0.95952$ & $-93.8153$ & $0.9098$ & $-0.1241$ & $0.92019$ & $-185.8273$ & $2.3960$ & $-0.3275$ & $ 0.93339$ \\
      4-methylphenol (Tyr) &$-339.6722$& $6.4313$& $-0.9432$ & $0.93764$ & $-55.0555$ & $-0.6341$ & $0.1184$ & $0.95091$ & $307.8364$ & $-9.0299$& $1.3837$ & $0.94686$ \\
      Toluene (Phe)        & $-52.8578$ & $0.2420$ & $-0.0239$ & $0.93993$ & $-243.1799$ & $4.8563$ & $-0.7198$ & $0.80403$ & $-35.5509$ & $0.0252$ & $0.0130$ & $0.90851$ \\   
      \hline
     Polar & \textit{a} & \textit{b} & \textit{c} & $R^\textrm{2}$ & \textit{a} & \textit{b} & \textit{c} & $R^\textrm{2}$ & \textit{a} & \textit{b} & \textit{c} & $R^\textrm{2}$\\
      \hline
      Methanol (Ser) & $-27.2782$ & $0.3709$ & $-0.0518$ & $0.96906$ & $-11.5717$ & $-0.6459$ &$0.1063$ & $0.99971$ & $-35.3124$ & $-0.0971$ & $0.0249$ & $0.99909$ \\
      Ethanol (Thr) & $-15.8606$ & $0.0431$ & $-0.0029$ & $0.97949$ & $14.7211$ & $-1.3157$ & $0.2072$ & $0.98123$ & $-110.9791$ & $1.5442$ & $-0.2182$ & $0.98312$ \\
      Acetamide (Asn) & $55.8768$ & $-1.6963$ & $0.2574$ & $0.94491$ & $89.5388$ & $	-3.5109$ & $0.5374$ & $0.97194$ & $8.7707$ & $-1.6703$ & $0.2635$ & $0.98706$ \\
      Propionamide (Gln) & $50.8798$ & $-1.6739$ & $0.2548$ & $0.93166$ & $-39.9522$ & $-0.8732$ & $0.1482$ & $0.95726$ & $-105.0984$ & $ 0.8572$ & $-0.1151$ & $0.92632$	 \\
      Methanethiol (Cys) & $-52.2218$ & $0.8032$ & $-0.1153$ & $0.95474$ & $-68.3014$ & $0.8724$& $-0.1221$ & $0.98555$ & $-116.9340$ & $2.1153$ & $ -0.3076$ & $0.98041$ \\     
      Methylimidazole (His) & $-64.2742$ & $0.7870$ & $-0.1117$ & $0.98703$ & $-108.2614$ & $1.2879$ & $-0.1813$ & $0.90130$ & $-517.3281$ & $10.3350$ & $-1.5278$ & $0.83670$ \\
      n-butylamine (Lys) & $-13.7771$ & $-0.2279$ & $0.0400$ & $0.93944$ & $-221.8295$ & $4.1868$ & $-0.6171$ & $0.94929$ & $-206.4630$ & $3.6800$ & $-0.5349$ & $0.91925$  \\
      n-propylguanidine (Arg) & $-941.5227$ & $20.1262$ & $-2.9884$ & $0.90171$ & $--$ & $--$ & $--$ & $--$ & $526.5740$& $-14.5680$ & $2.2182$ & $0.84896$ \\
      Acetic acid (Asp) & $104.4444$ & $1.79371$ & $-0.26170$ & $0.98557$ & $-129.4953$ & $1.6740$ & $-0.2374$ & $0.99669$ & $-90.2838$ & $0.9694$ & $-0.1338$ & $0.98984$ \\
      Propionic acid (Glu) & $34.5187$ & $-1.4269$ & $0.2189$ & $0.94840$ & $47.2406$ & $-2.4766$ & $0.3847$ & $0.99048$ & $-216.3530$ & $3.6654$ & $-0.5345$ & $0.92398$ \\    
    \end{tabular}
\end{table}%
\end{turnpage}


\clearpage
\newpage
\section{Supplementary Figures}

\begin{figure}[htpb]
\centering
 \captionsetup{justification=raggedright,width=\linewidth}
  \begin{subfigure}{8cm}
     \includegraphics[width=\linewidth]{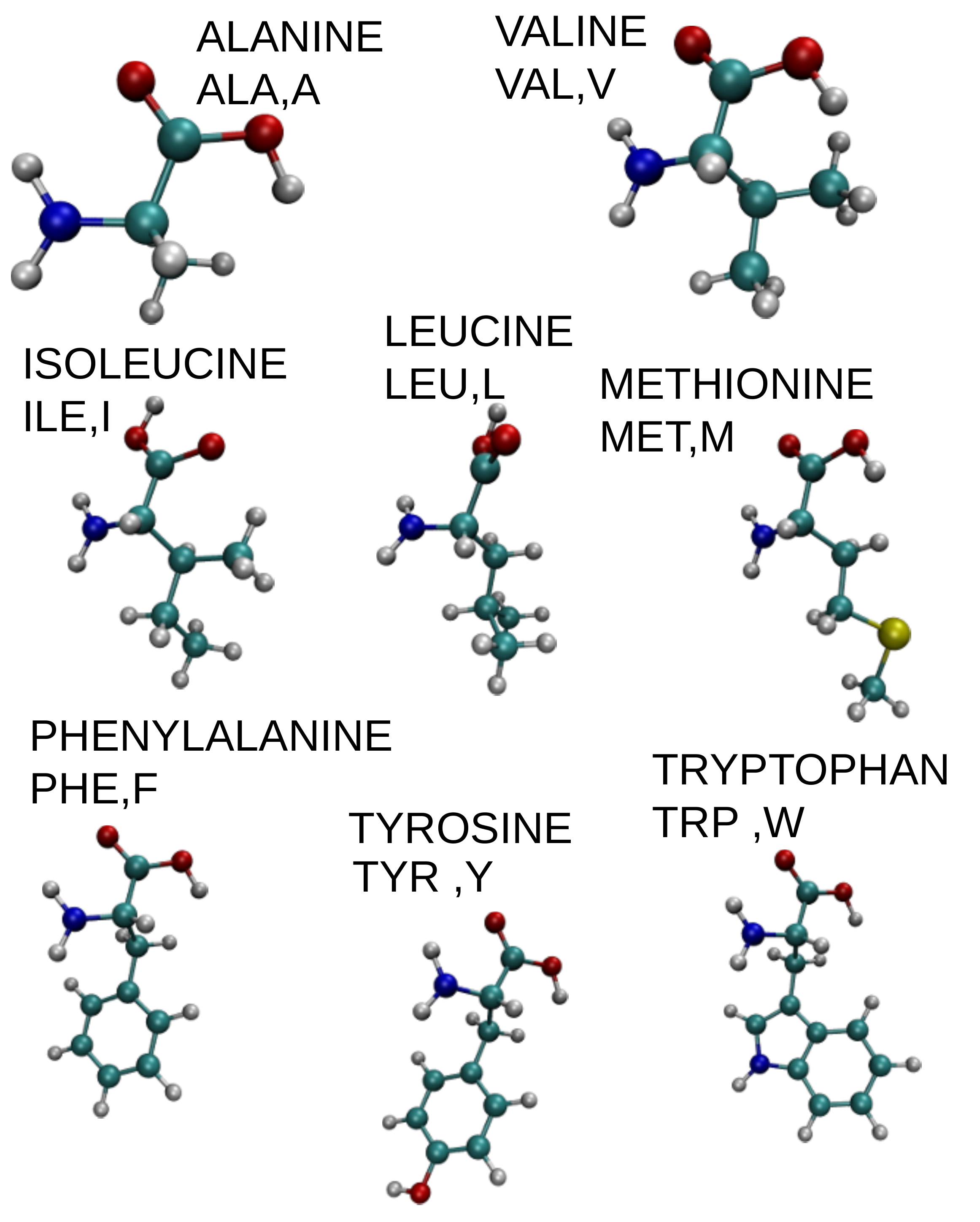}
    \caption{}\label{fig:aminoacids_a}
  \end{subfigure}
  \begin{subfigure}{8cm}
     \includegraphics[width=\linewidth]{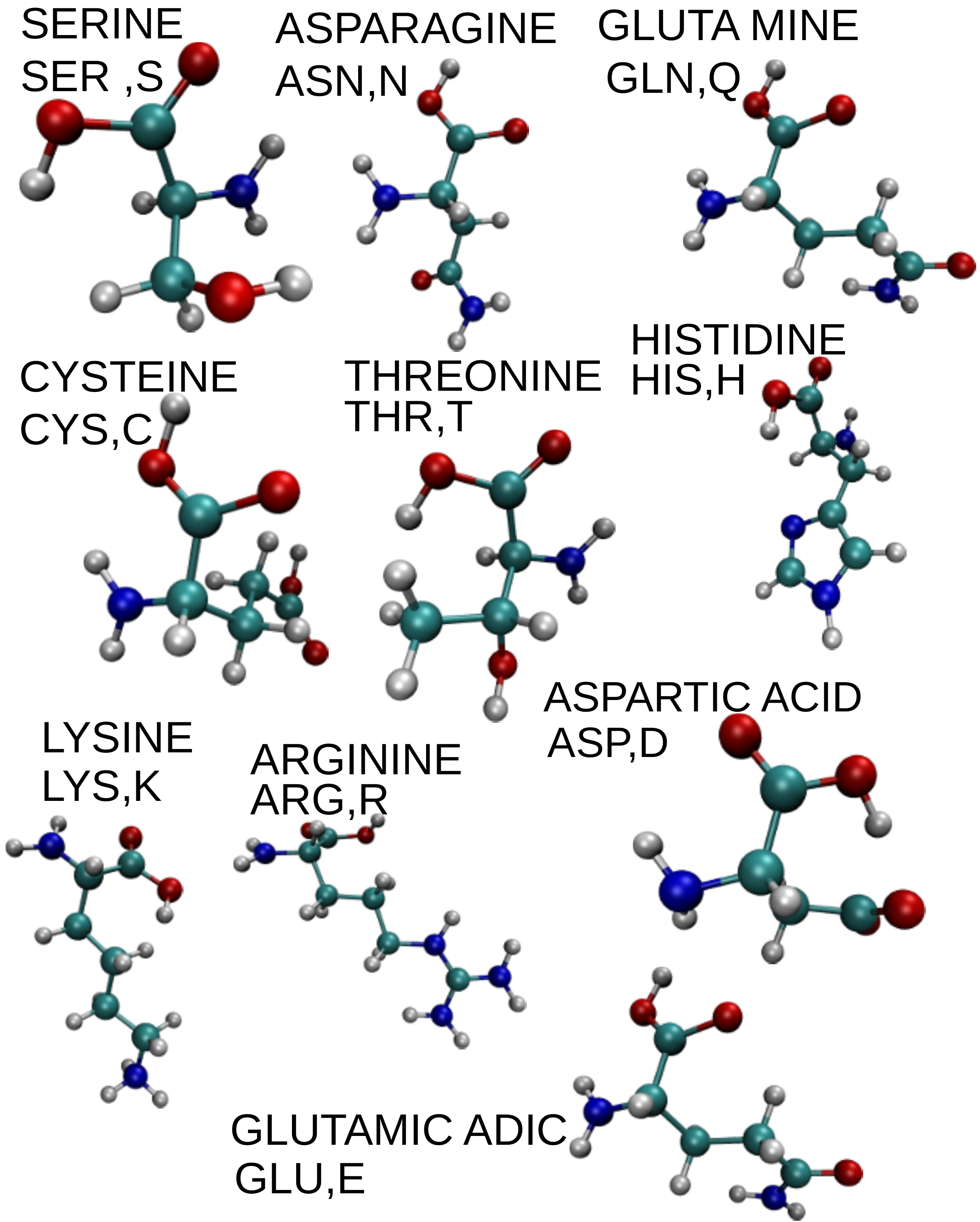}
    \caption{}\label{fig:aminoacids_b}
  \end{subfigure}
  \caption{(a) Hydrophobic amino acids. (b) Polar amino acids.} 
  \label{fig:aminoacids}
\end{figure}
\begin{figure}[h]
\centering
 \captionsetup{justification=raggedright,width=1.0\linewidth}
  \includegraphics[width=0.8\linewidth,trim=0cm 0cm 0cm 0cm, angle=0]{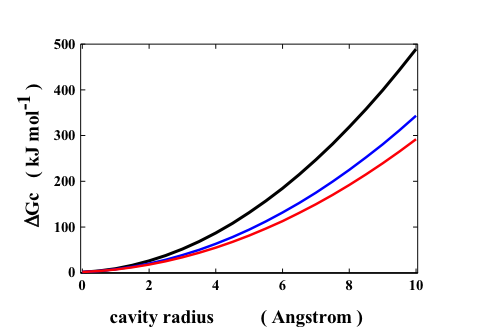}
  \caption{Trend of the free energy of cavity creation in the three liquids versus the radius of the spherical cavity, calculated by means of classic SPT at 28$^\circ$C and 1 atm; black line refers to water \ce{H2O}, blue line refers to ethanol \ce{EtOH} , and red line refers to cyclohexane \ce{cC6H12}.}
  \label{fig:spt}
\end{figure}

\begin{figure}[h]
\centering
 \captionsetup{justification=raggedright,width=\linewidth}
  \includegraphics[width=0.7\linewidth,trim=0cm 0cm 0cm 0cm, angle=0]{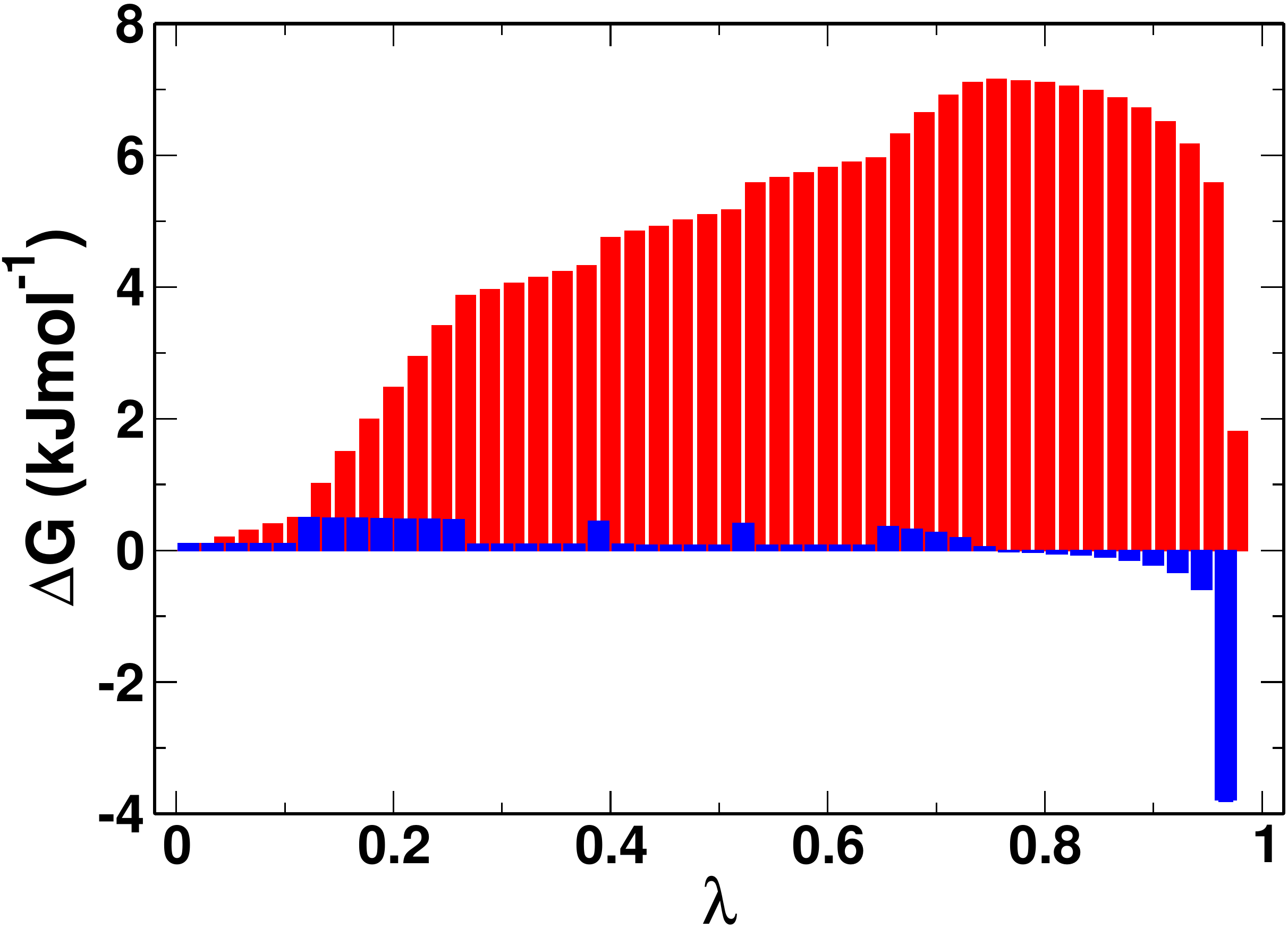}
  \caption{Illustrative case of the decoupling process for Methanol, SER  in cyclohexane, cC6H12. \textbf{Blue} histograms show the free energy difference between two consecutive lambda points while \textbf{red} ones display the integral i.e. the cumulative free energy change as a function of lambda. While throughout the work 21 lambda points were used, in this particular case the plot is displayed for 45 lambda points.}
  \label{fig:decoupling_process}
\end{figure}

\begin{figure}[htpb]
\centering
 \captionsetup{justification=raggedright,width=\linewidth}
  \begin{subfigure}{10cm}
     \includegraphics[width=\linewidth]{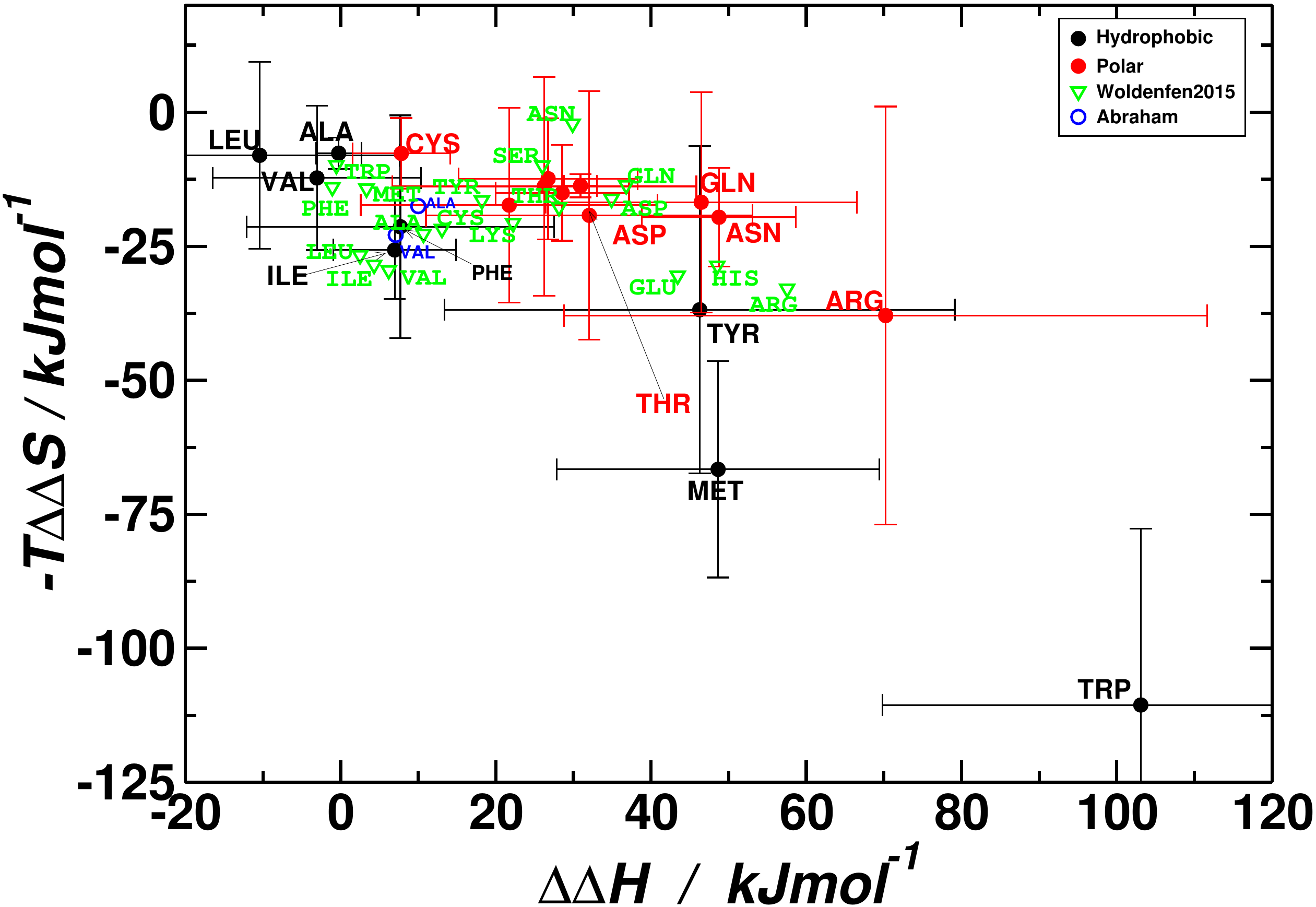}
    \caption{}\label{fig:compoensation_transfer_a}
  \end{subfigure}
  \begin{subfigure}{10cm}
     \includegraphics[width=\linewidth]{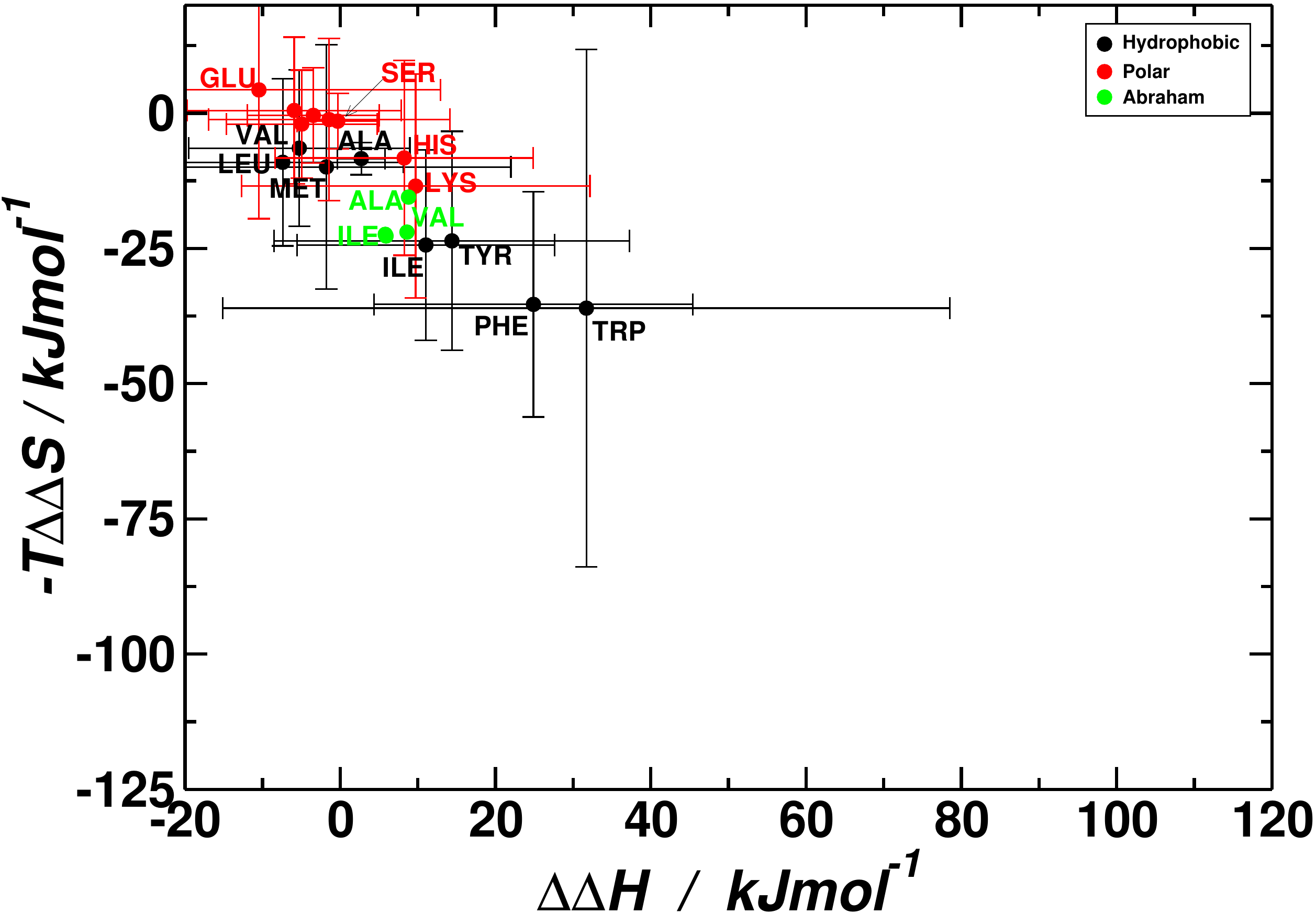}
    \caption{}\label{fig:compensation_transfer_b}
  \end{subfigure}
  \caption{Change in the entropic term $-T \Delta \Delta S$ as a function of the change in the entropic part $\Delta \Delta H$ in the case of  (a) water to cyclohexane; (b) water to  ethanol . In the case of water to cyclohexane, results from Wolfenden \textit{et al} are also included. Units are in \SI{}{\kilo\joule\per\mole}).}
 \label{fig:compensation_transfer}
\end{figure}

\begin{figure}[h]
\centering
 \captionsetup{justification=raggedright,width=\linewidth}
   \begin{overpic}[scale=0.6]{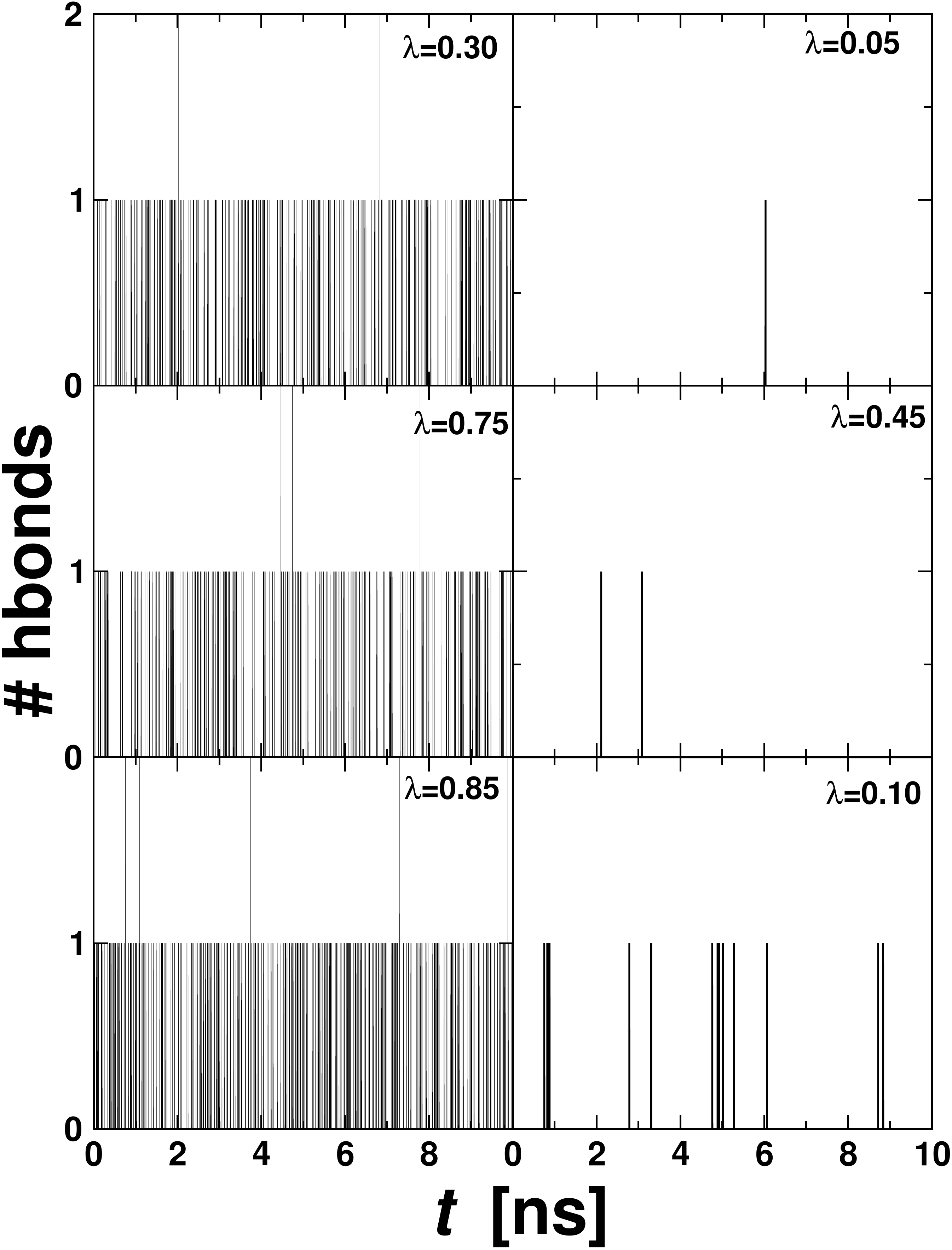}
       \put(15,85){\includegraphics[scale=0.18]{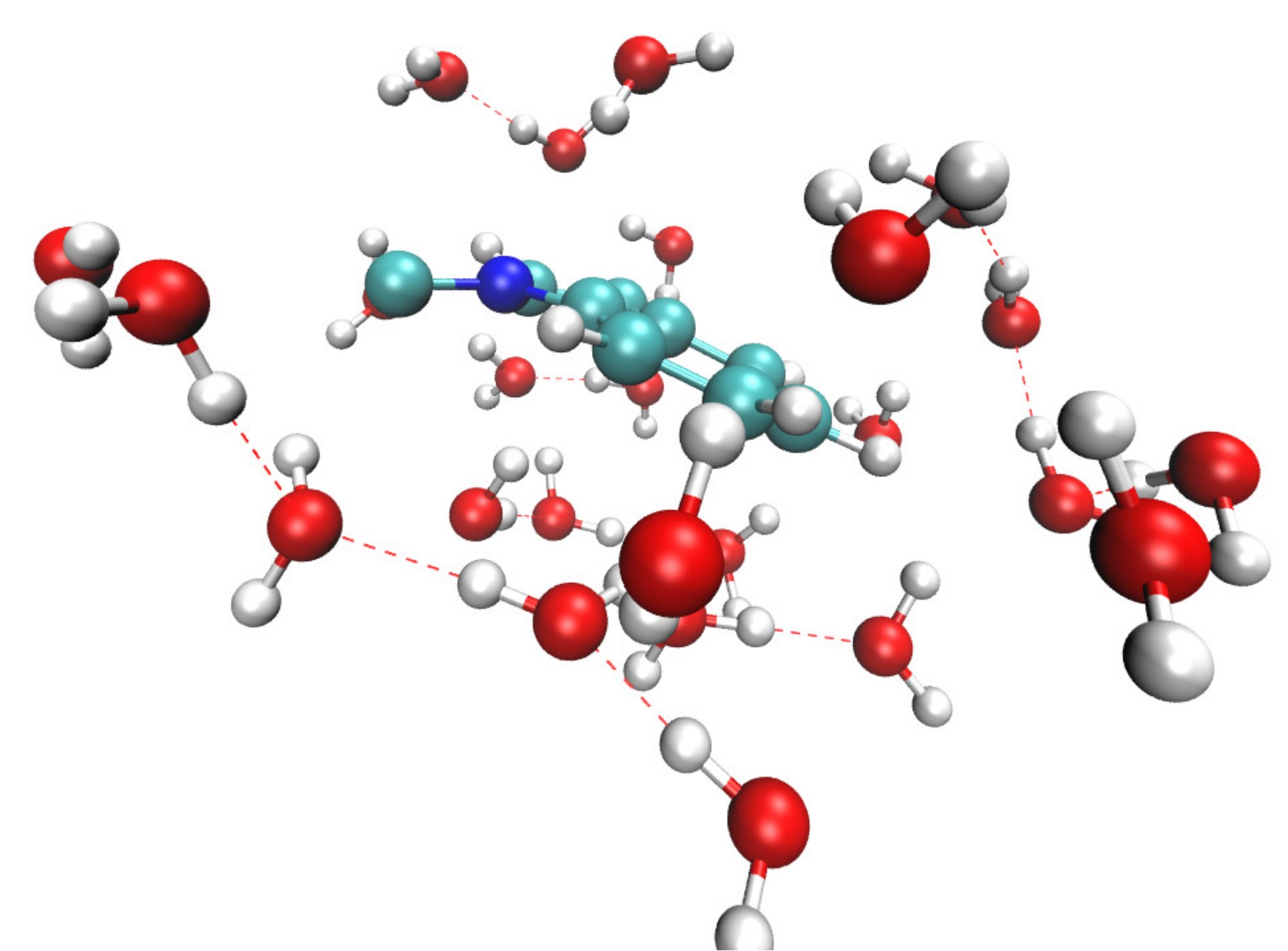}}  
       \put(44,78){\includegraphics[scale=0.1]{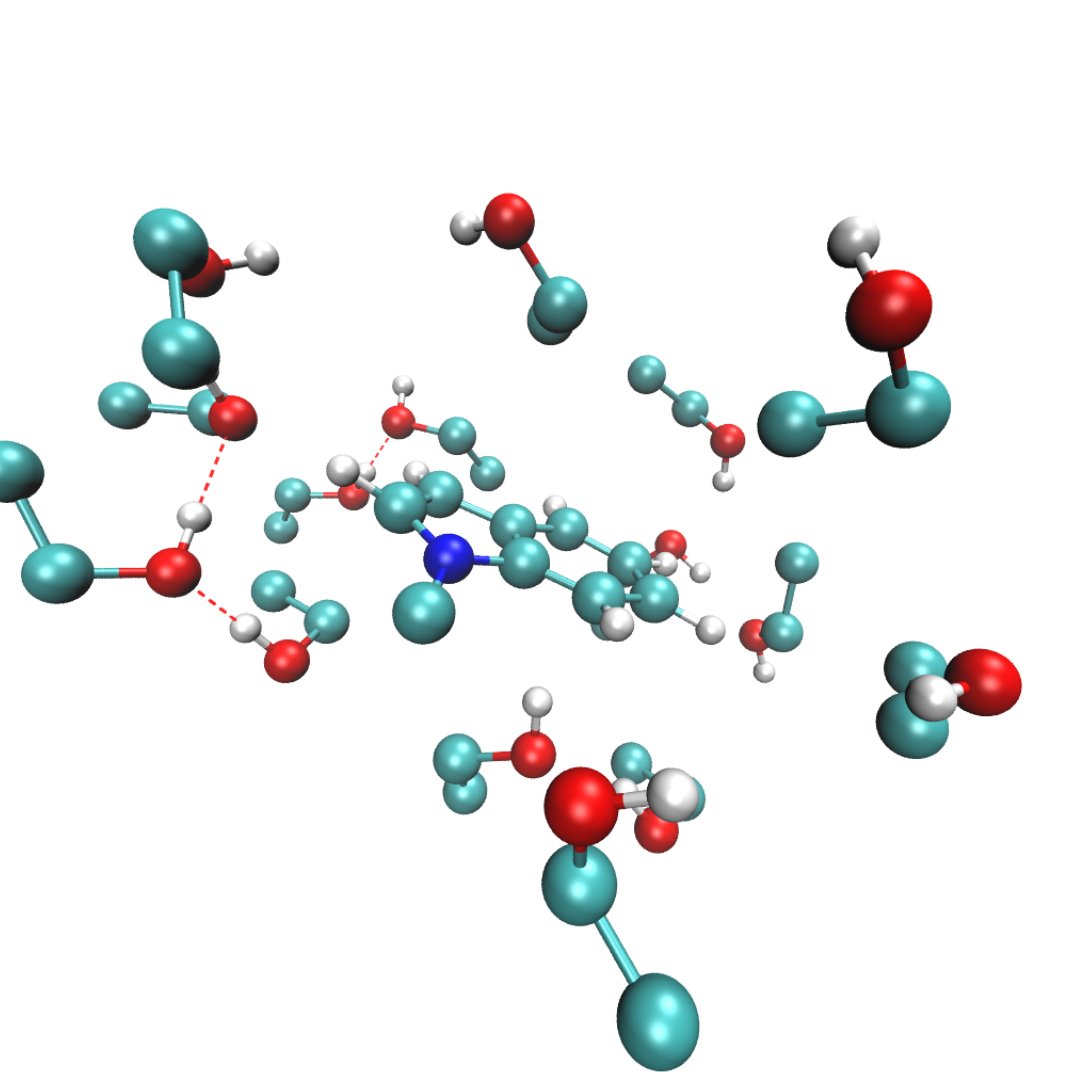}}
    \end{overpic}
  \caption{Time-based number of hydrogen bonds change for 3-methylindole in water, \ce{H2O} (\textbf{left panel}) and in ethanol, \ce{EtOH} (\textbf{right panel}) at three different temperatures 280, 290 and 300 K upon moving from the \textbf{top} to the \textbf{bottom}, respectively. Insets are representative snapshots. Hydrogen bonds are computed using \textit{gmx hbond} tool of Gromacs package implying that both faces of the phenyl rings are potentially involved in the geometric consideration for Hbond existence.}
  \label{fig:hbonds}
\end{figure}




\begin{turnpage}
\begin{figure}[h]
\centering
\begin{tabular}[c]{cccccc}
\begin{subfigure}[c]{0.22\textwidth}
\includegraphics[height=4.5cm, width=\textwidth]{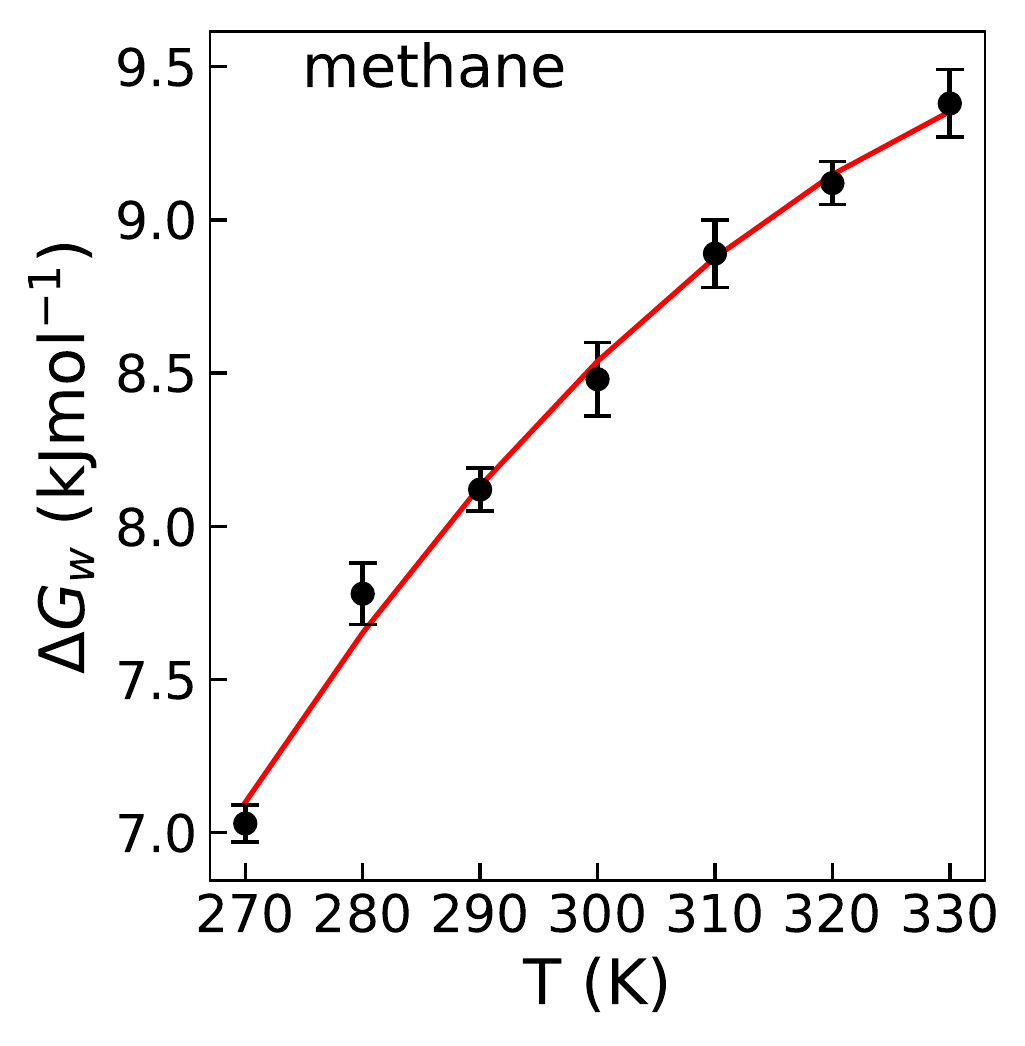}
\end{subfigure}&
\begin{subfigure}[c]{0.22\textwidth}
\includegraphics[height=4.5cm, width=\textwidth]{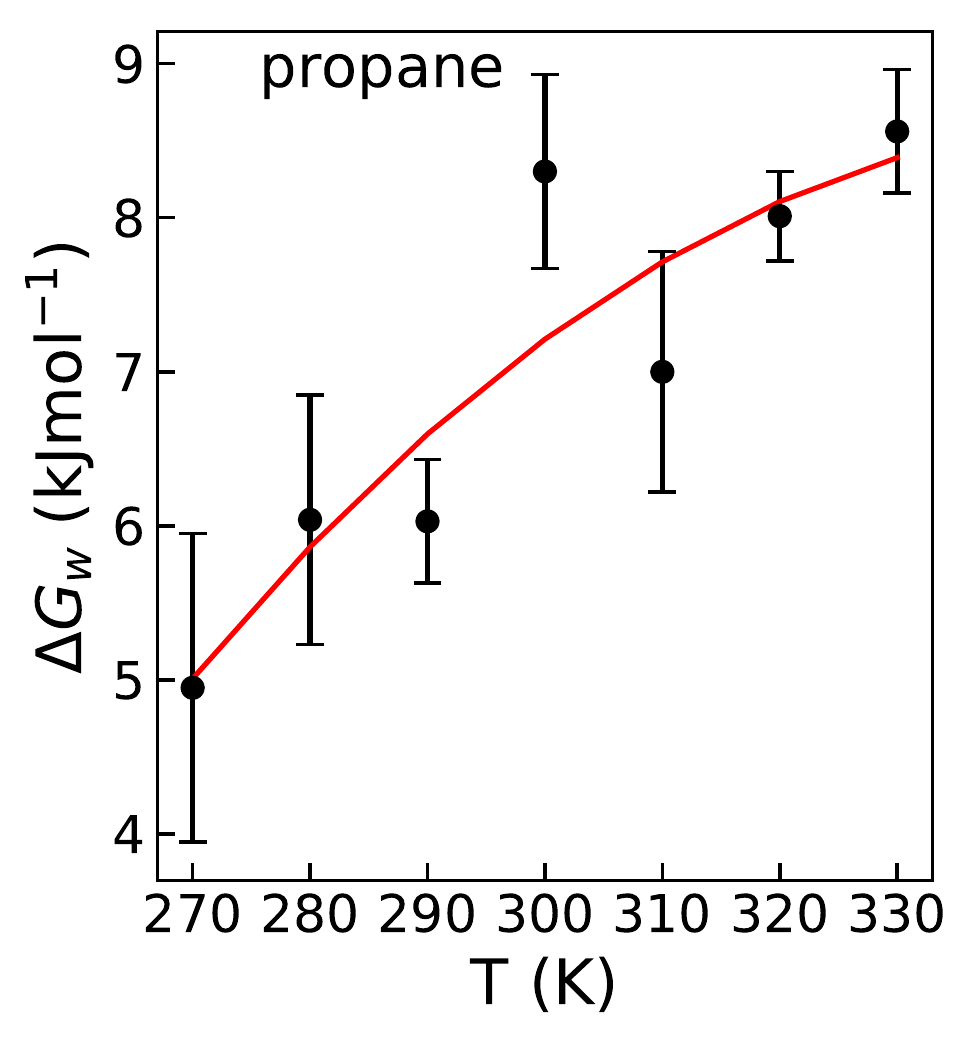}
\end{subfigure}&
\begin{subfigure}[c]{0.22\textwidth}
\includegraphics[height=4.5cm, width=\textwidth]{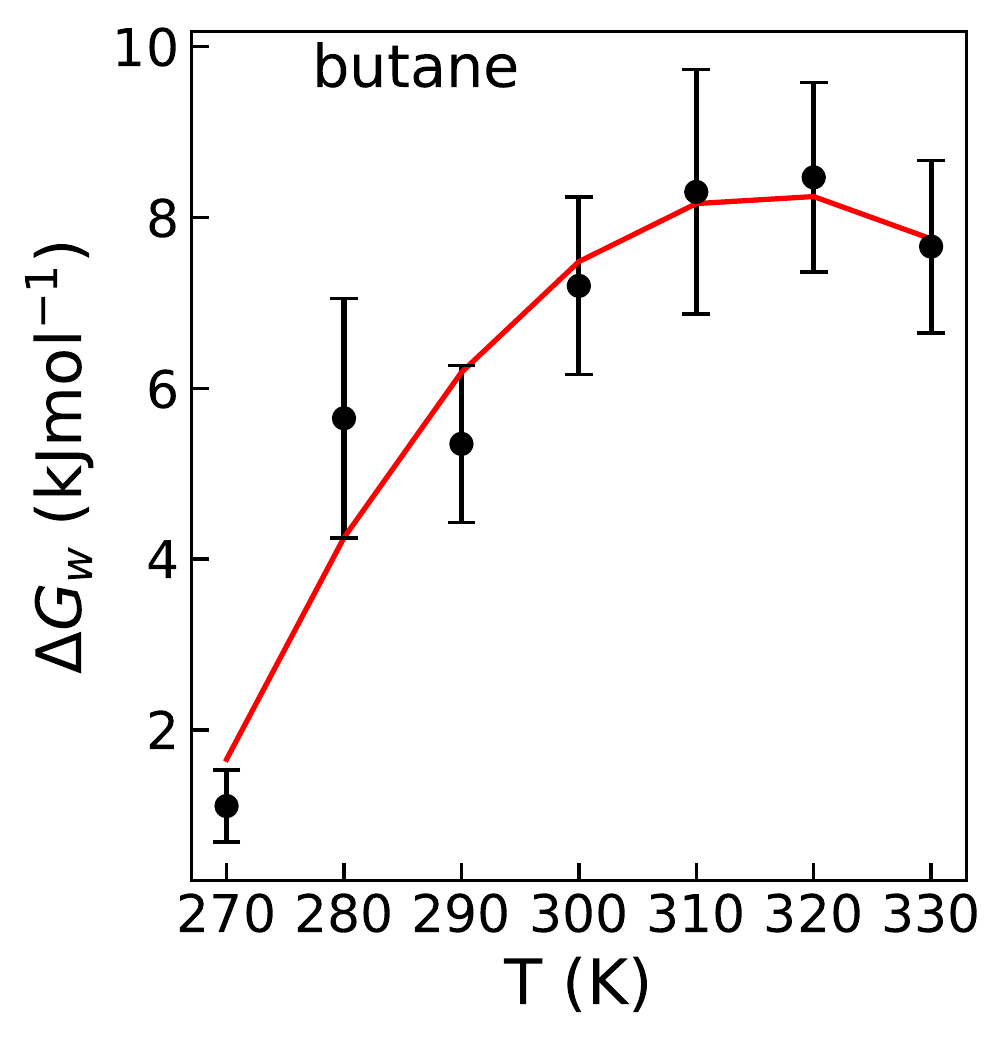}
\end{subfigure}&
\begin{subfigure}[c]{0.22\textwidth}
\includegraphics[height=4.5cm, width=\textwidth]{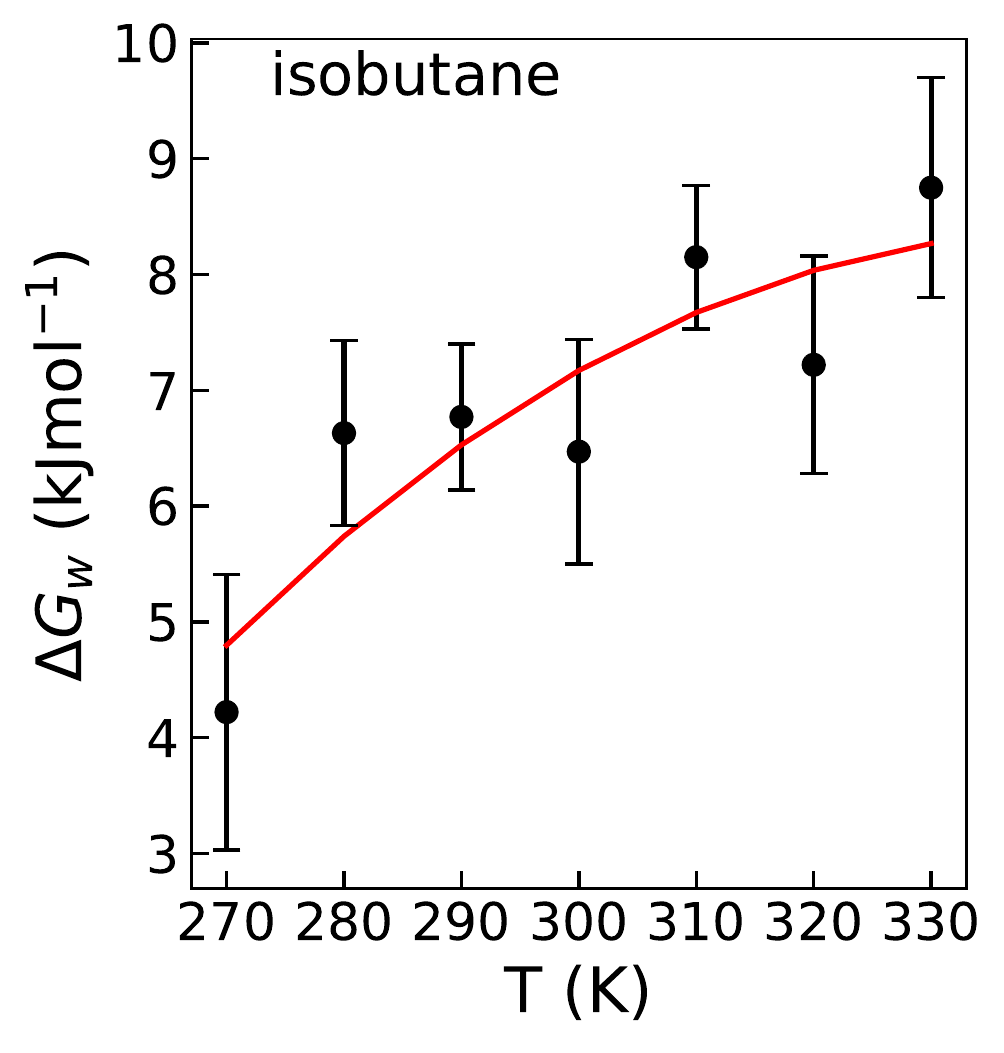}
\end{subfigure}&
\begin{subfigure}[c]{0.22\textwidth}
\includegraphics[height=4.5cm, width=\textwidth]{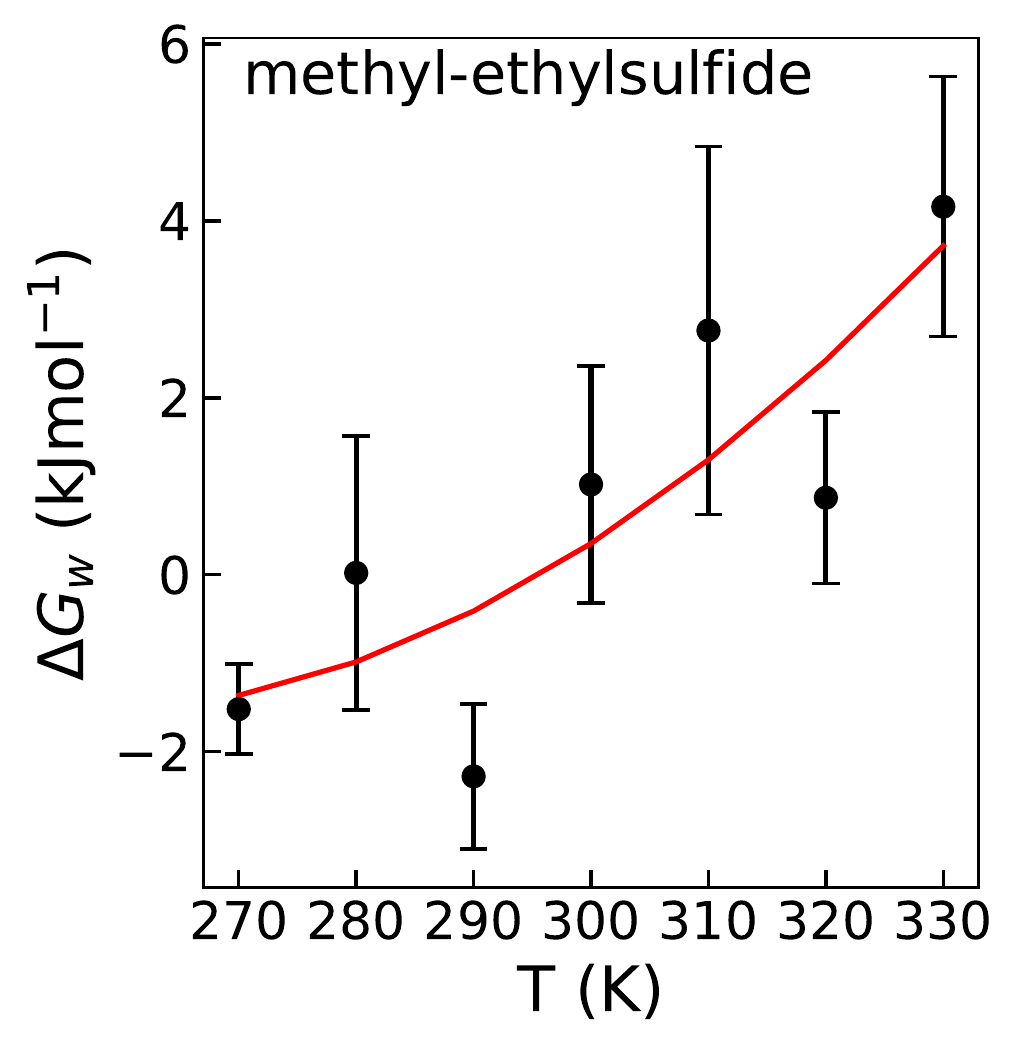}
\end{subfigure}&
\begin{subfigure}[c]{0.22\textwidth}
\includegraphics[height=4.5cm, width=\textwidth]{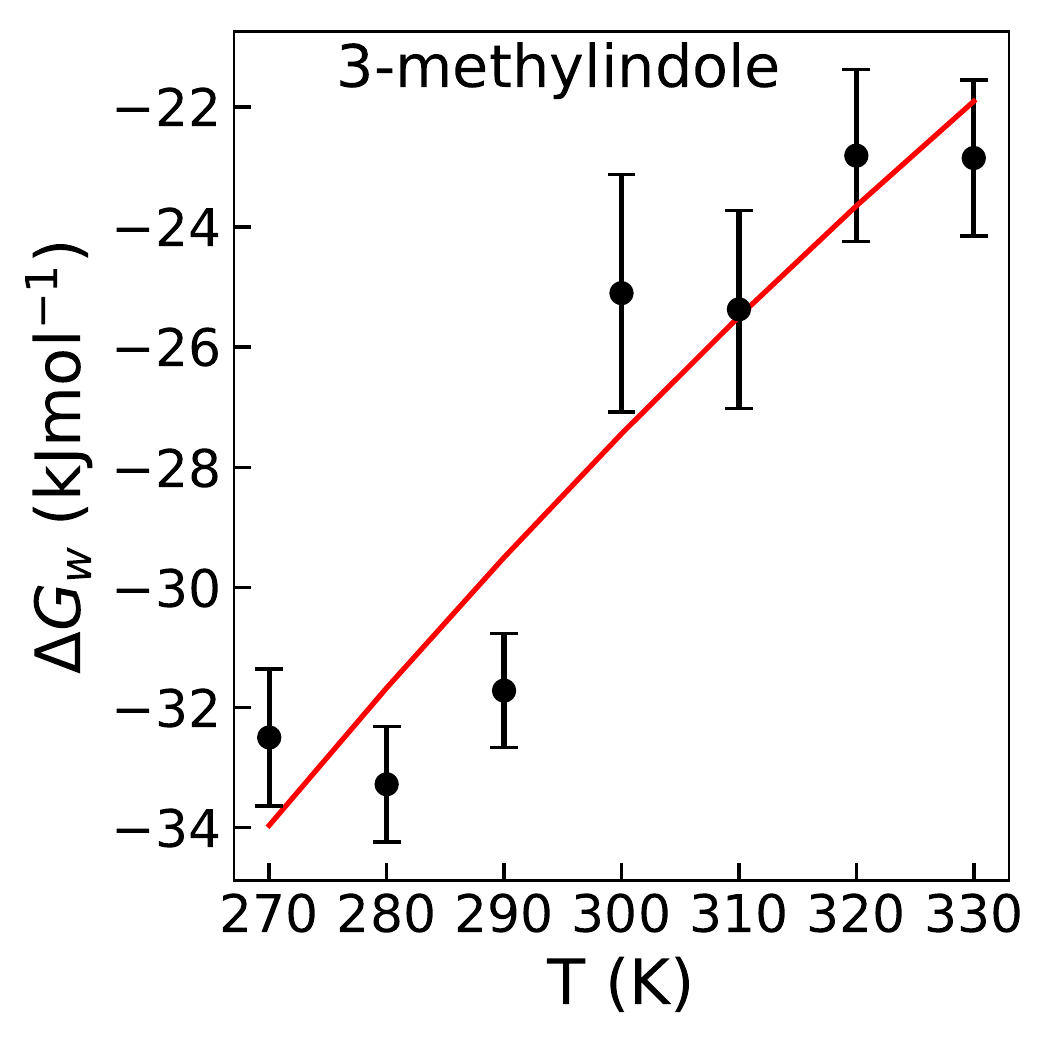}
\end{subfigure}\\  
\begin{subfigure}[c]{0.22\textwidth}
\includegraphics[height=4.5cm, width=\textwidth]{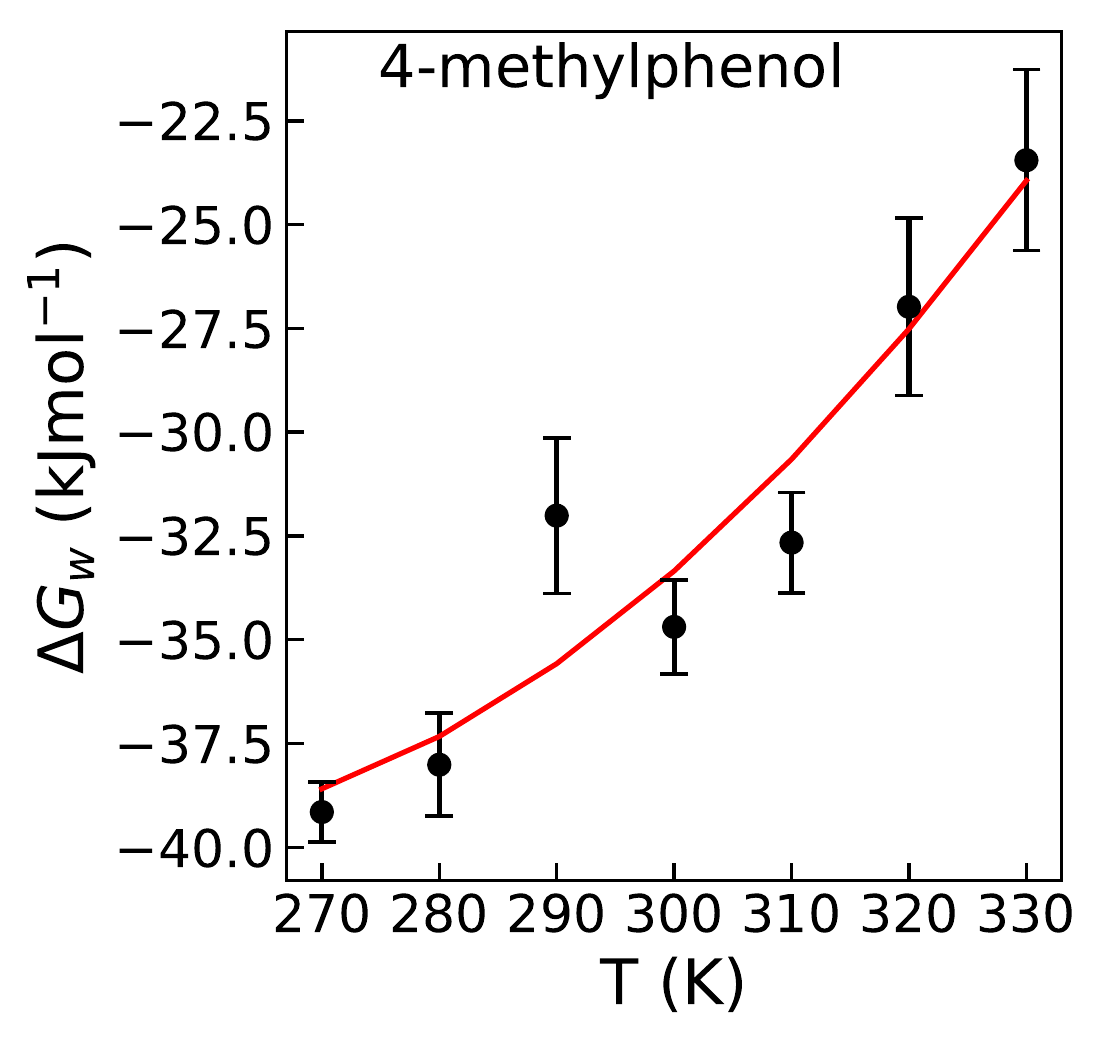}
\end{subfigure}&
\begin{subfigure}[c]{0.22\textwidth}
\includegraphics[height=4.5cm, width=\textwidth]{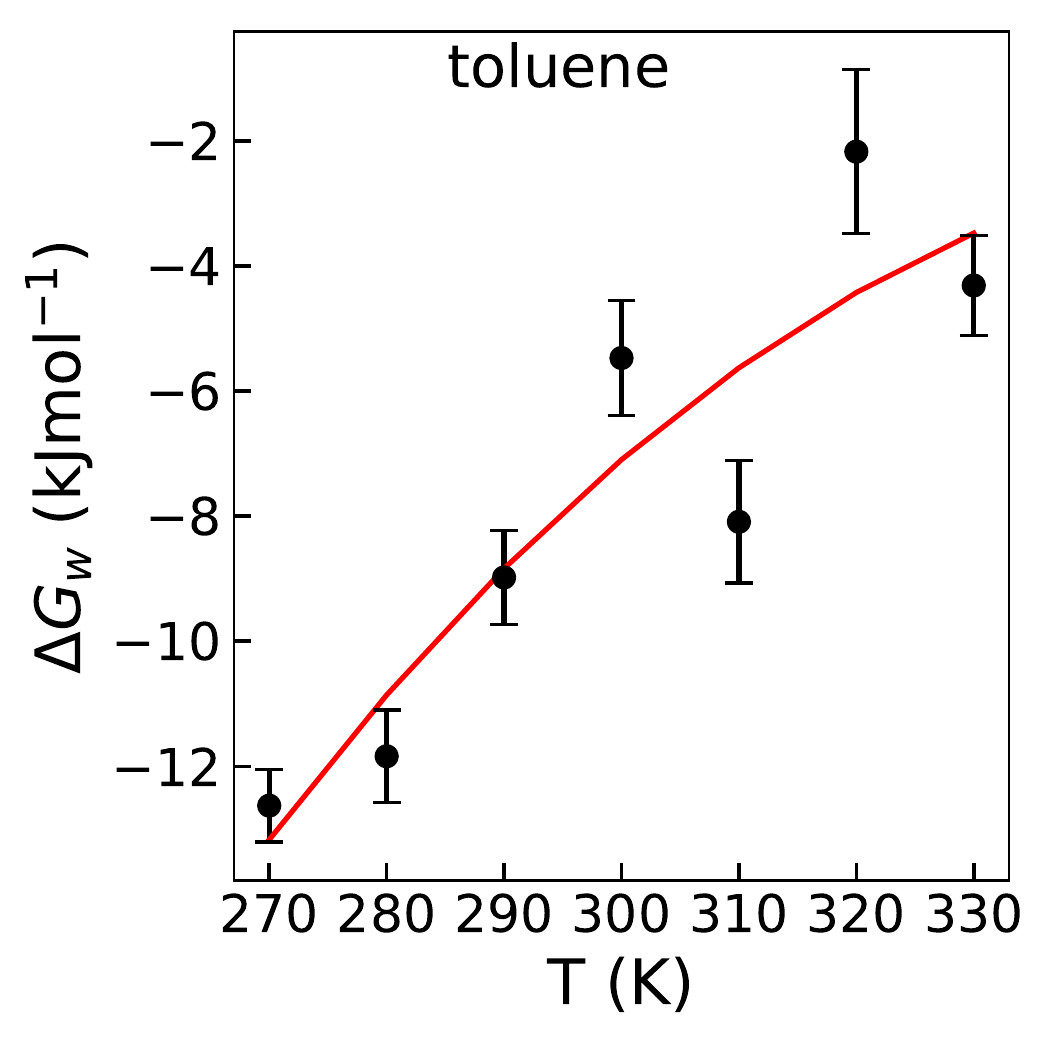}
\end{subfigure}&
\begin{subfigure}[c]{0.22\textwidth}
\includegraphics[height=4.5cm, width=\textwidth]{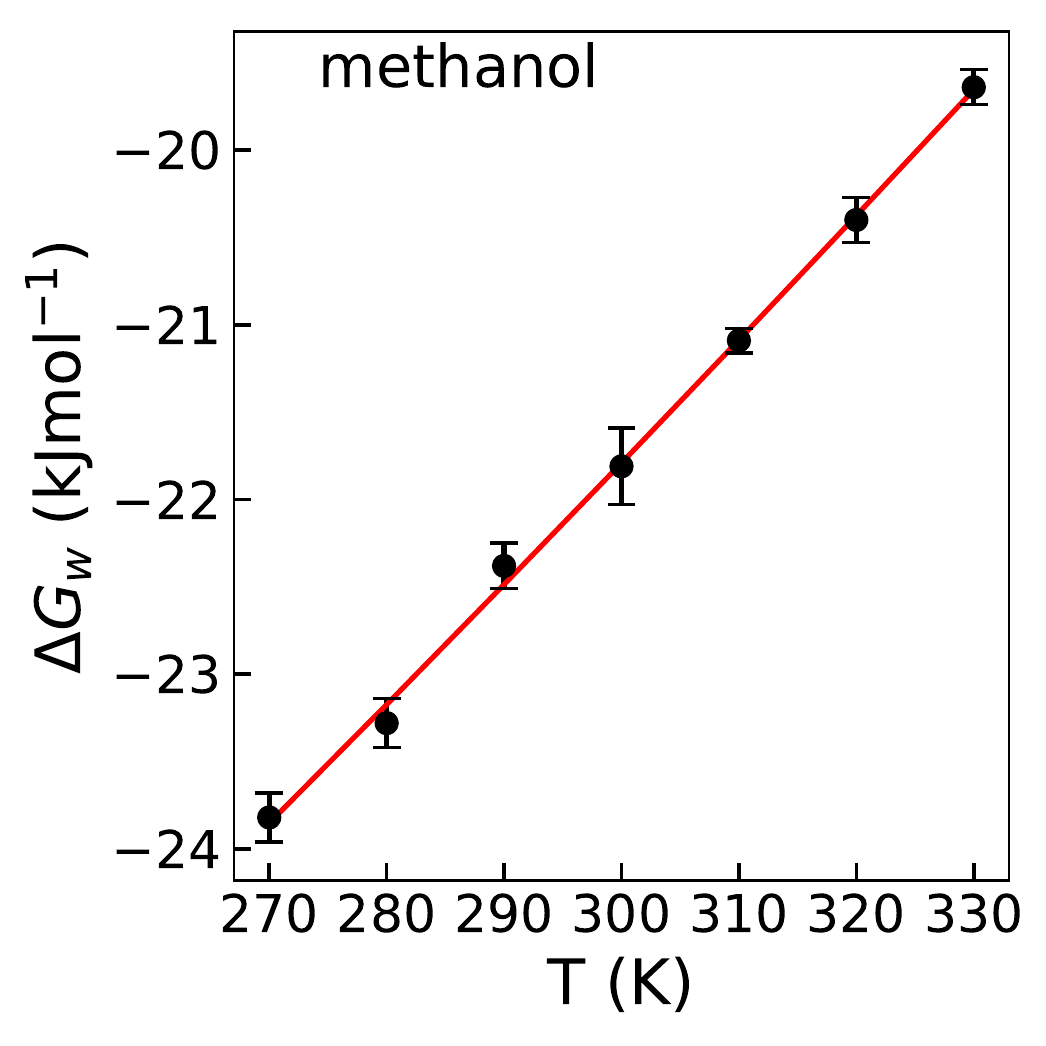}
\end{subfigure}&
\begin{subfigure}[c]{0.22\textwidth}
\includegraphics[height=4.5cm, width=\textwidth]{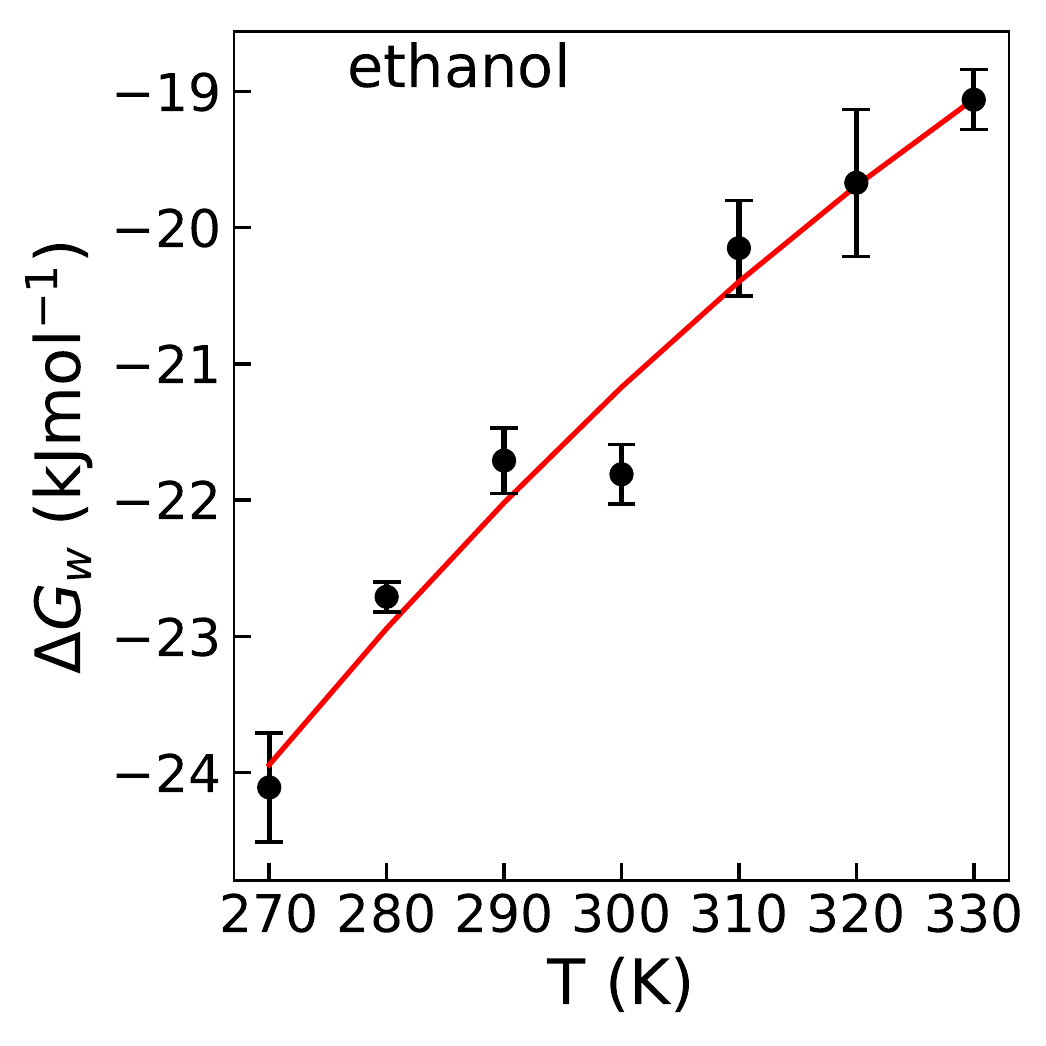}
\end{subfigure}&
\begin{subfigure}[c]{0.22\textwidth}
\includegraphics[height=4.5cm, width=\textwidth]{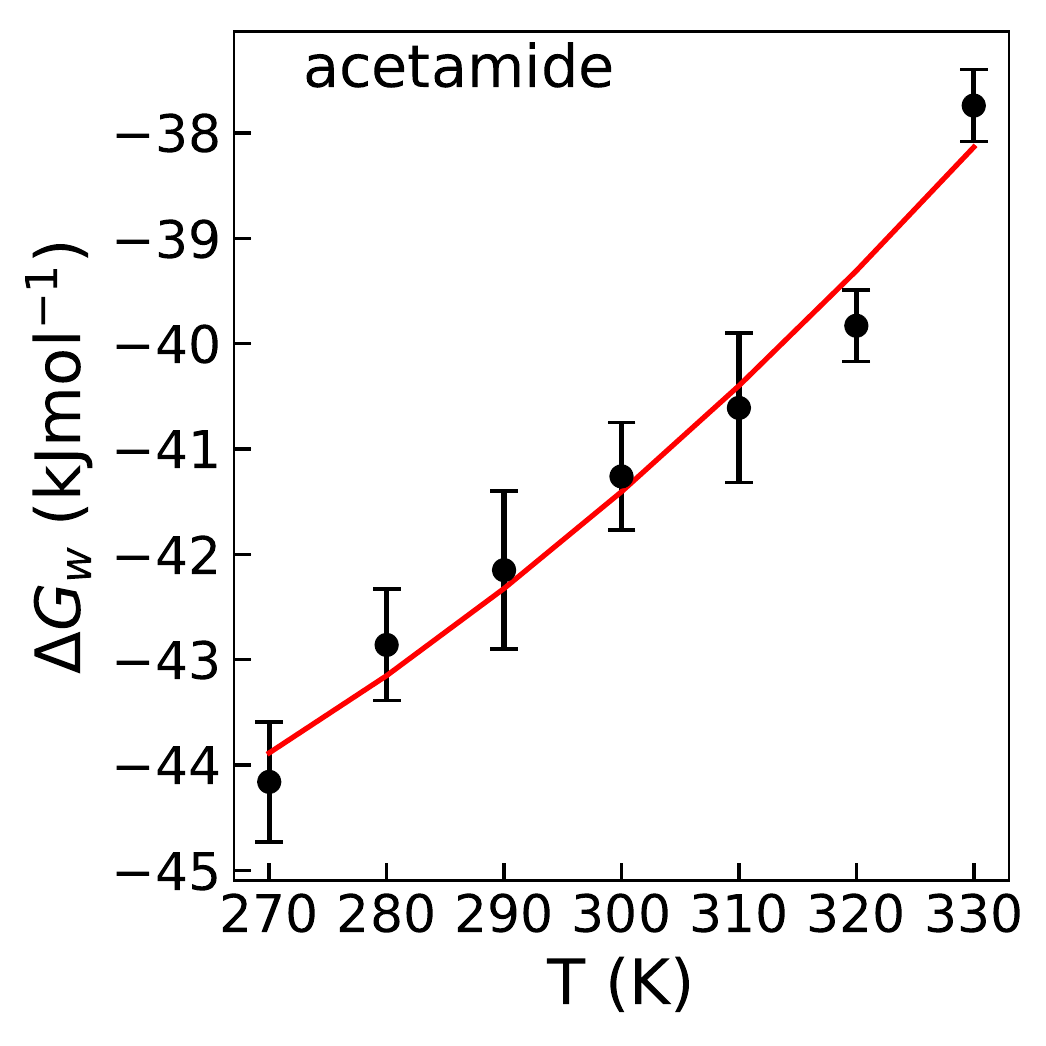}
\end{subfigure}&
\begin{subfigure}[c]{0.22\textwidth}
\includegraphics[height=4.5cm, width=\textwidth]{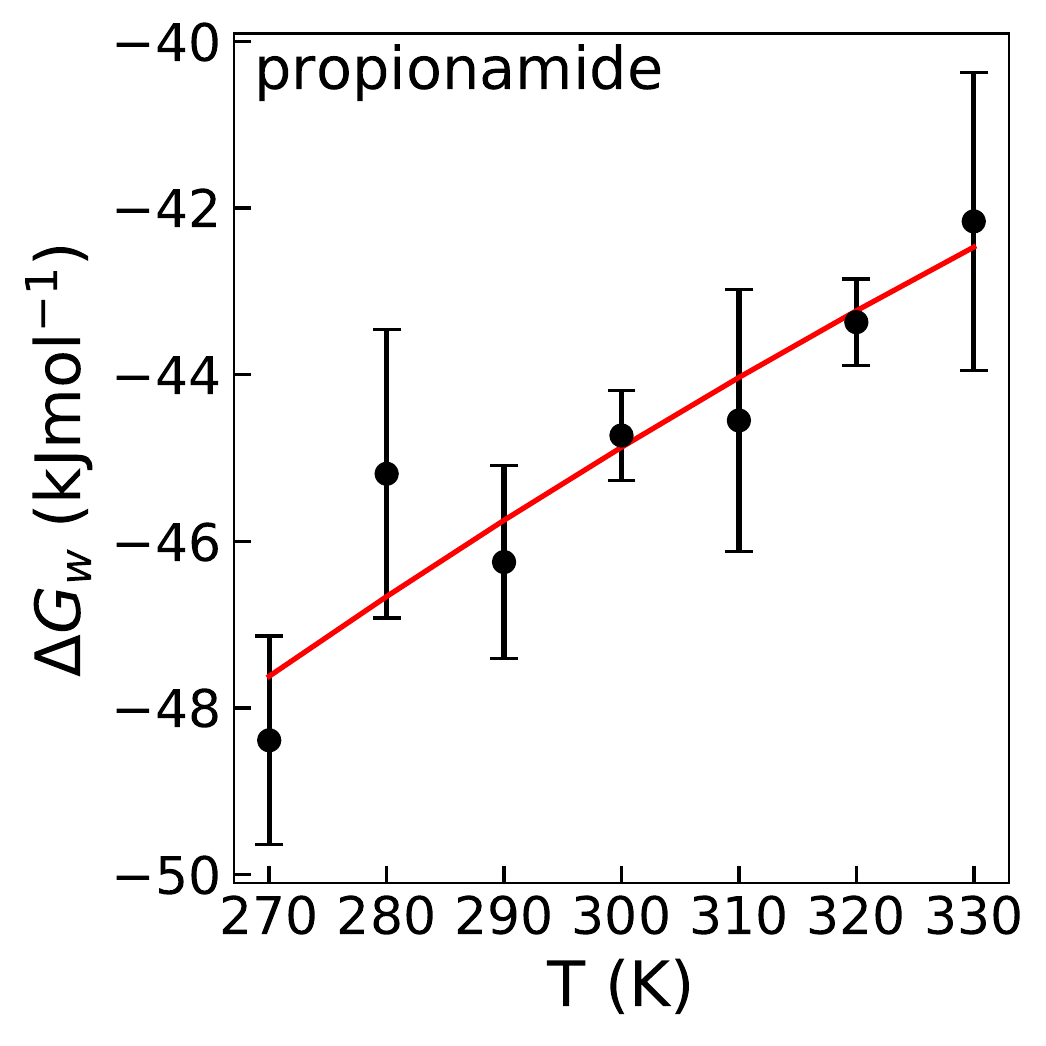}
\end{subfigure}\\
\begin{subfigure}[c]{0.22\textwidth}
\includegraphics[height=4.5cm, width=\textwidth]{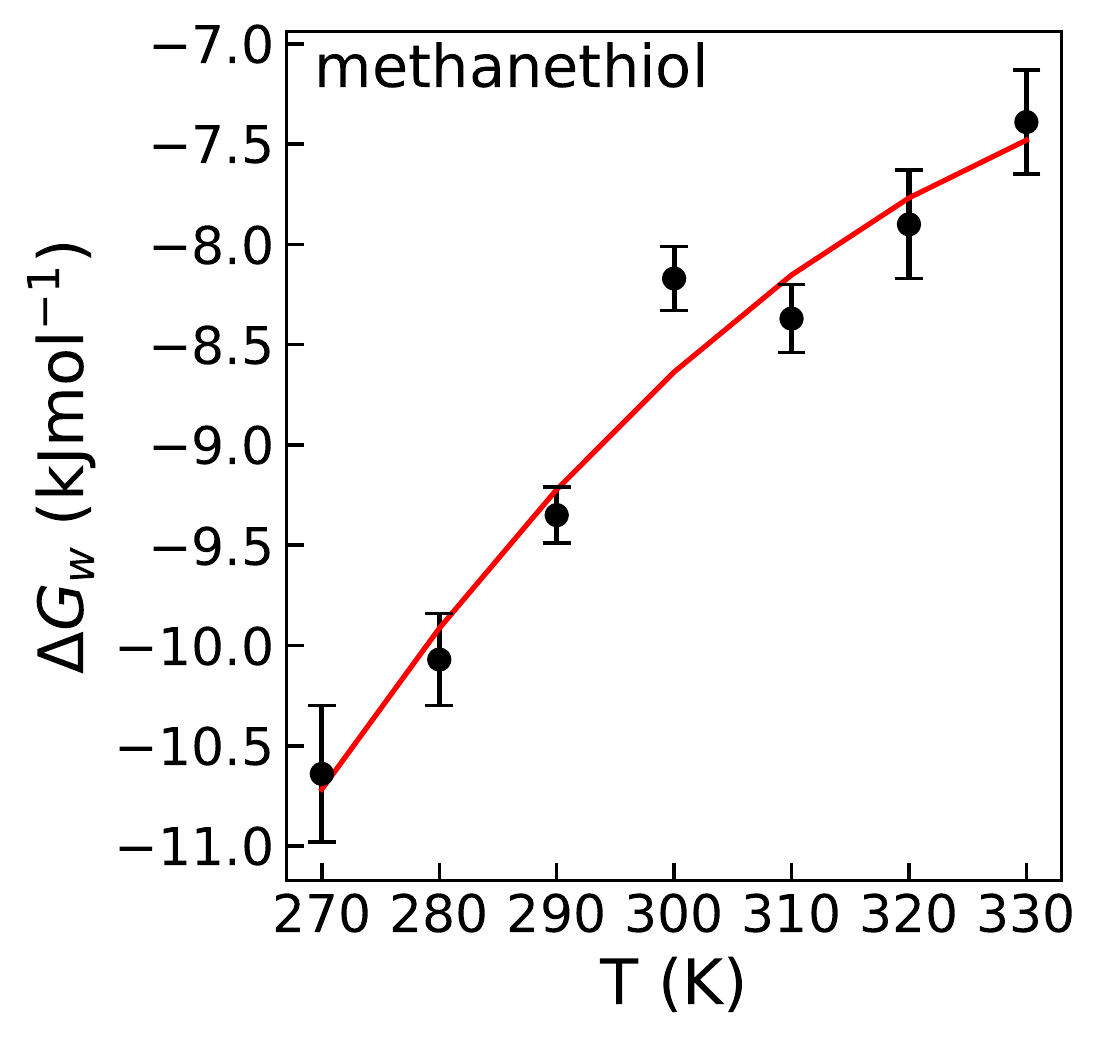}
\end{subfigure}&
\begin{subfigure}[c]{0.22\textwidth}
\includegraphics[height=4.5cm, width=\textwidth]{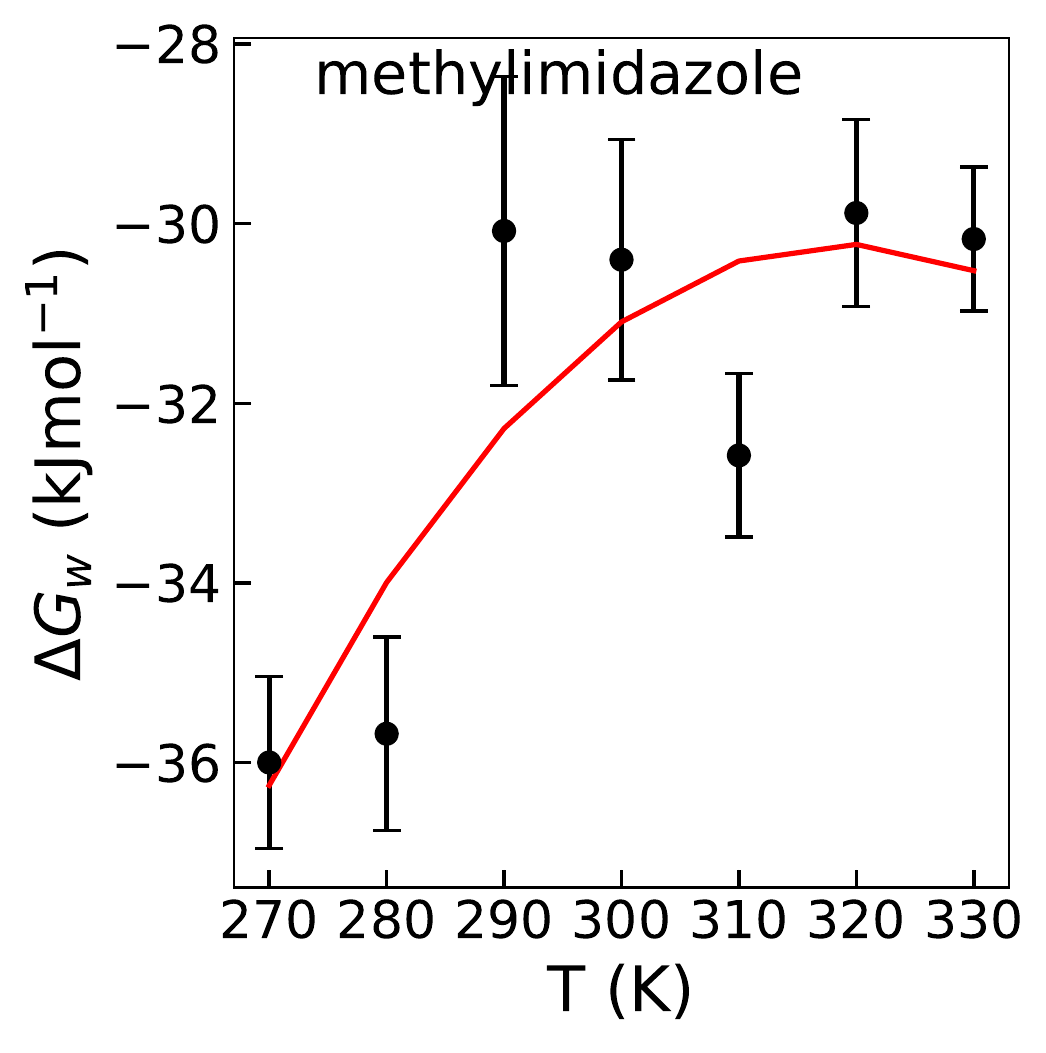}
\end{subfigure}&
\begin{subfigure}[c]{0.22\textwidth}
\includegraphics[height=4.5cm, width=\textwidth]{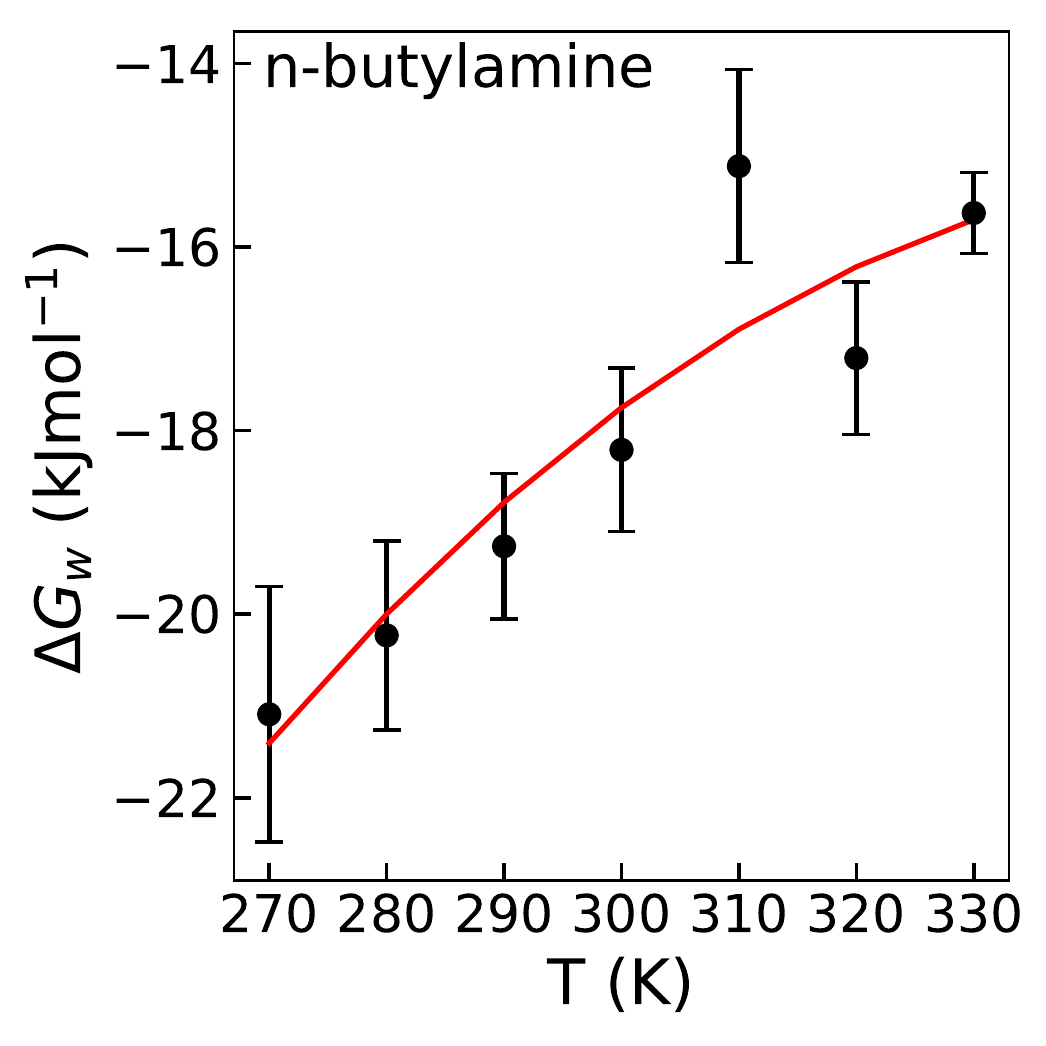}
\end{subfigure}&
\begin{subfigure}[c]{0.22\textwidth}
\includegraphics[height=4.5cm, width=\textwidth]{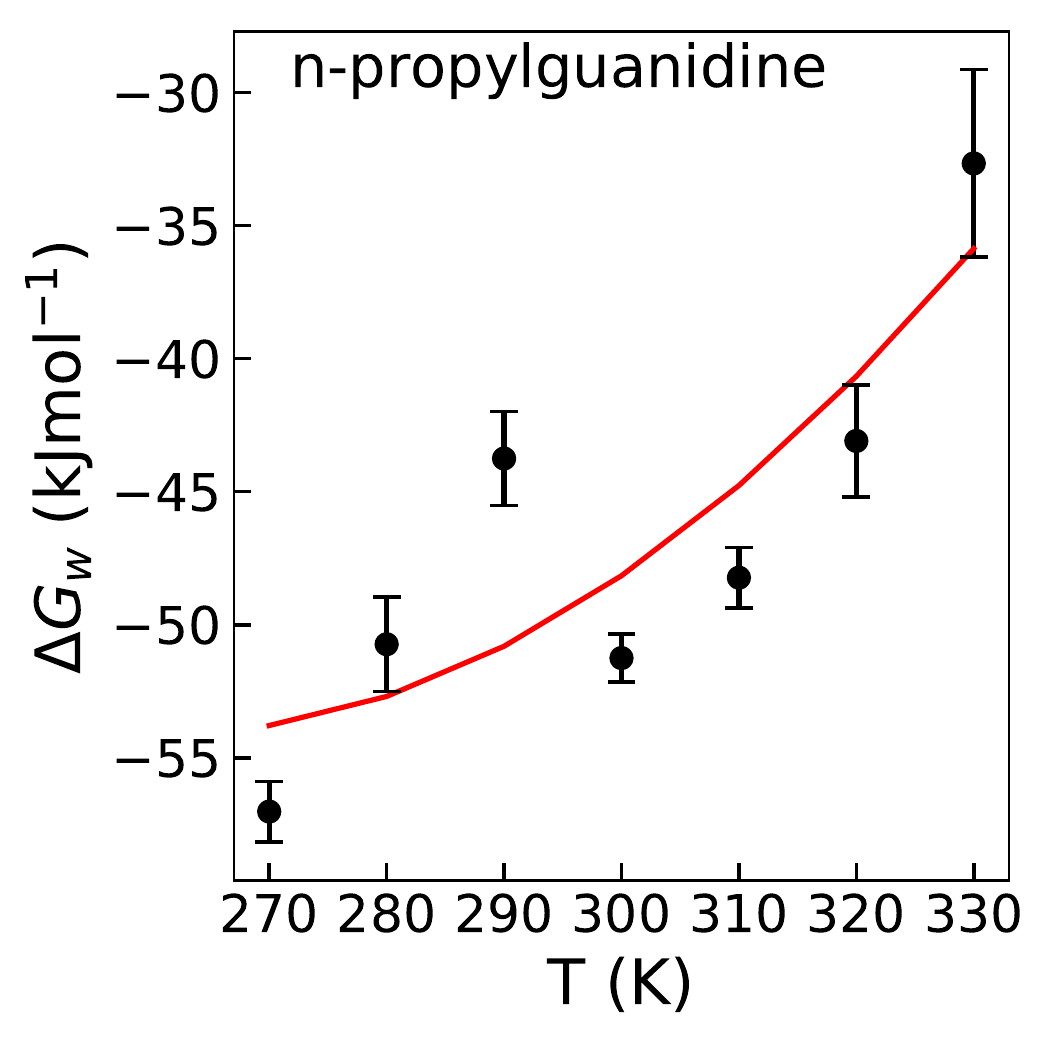}
\end{subfigure}&
\begin{subfigure}[c]{0.22\textwidth}
\includegraphics[height=4.5cm, width=\textwidth]{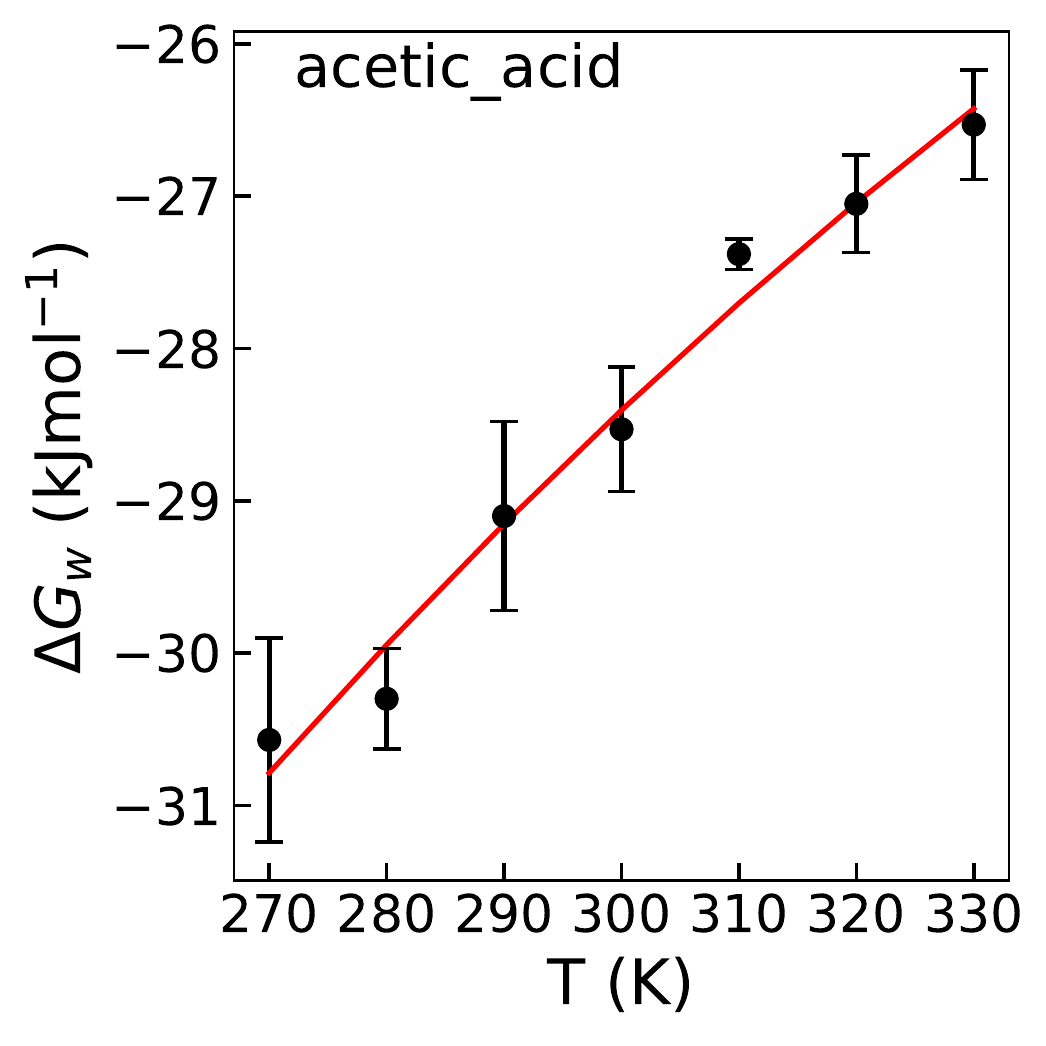}
\end{subfigure}&
\begin{subfigure}[c]{0.22\textwidth}
\includegraphics[height=4.5cm, width=\textwidth]{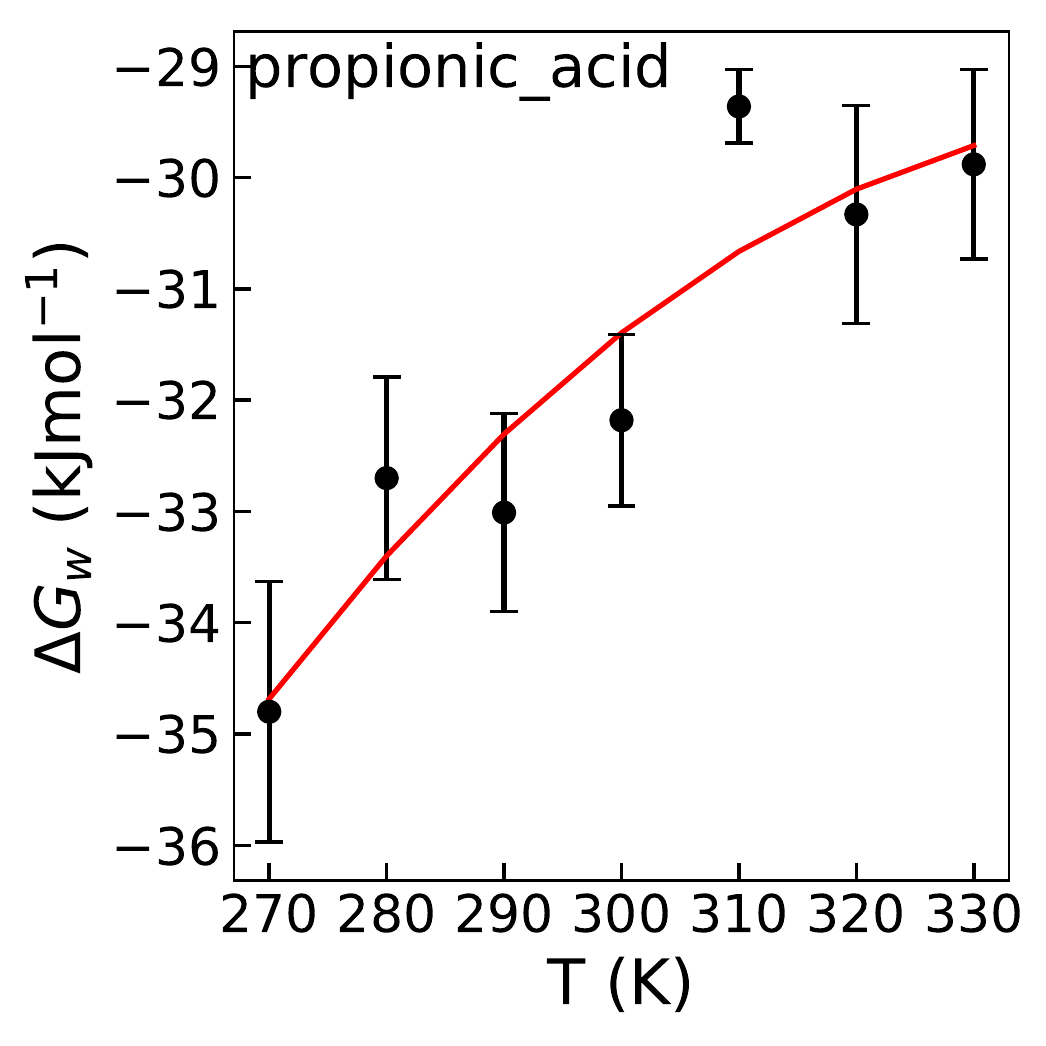}
\end{subfigure}\\        
\end{tabular}    
\caption{Temperature dependence of the solvation free energy from gas to water \ce{H2O}, $\Delta G_{w}$.}
\label{fig:water}
\end{figure}
\end{turnpage}

\begin{turnpage}
\begin{figure}[h]
\centering
\begin{tabular}[c]{cccccc}
\begin{subfigure}[c]{0.22\textwidth}
\includegraphics[height=4.5cm, width=\textwidth]{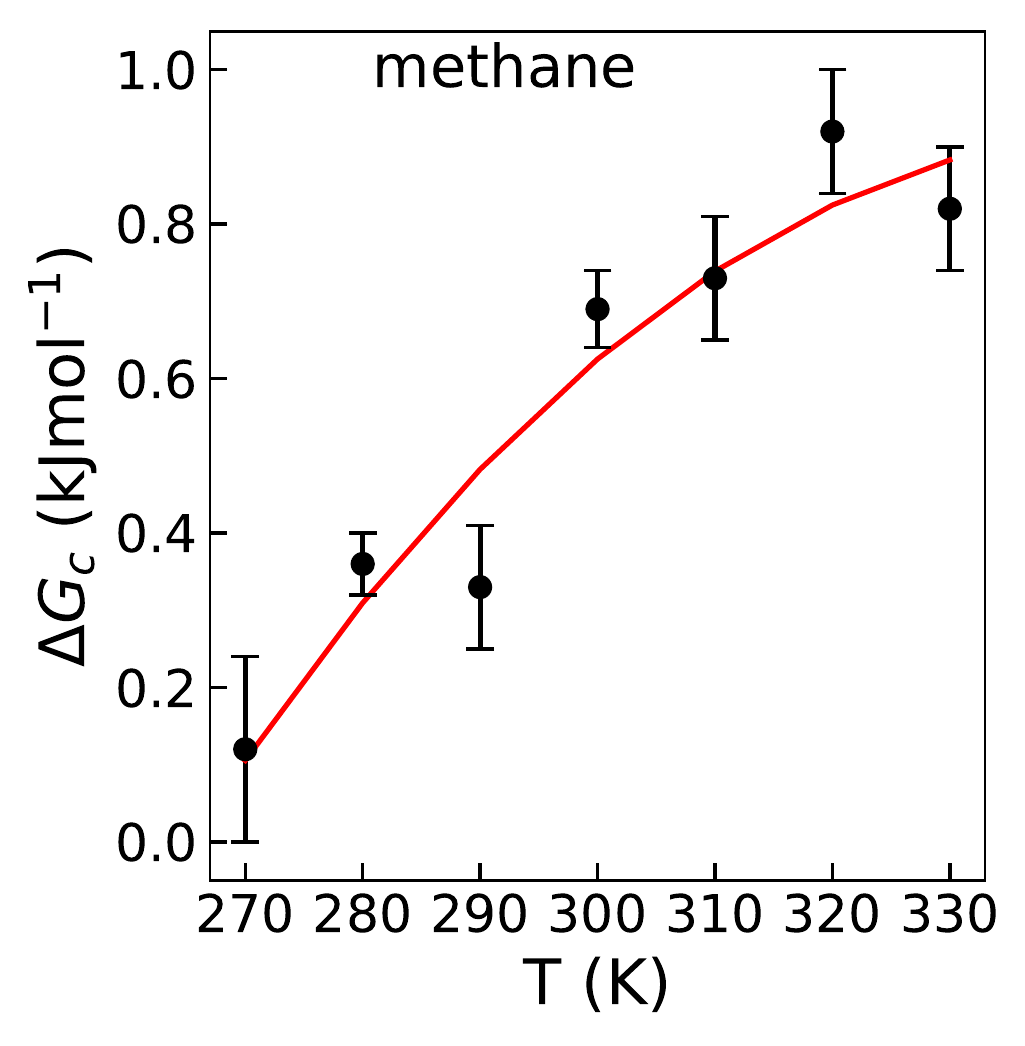}
\end{subfigure}&
\begin{subfigure}[c]{0.22\textwidth}
\includegraphics[height=4.5cm, width=\textwidth]{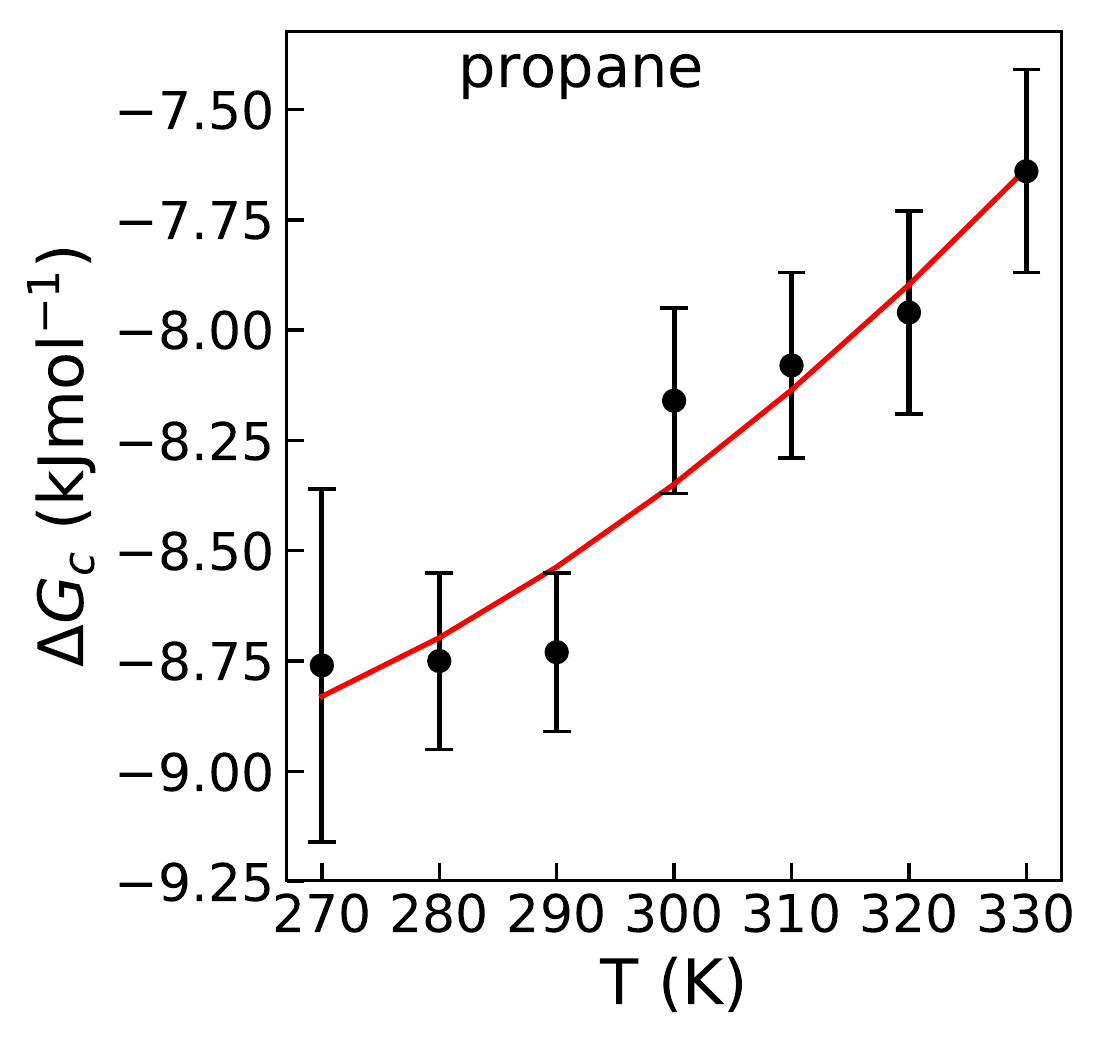}
\end{subfigure}&
\begin{subfigure}[c]{0.22\textwidth}
\includegraphics[height=4.5cm, width=\textwidth]{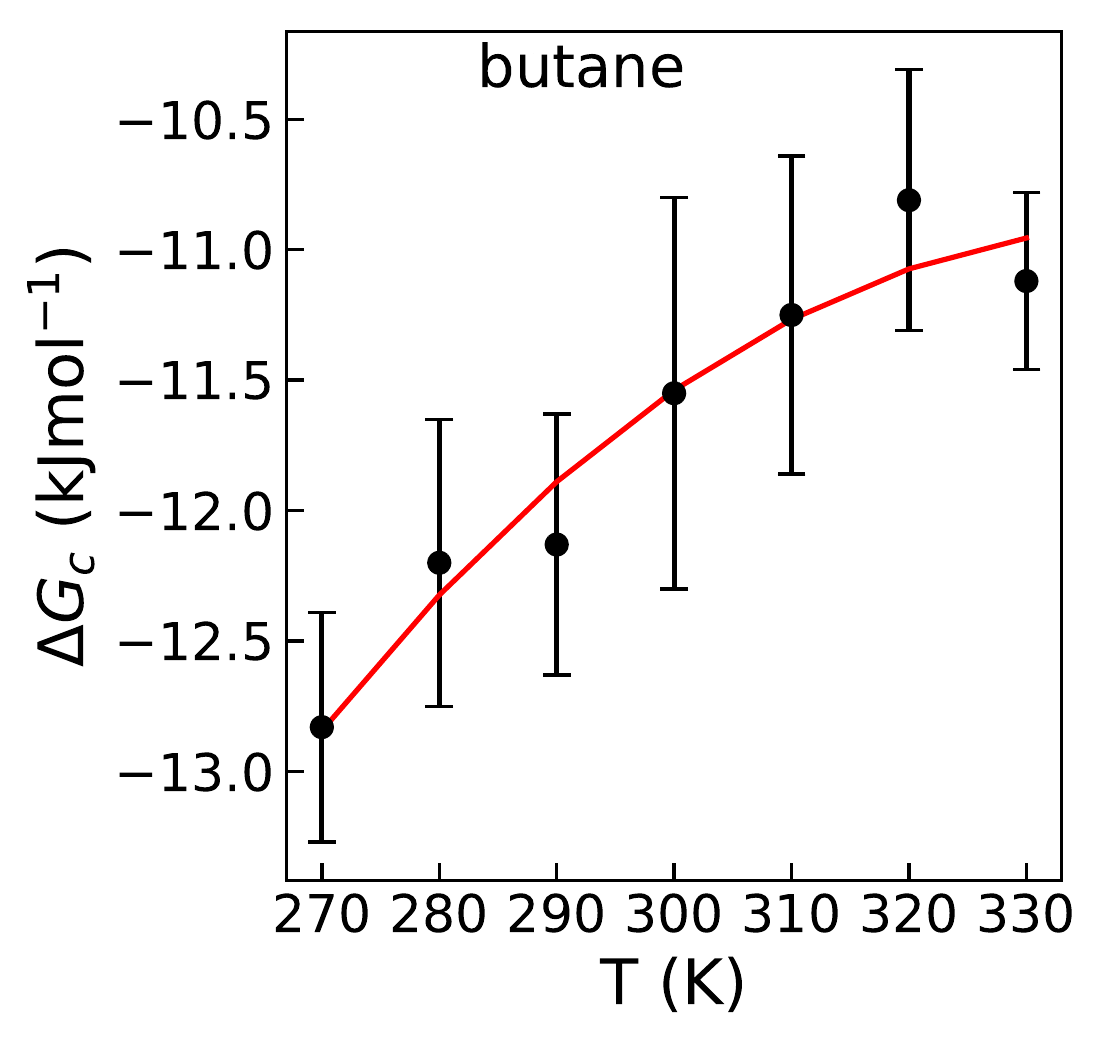}
\end{subfigure}&
\begin{subfigure}[c]{0.22\textwidth}
\includegraphics[height=4.5cm, width=\textwidth]{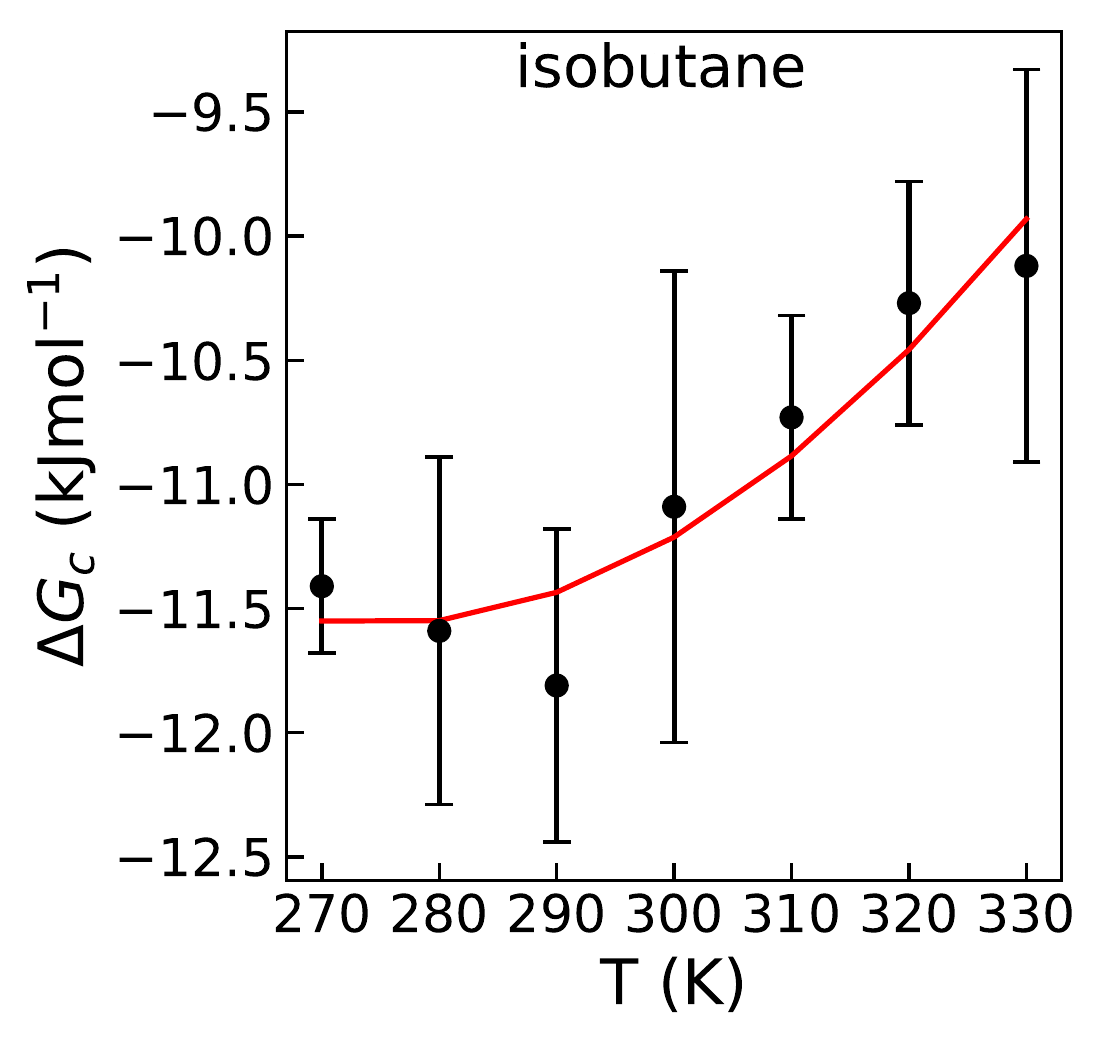}
\end{subfigure}&
\begin{subfigure}[c]{0.22\textwidth}
\includegraphics[height=4.5cm, width=\textwidth]{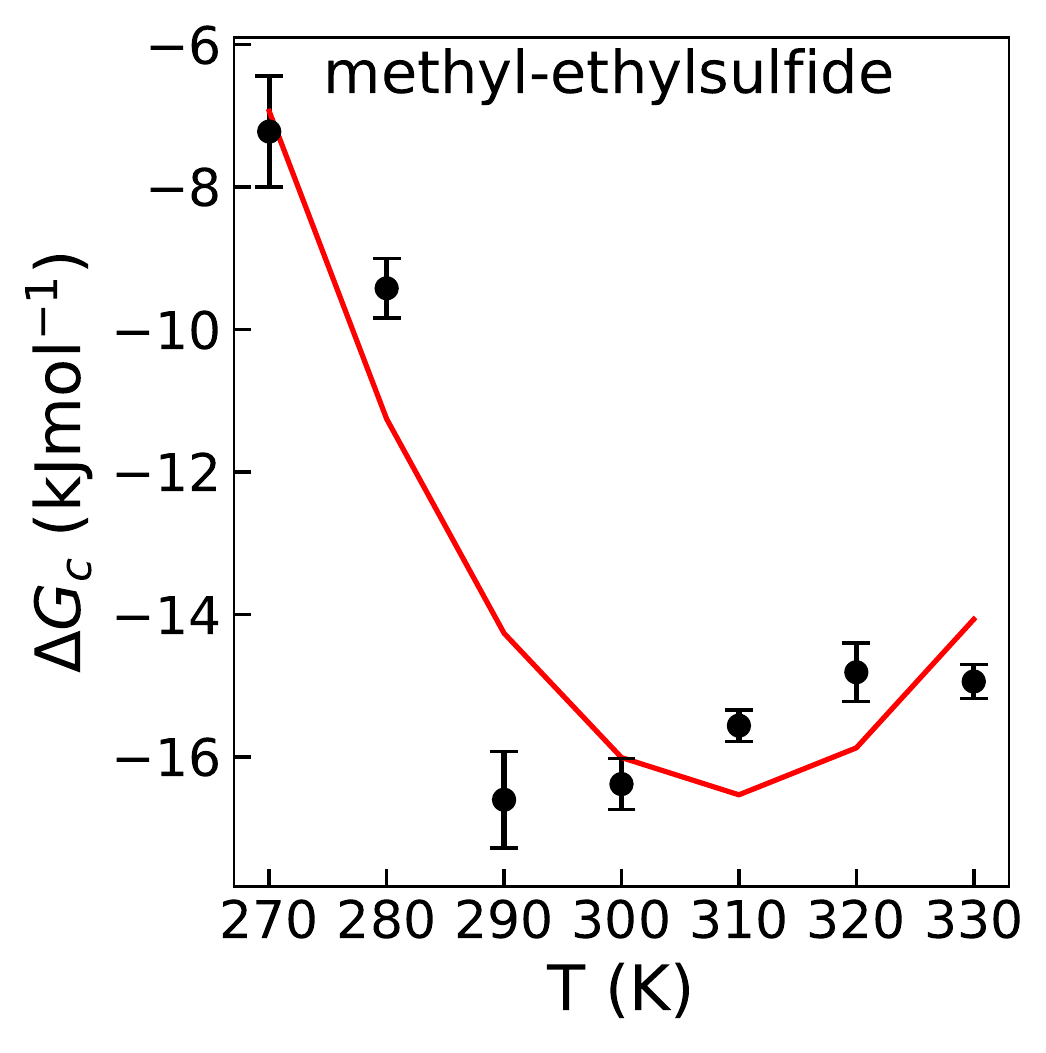}
\end{subfigure}&
\begin{subfigure}[c]{0.22\textwidth}
\includegraphics[height=4.5cm, width=\textwidth]{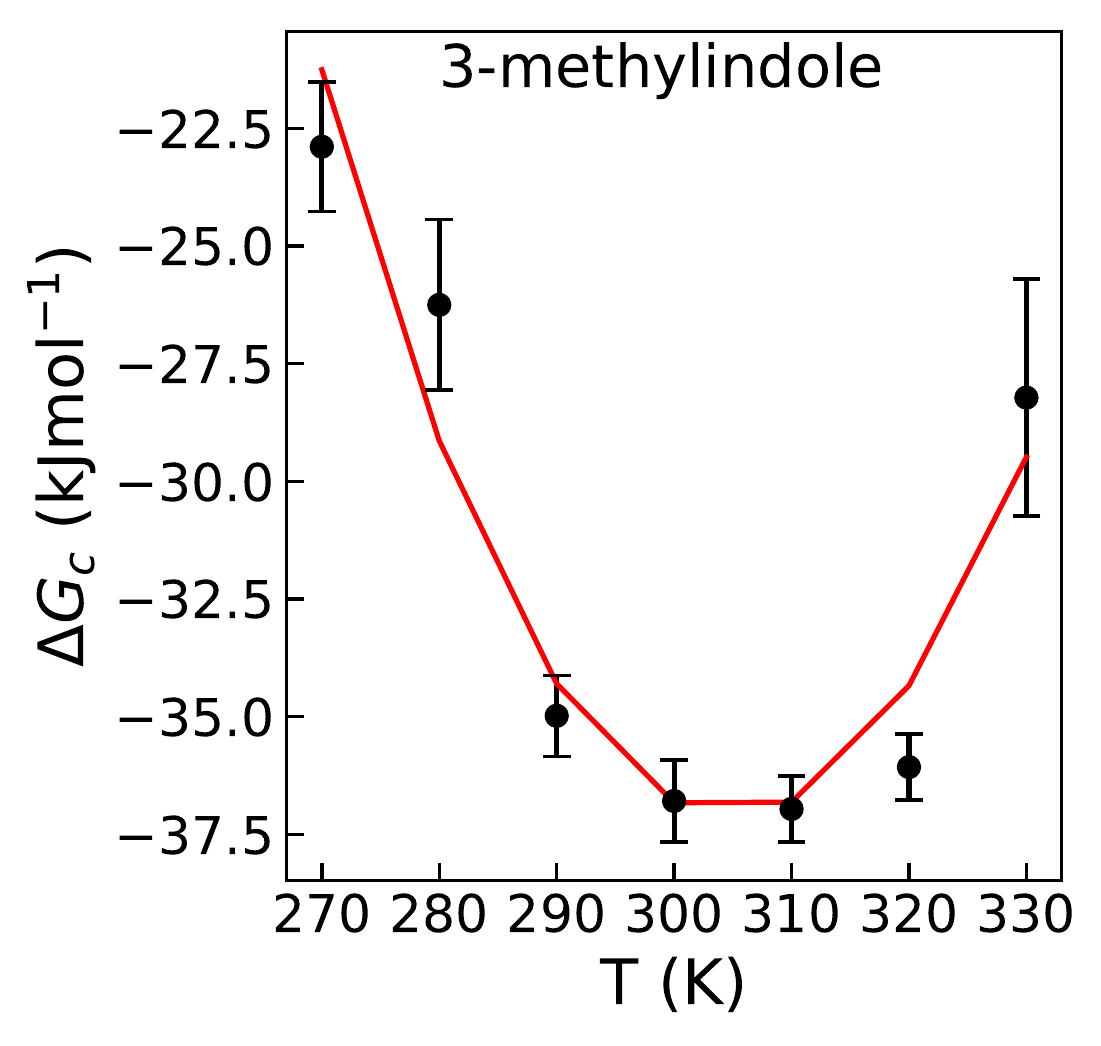}
\end{subfigure}\\  
\begin{subfigure}[c]{0.22\textwidth}
\includegraphics[height=4.5cm, width=\textwidth]{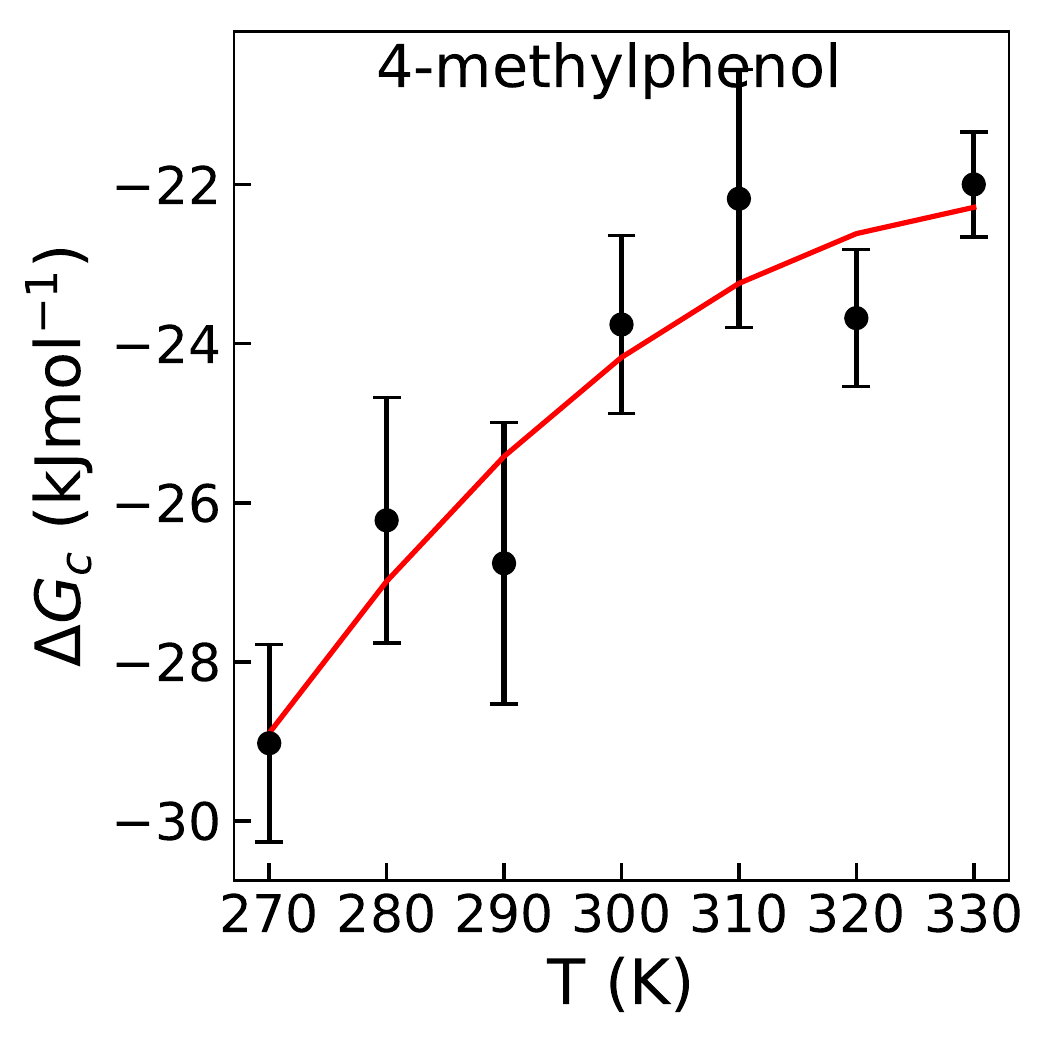}
\end{subfigure}&
\begin{subfigure}[c]{0.22\textwidth}
\includegraphics[height=4.5cm, width=\textwidth]{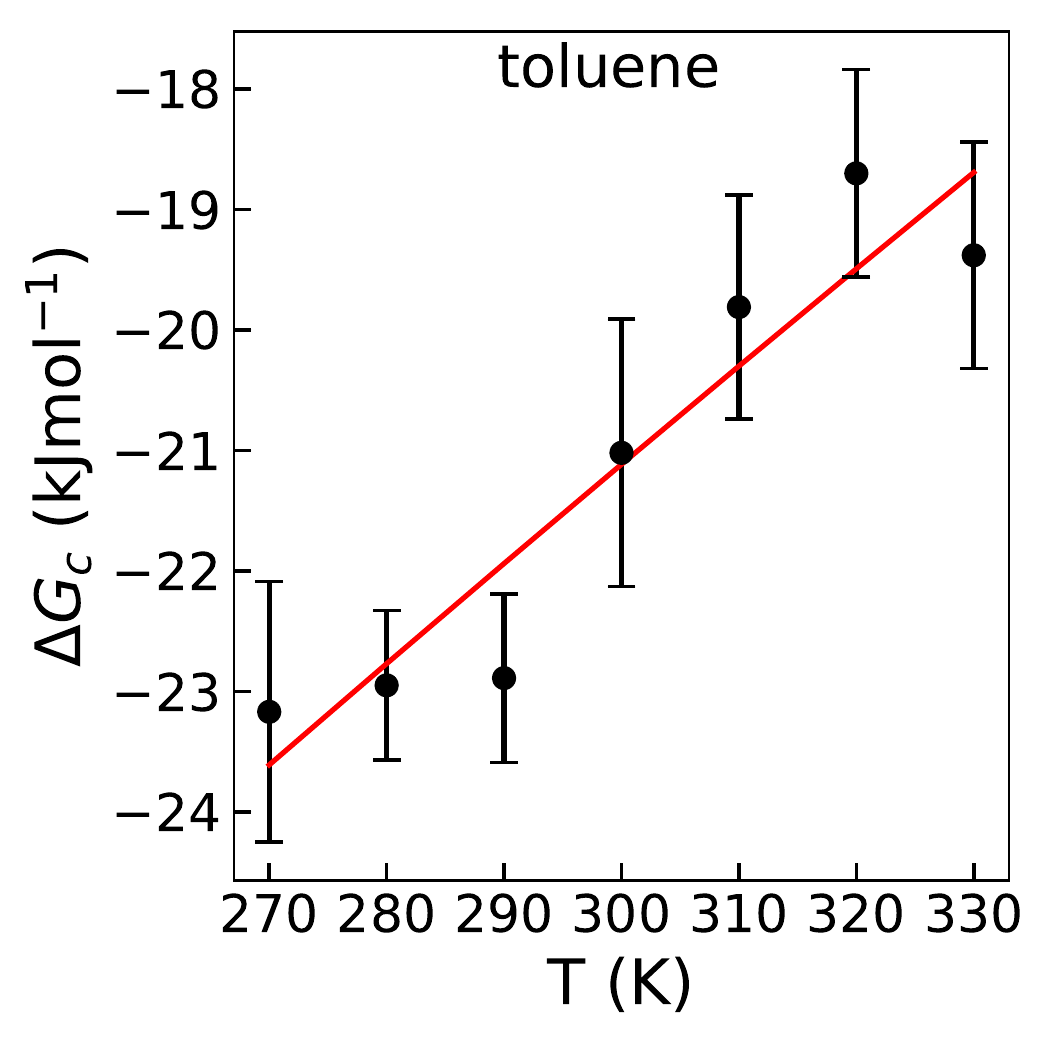}
\end{subfigure}&
\begin{subfigure}[c]{0.22\textwidth}
\includegraphics[height=4.5cm, width=\textwidth]{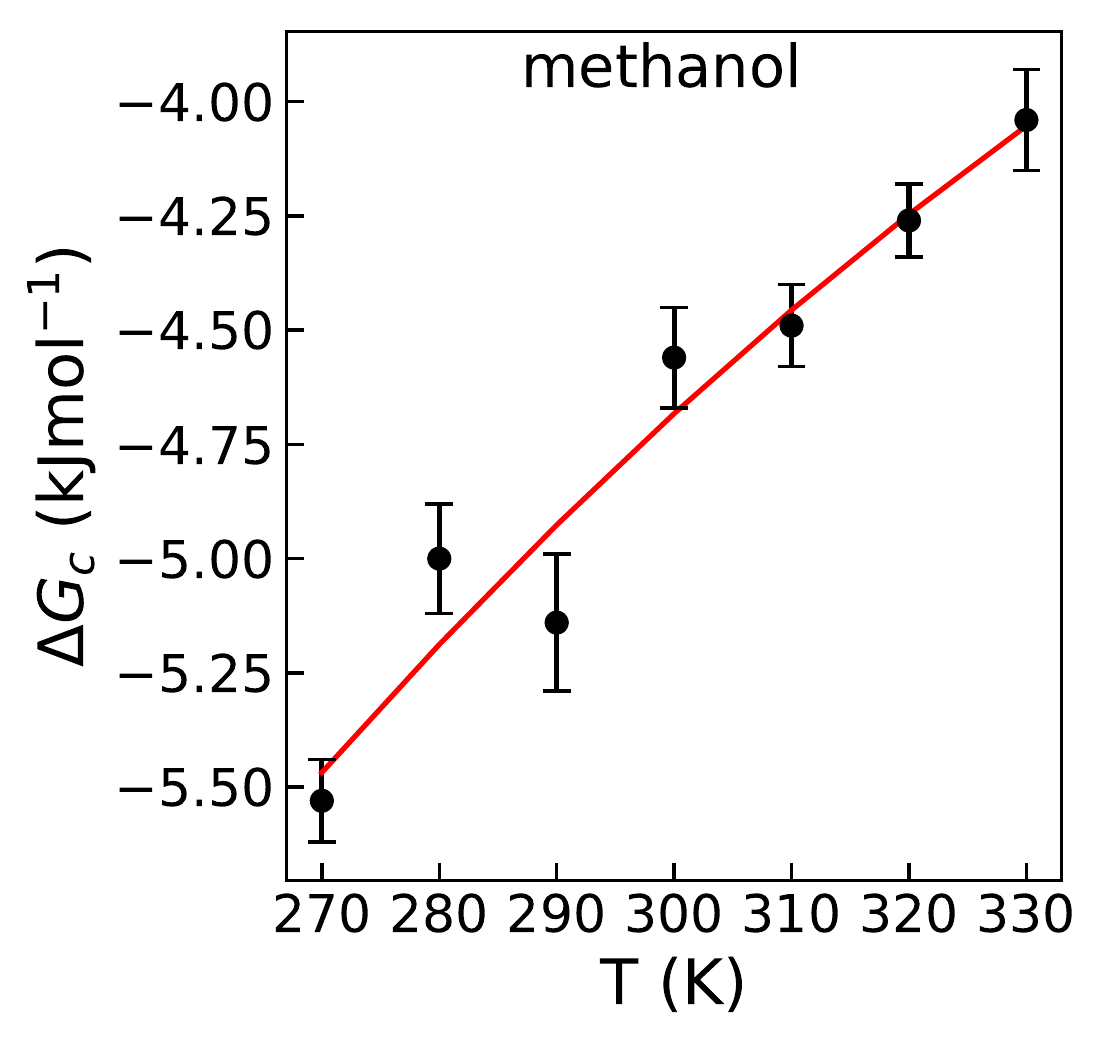}
\end{subfigure}&
\begin{subfigure}[c]{0.22\textwidth}
\includegraphics[height=4.5cm, width=\textwidth]{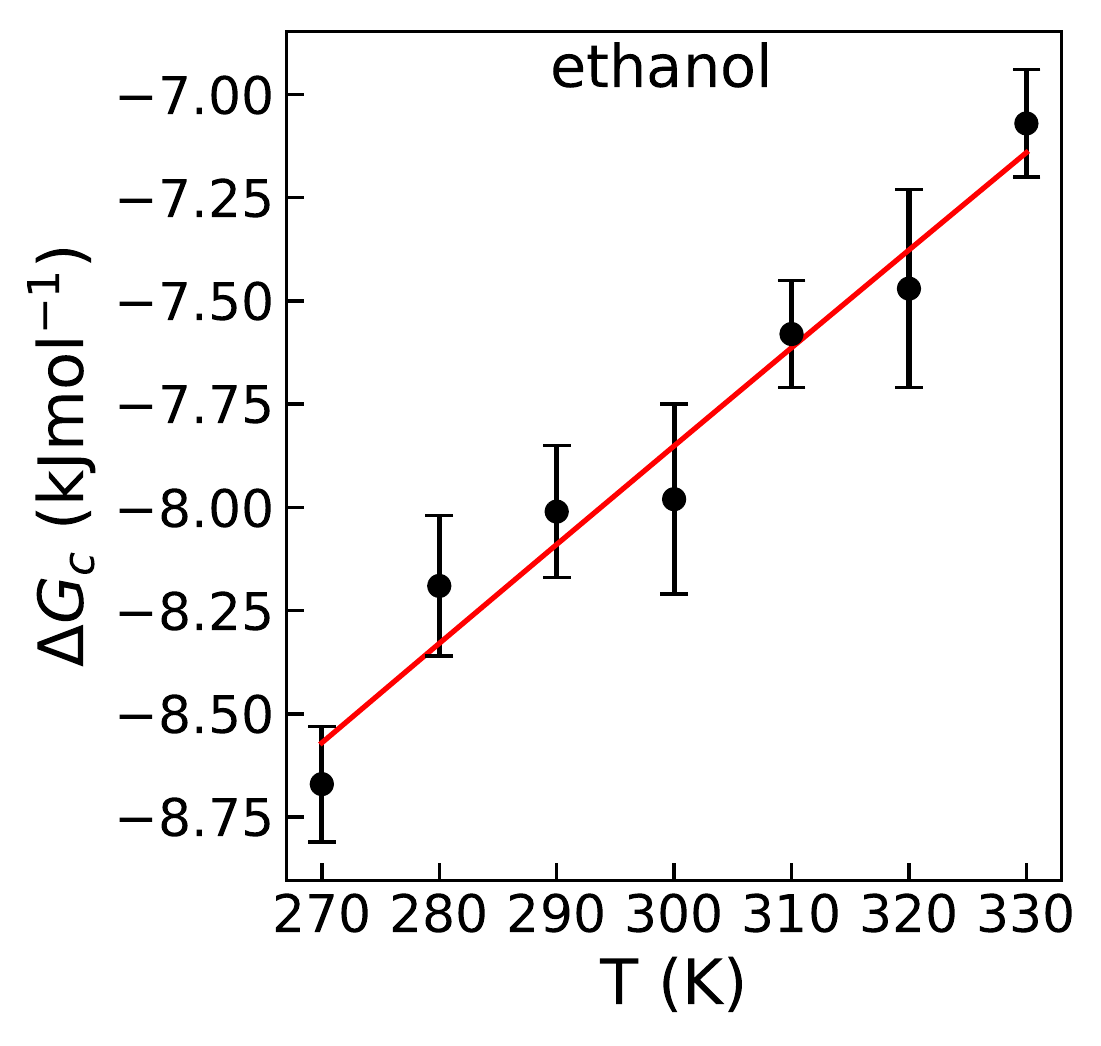}
\end{subfigure}&
\begin{subfigure}[c]{0.22\textwidth}
\includegraphics[height=4.5cm, width=\textwidth]{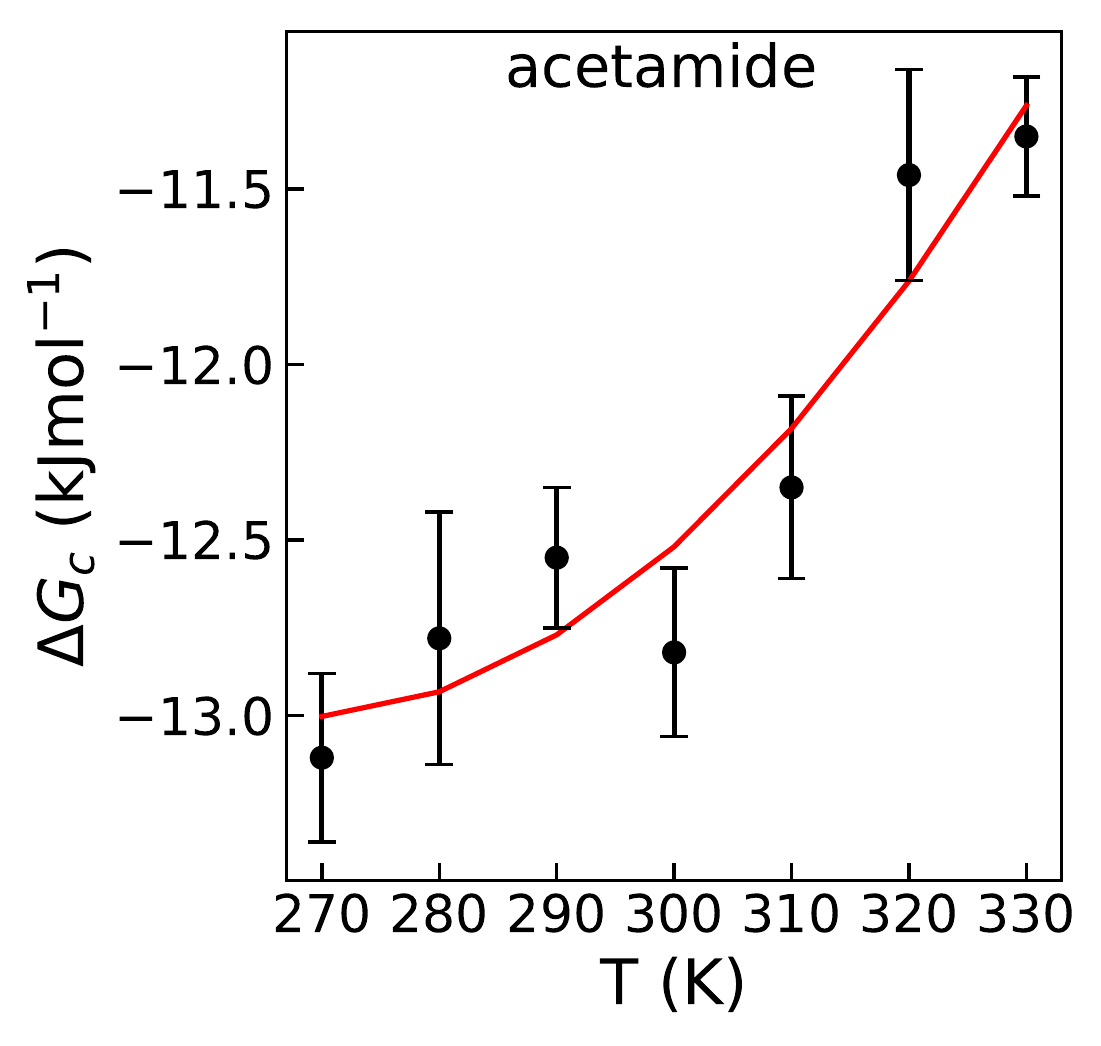}
\end{subfigure}&
\begin{subfigure}[c]{0.22\textwidth}
\includegraphics[height=4.5cm, width=\textwidth]{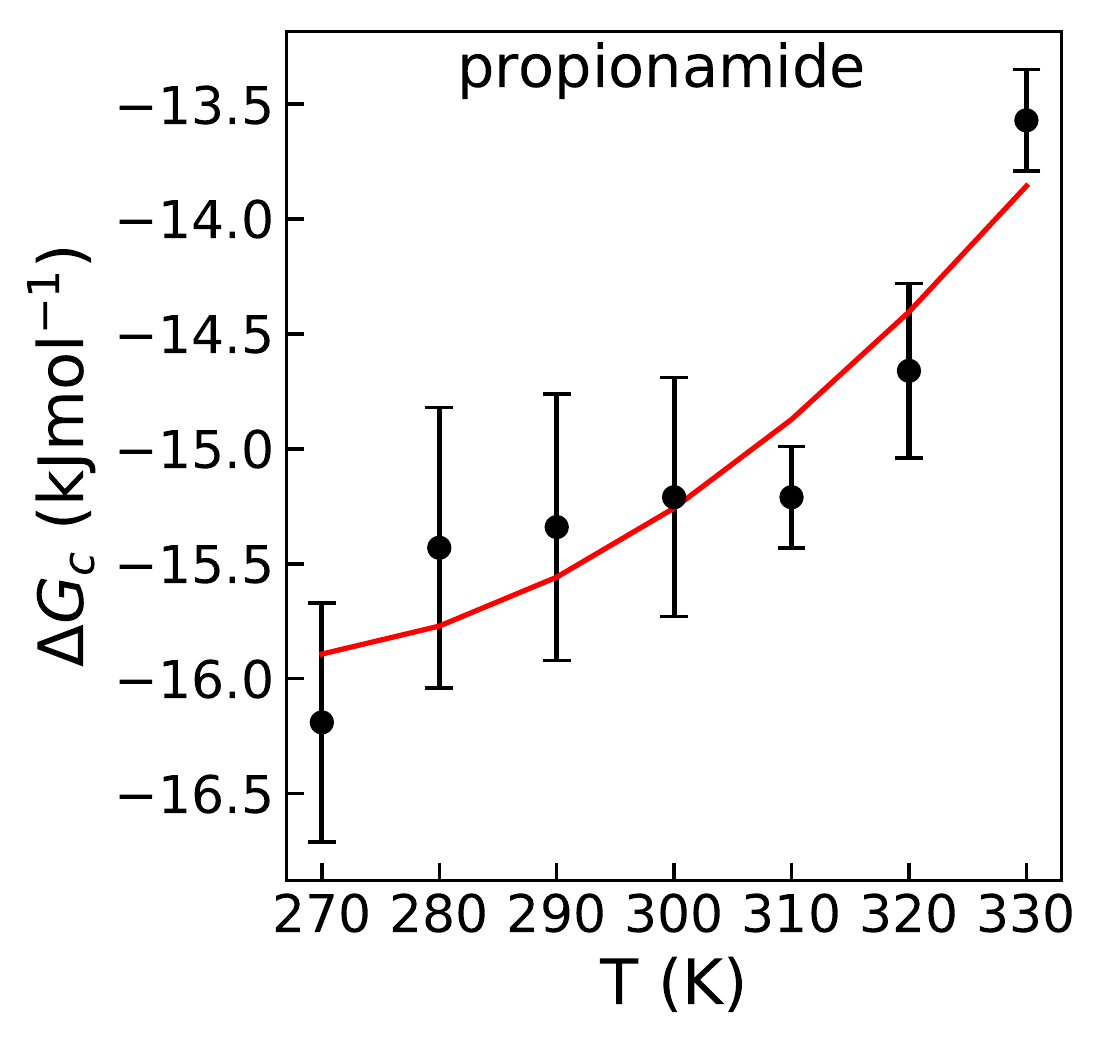}
\end{subfigure}\\
\begin{subfigure}[c]{0.22\textwidth}
\includegraphics[height=4.5cm, width=\textwidth]{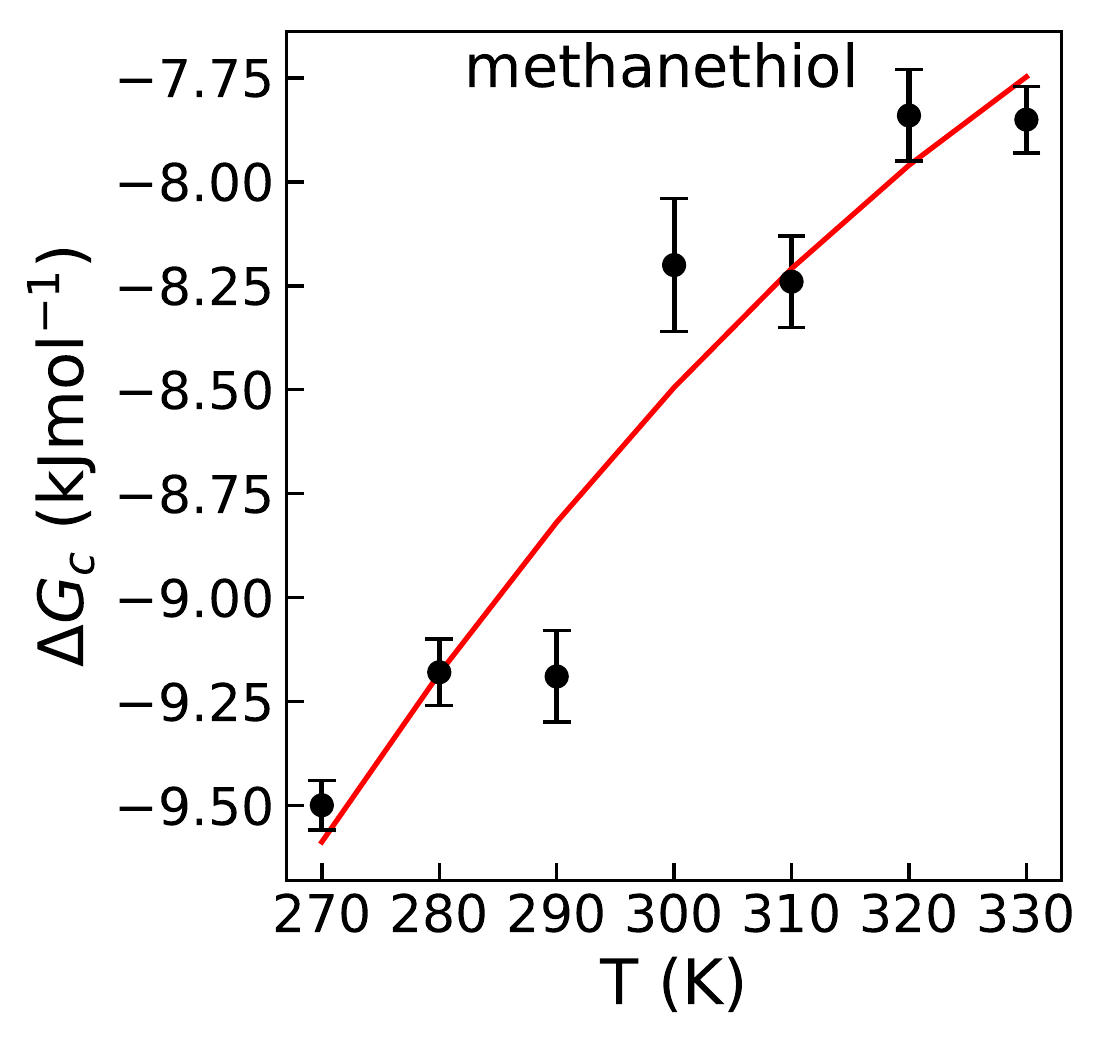}
\end{subfigure}&
\begin{subfigure}[c]{0.22\textwidth}
\includegraphics[height=4.5cm, width=\textwidth]{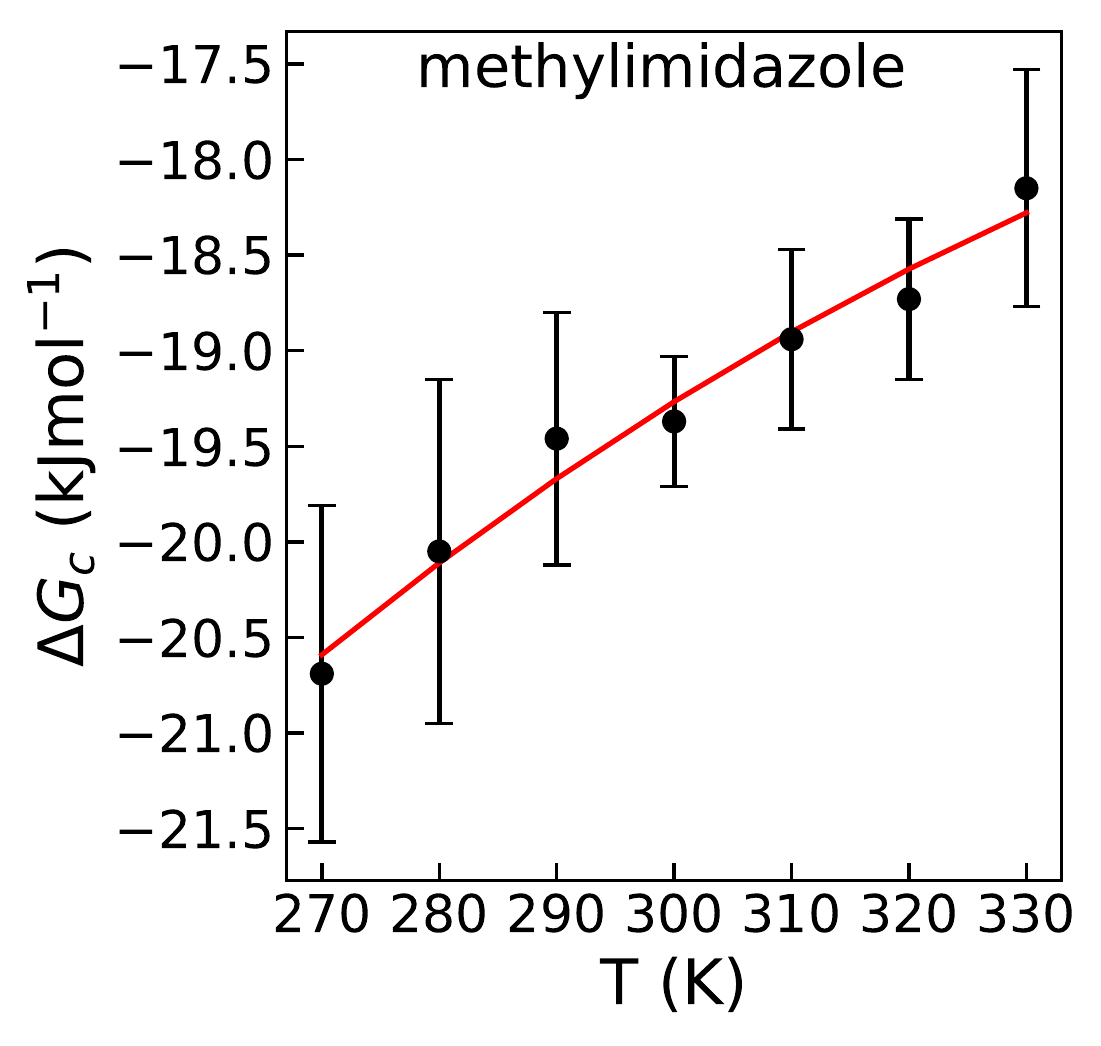}
\end{subfigure}&
\begin{subfigure}[c]{0.22\textwidth}
\includegraphics[height=4.5cm, width=\textwidth]{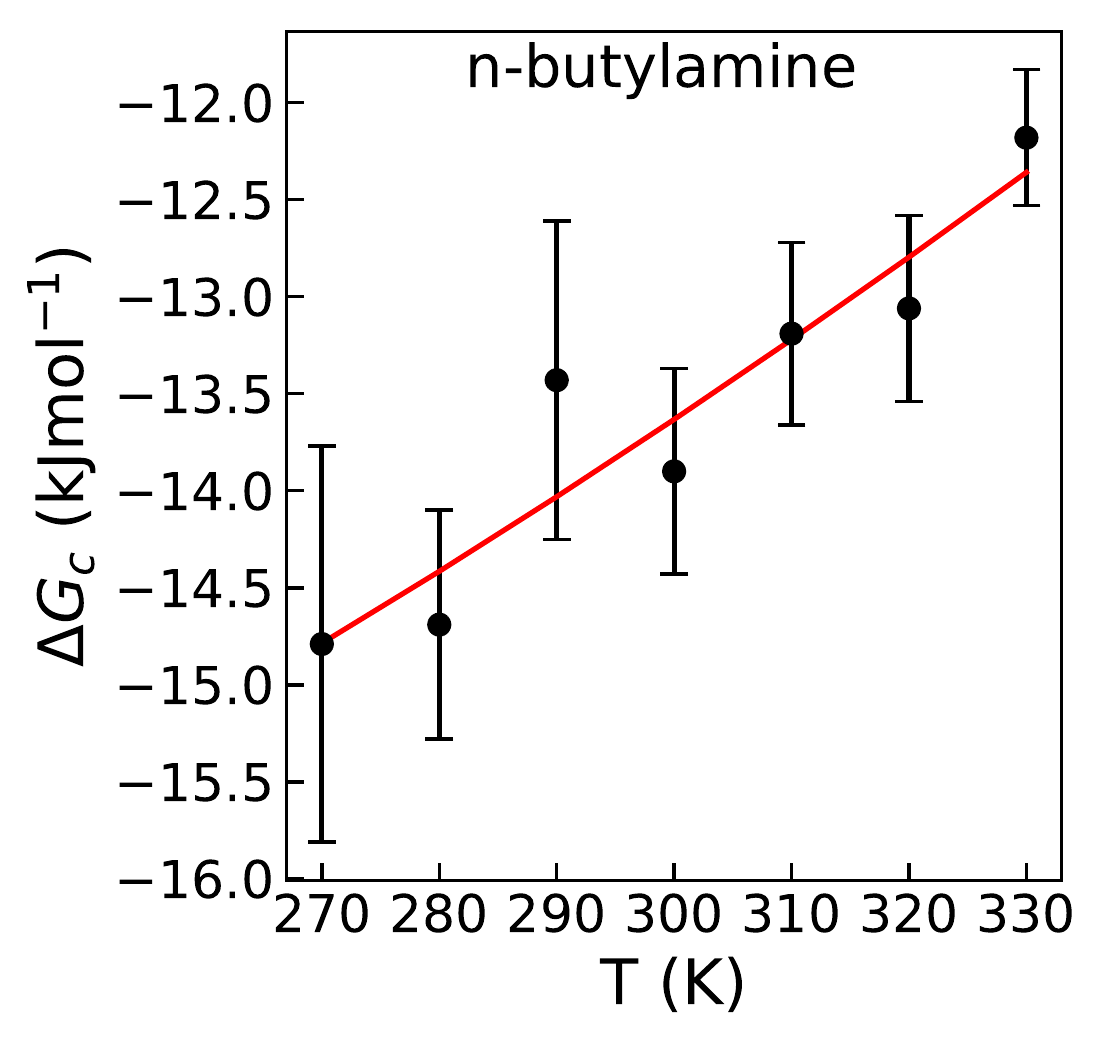}
\end{subfigure}&
\begin{subfigure}[c]{0.22\textwidth}
\includegraphics[height=4.5cm, width=\textwidth]{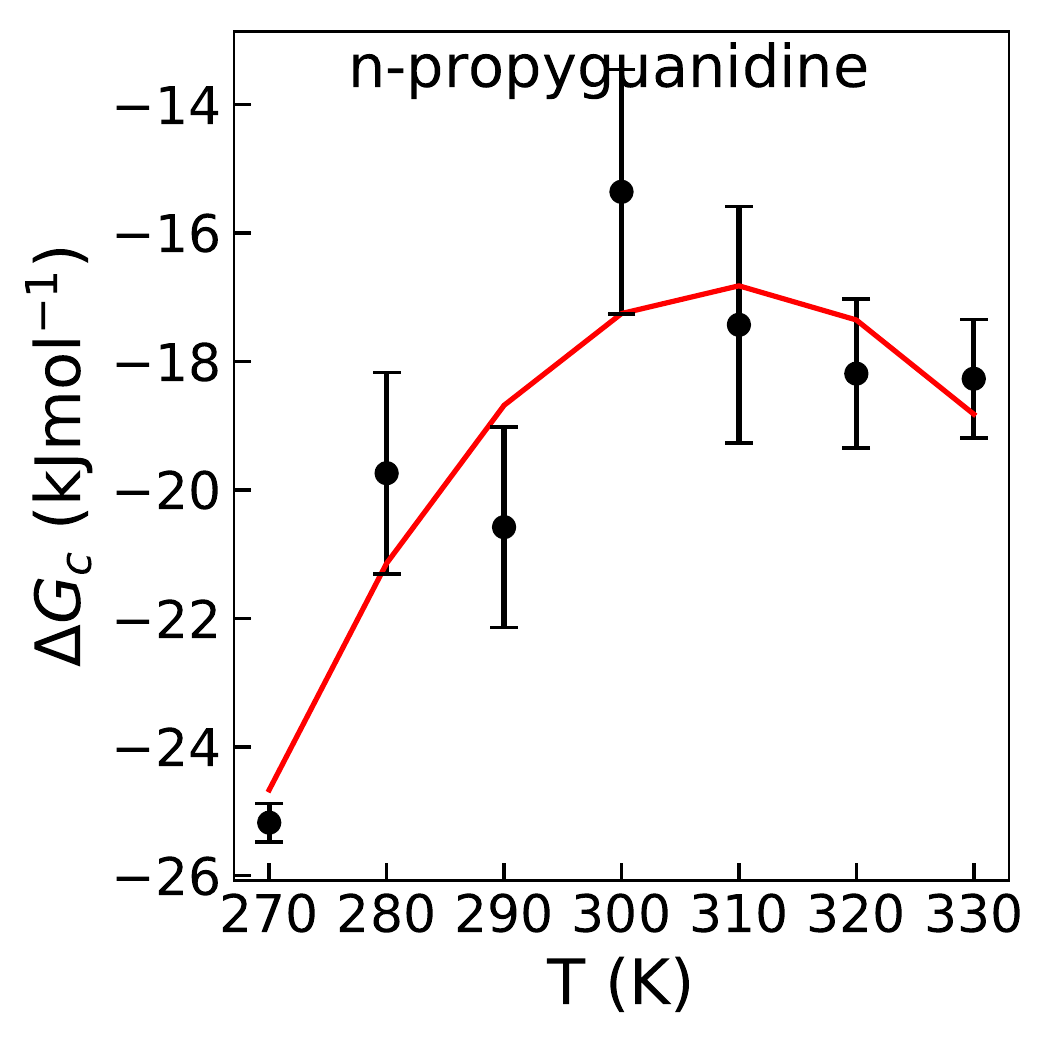}
\end{subfigure}&
\begin{subfigure}[c]{0.22\textwidth}
\includegraphics[height=4.5cm, width=\textwidth]{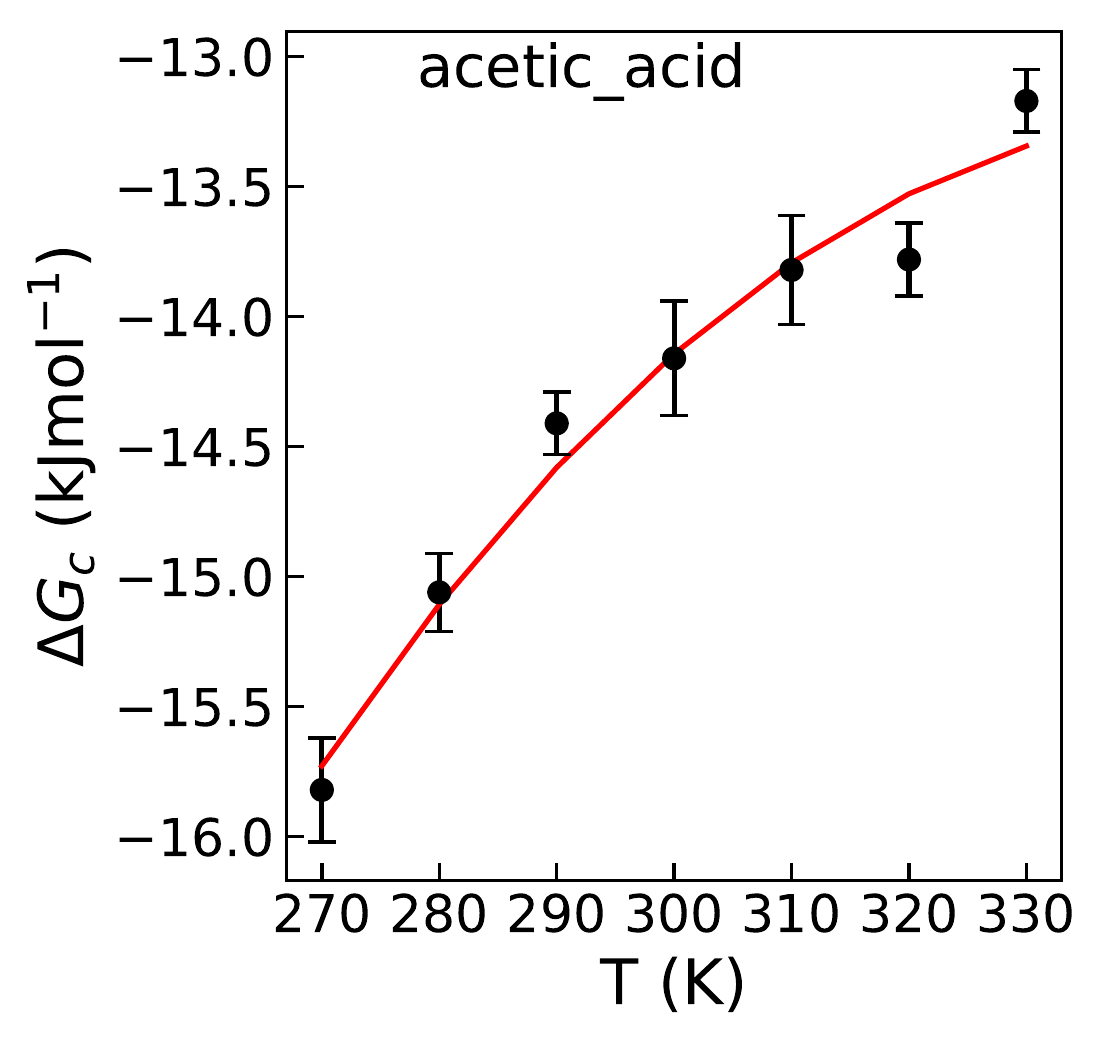}
\end{subfigure}&
\begin{subfigure}[c]{0.22\textwidth}
\includegraphics[height=4.5cm, width=\textwidth]{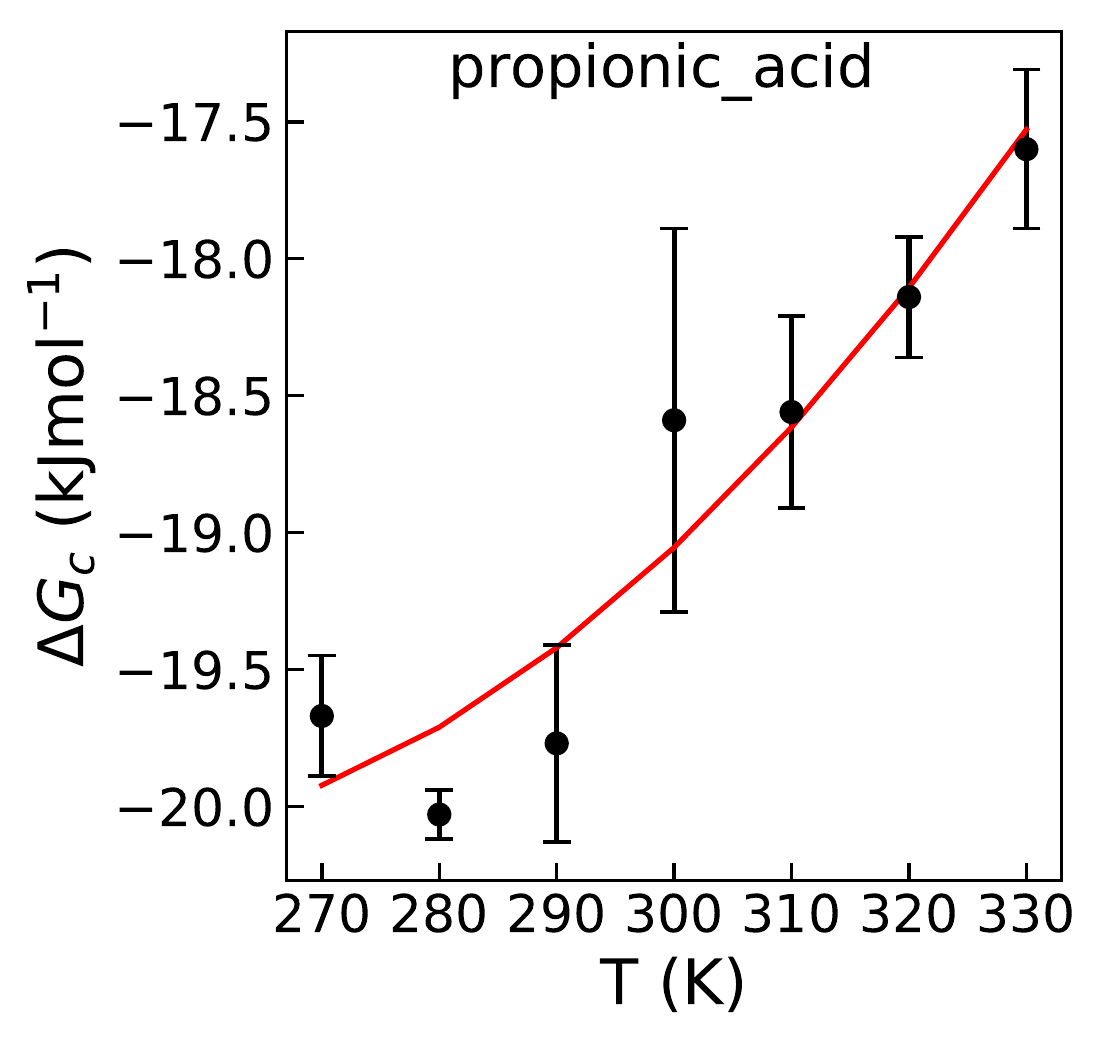}
\end{subfigure}\\        
\end{tabular}    
\caption{Temperature dependence of the solvation free energy from gas to cyclohexane \ce{cC6H12}, $\Delta G_{c}$}
\label{fig:chex}
\end{figure}
\end{turnpage}

\begin{turnpage}
\begin{figure}[h]
\centering
\begin{tabular}[c]{cccccc}
\begin{subfigure}[c]{0.22\textwidth}
\includegraphics[height=4.5cm, width=\textwidth]{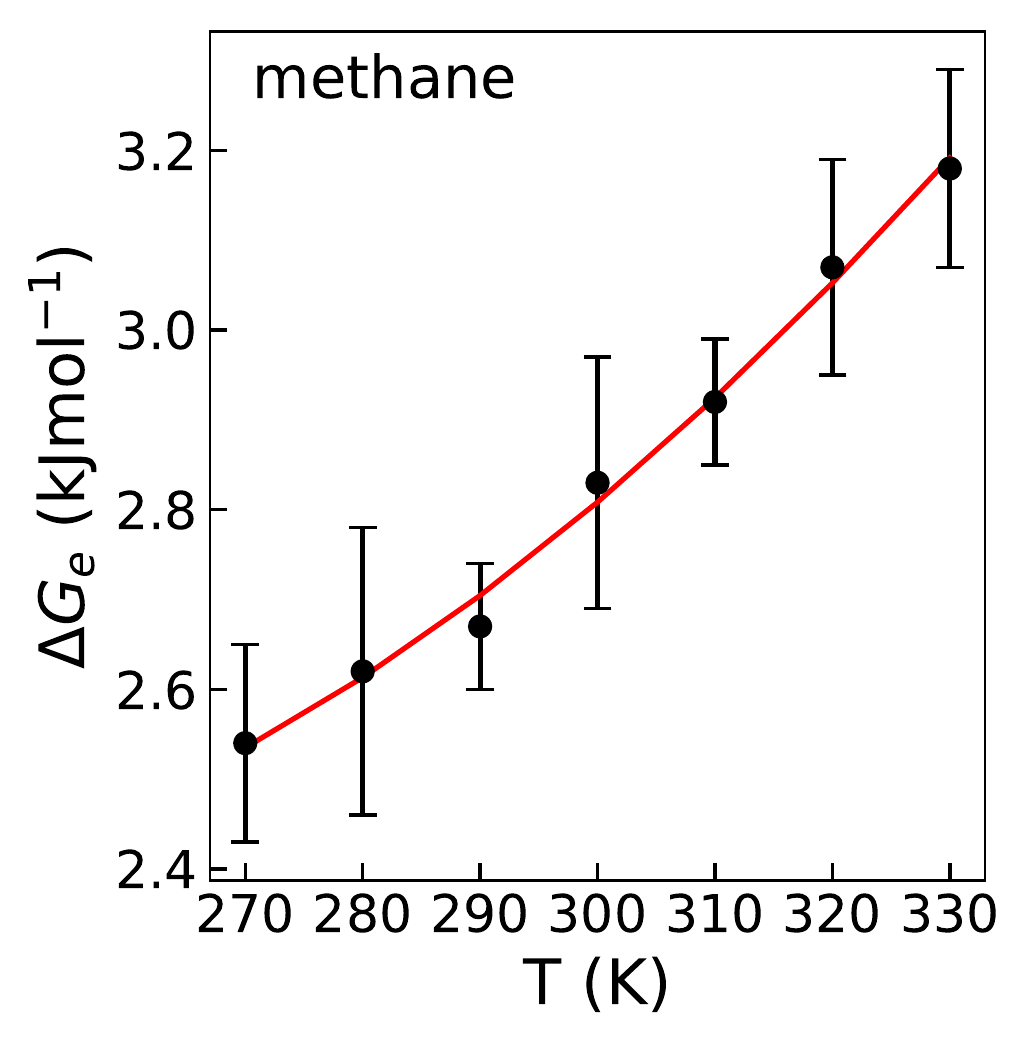}
\end{subfigure}&
\begin{subfigure}[c]{0.22\textwidth}
\includegraphics[height=4.5cm, width=\textwidth]{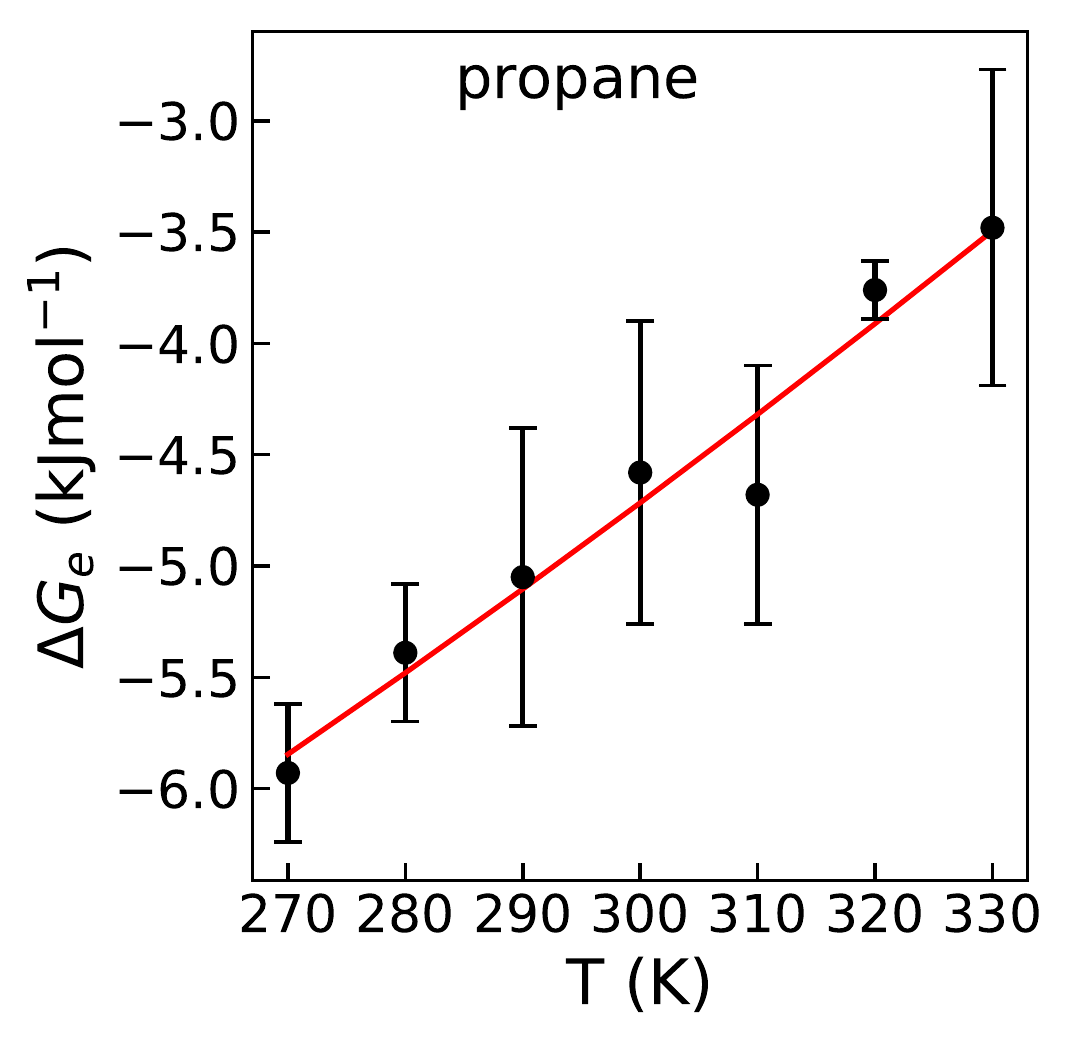}
\end{subfigure}&
\begin{subfigure}[c]{0.22\textwidth}
\includegraphics[height=4.5cm, width=\textwidth]{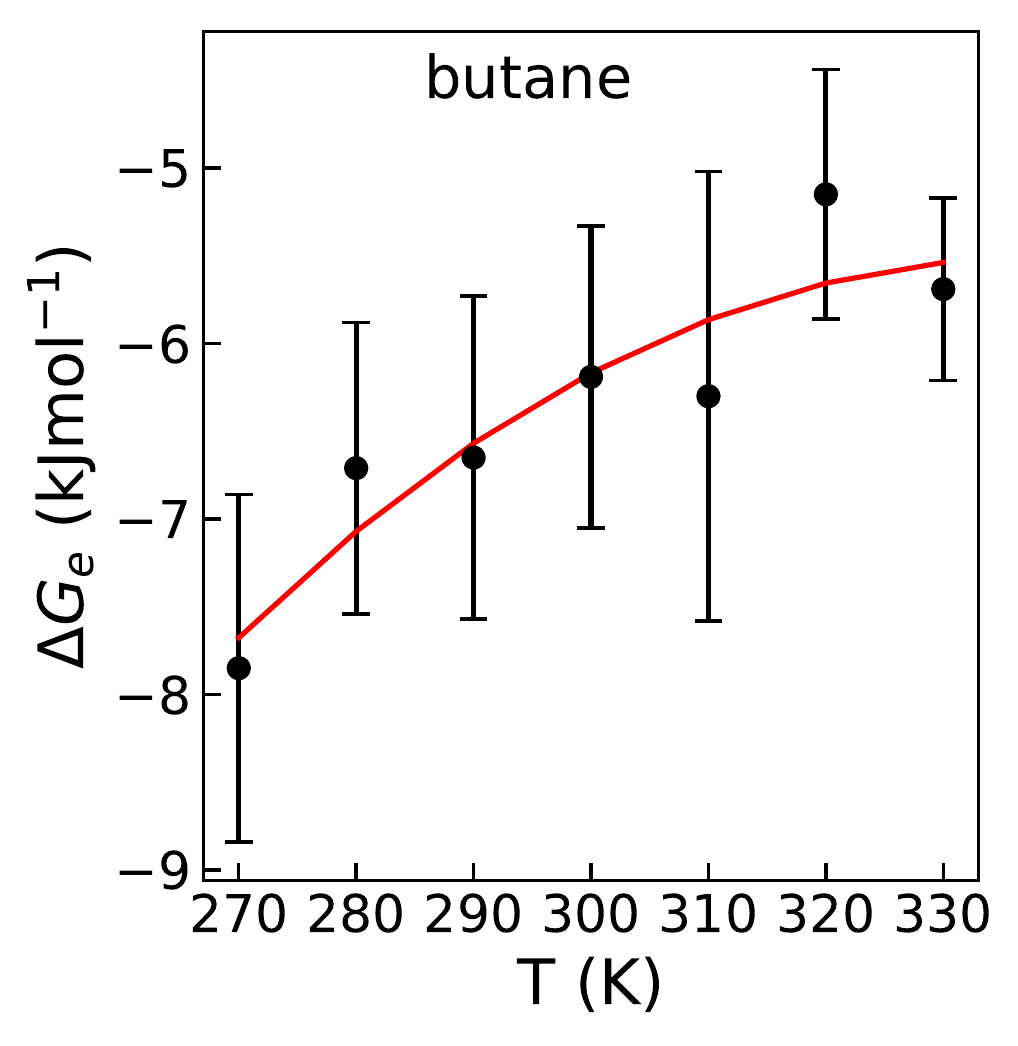}
\end{subfigure}&
\begin{subfigure}[c]{0.22\textwidth}
\includegraphics[height=4.5cm, width=\textwidth]{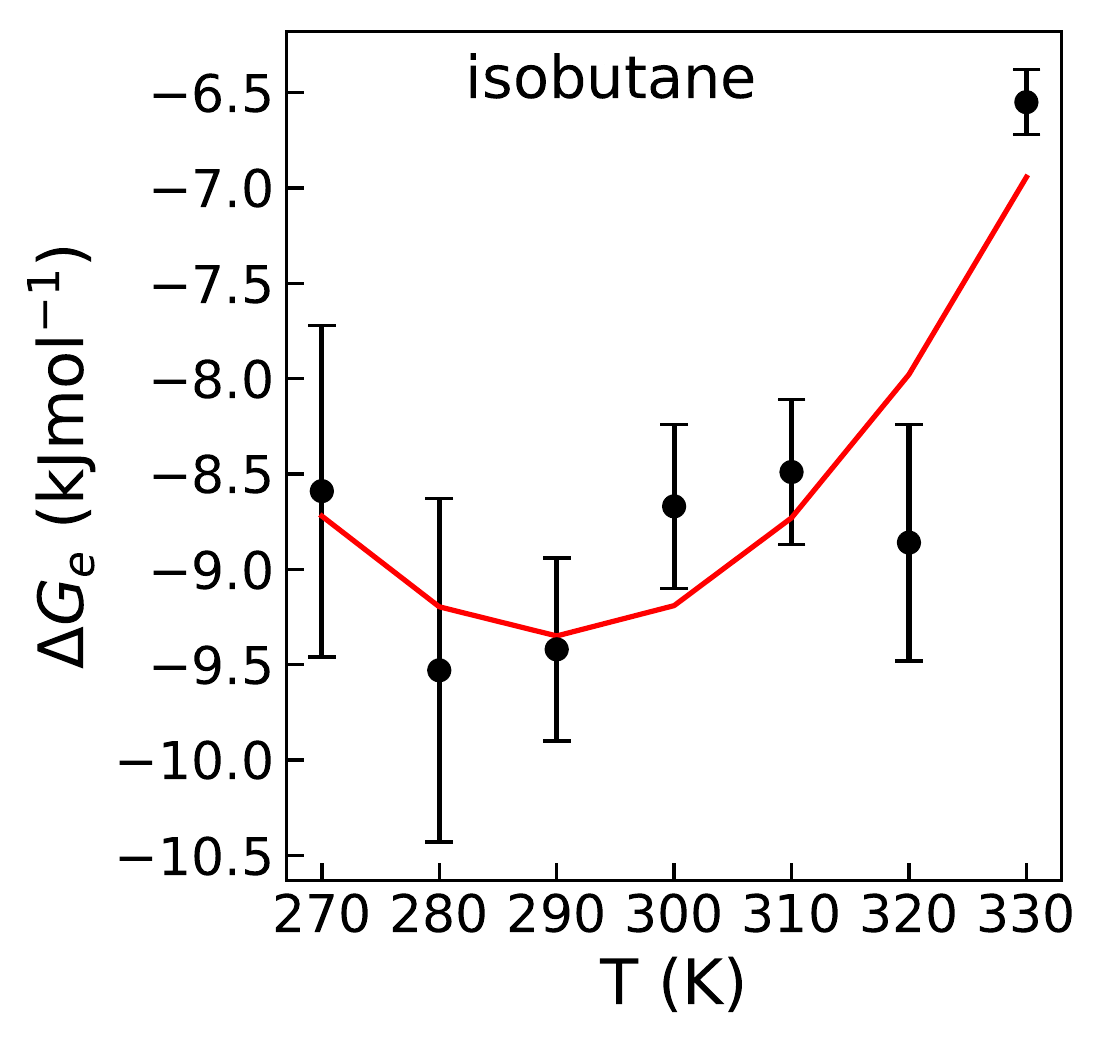}
\end{subfigure}&
\begin{subfigure}[c]{0.22\textwidth}
\includegraphics[height=4.5cm, width=\textwidth]{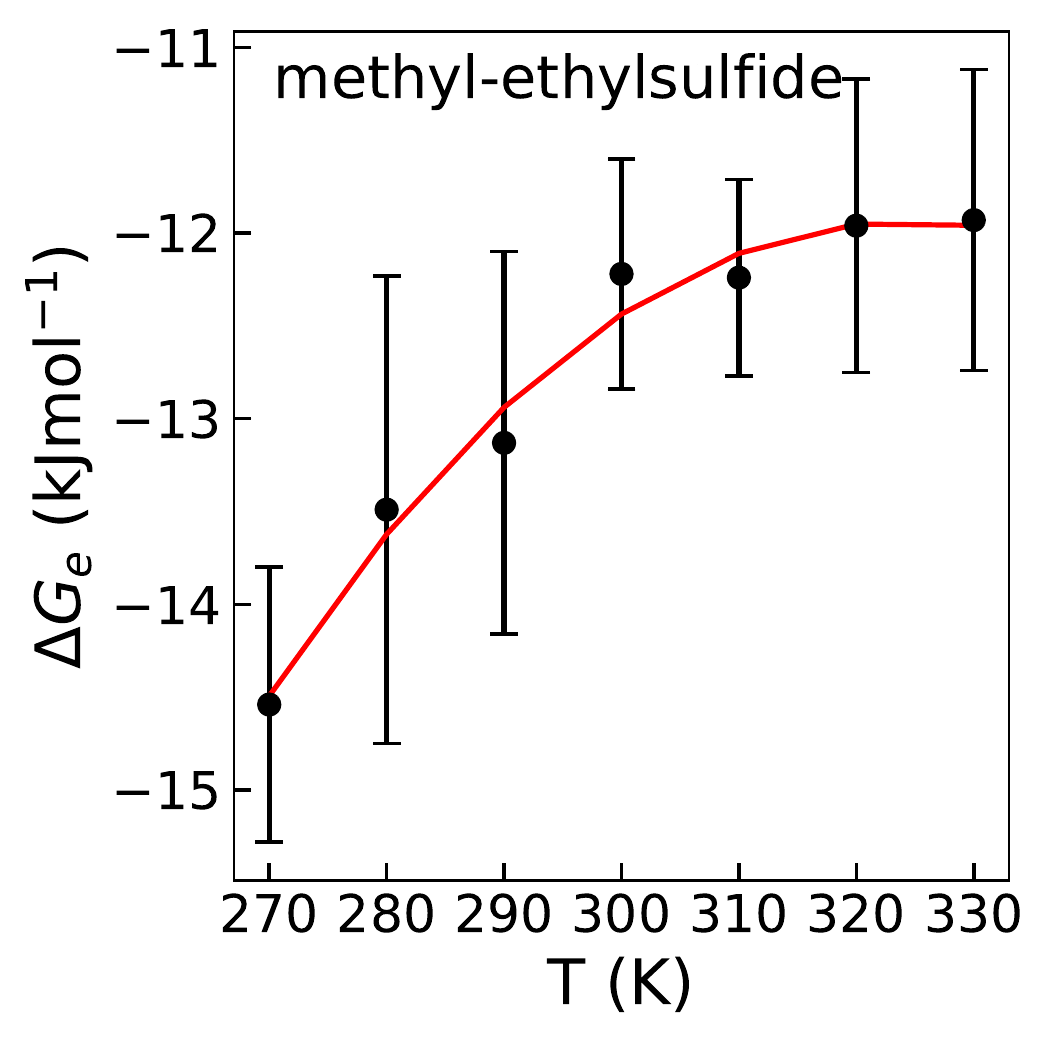}
\end{subfigure}&
\begin{subfigure}[c]{0.22\textwidth}
\includegraphics[height=4.5cm, width=\textwidth]{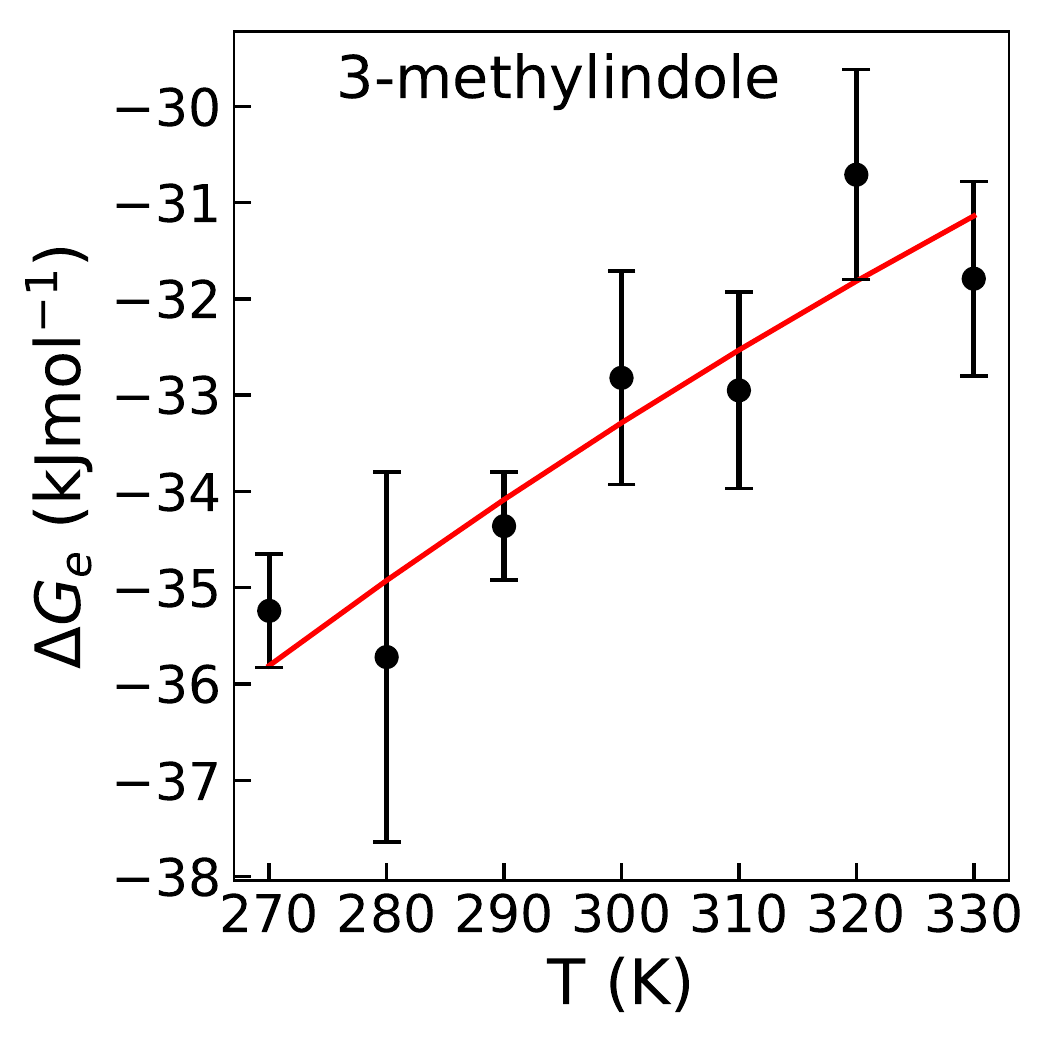}
\end{subfigure}\\  
\begin{subfigure}[c]{0.22\textwidth}
\includegraphics[height=4.5cm, width=\textwidth]{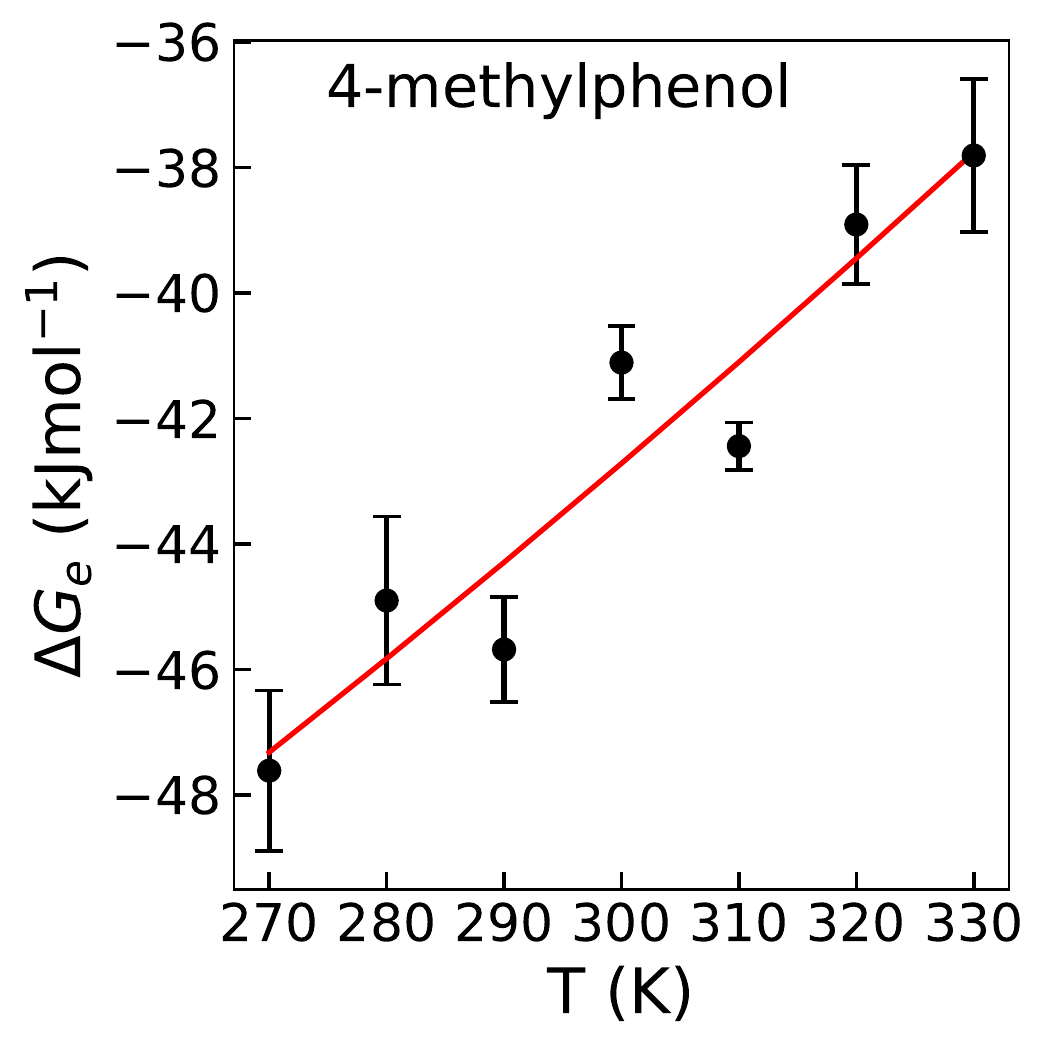}
\end{subfigure}&
\begin{subfigure}[c]{0.22\textwidth}
\includegraphics[height=4.5cm, width=\textwidth]{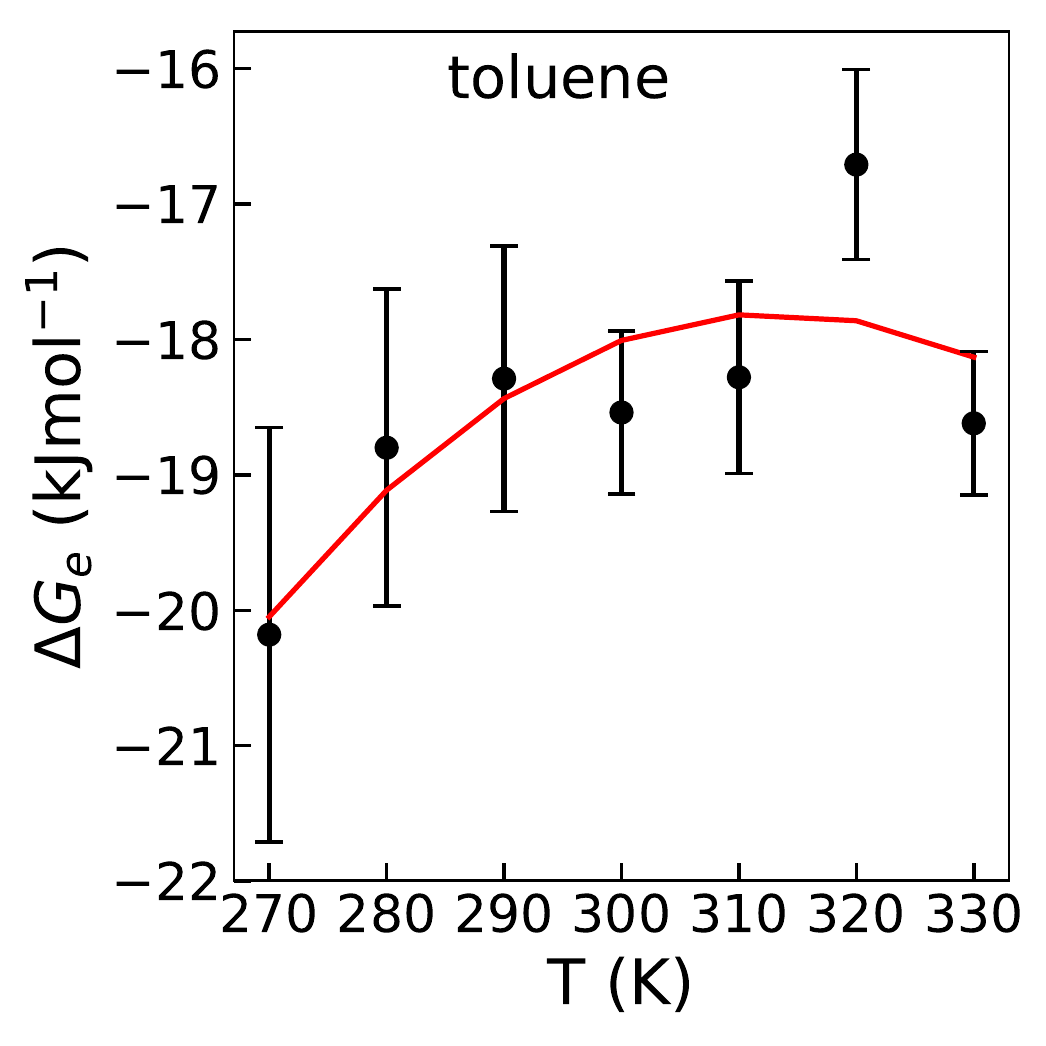}
\end{subfigure}&
\begin{subfigure}[c]{0.22\textwidth}
\includegraphics[height=4.5cm, width=\textwidth]{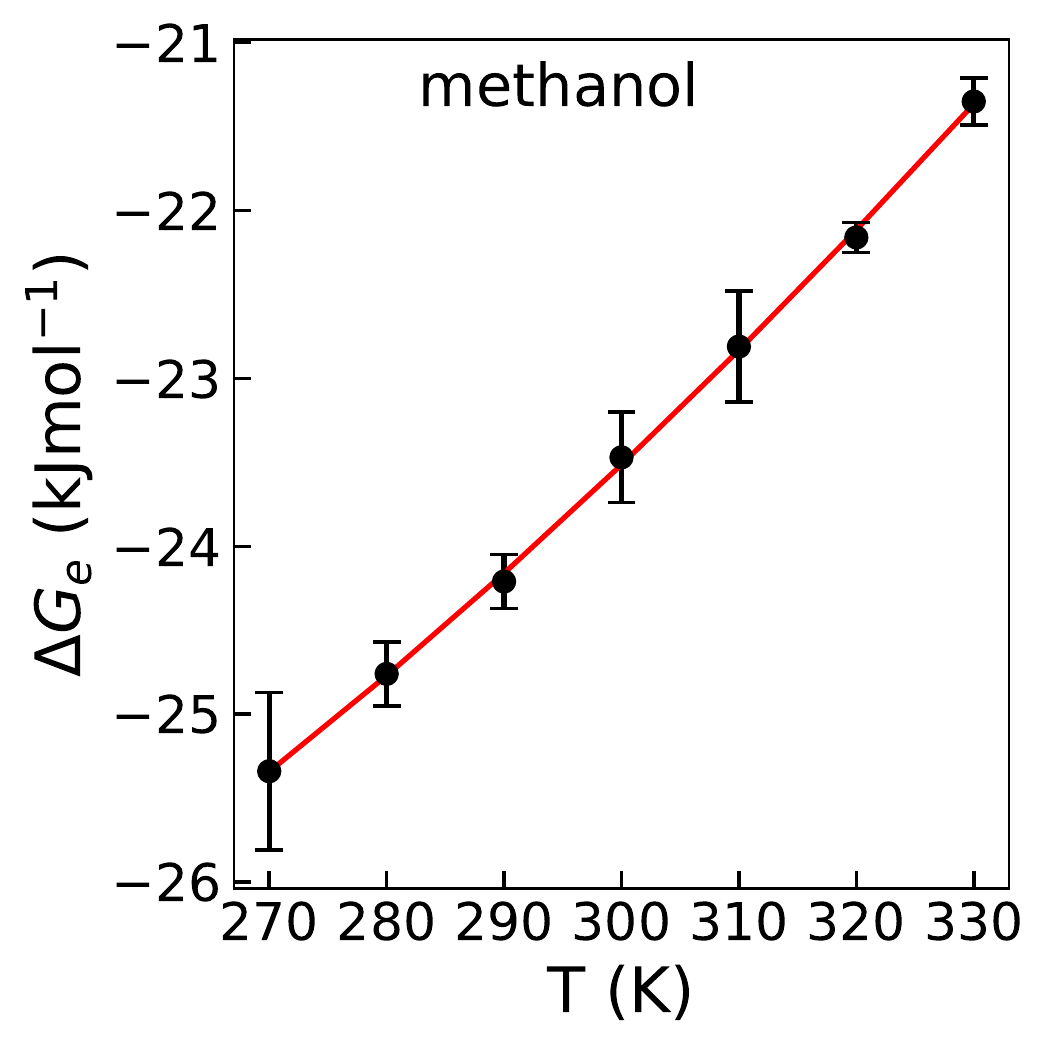}
\end{subfigure}&
\begin{subfigure}[c]{0.22\textwidth}
\includegraphics[height=4.5cm, width=\textwidth]{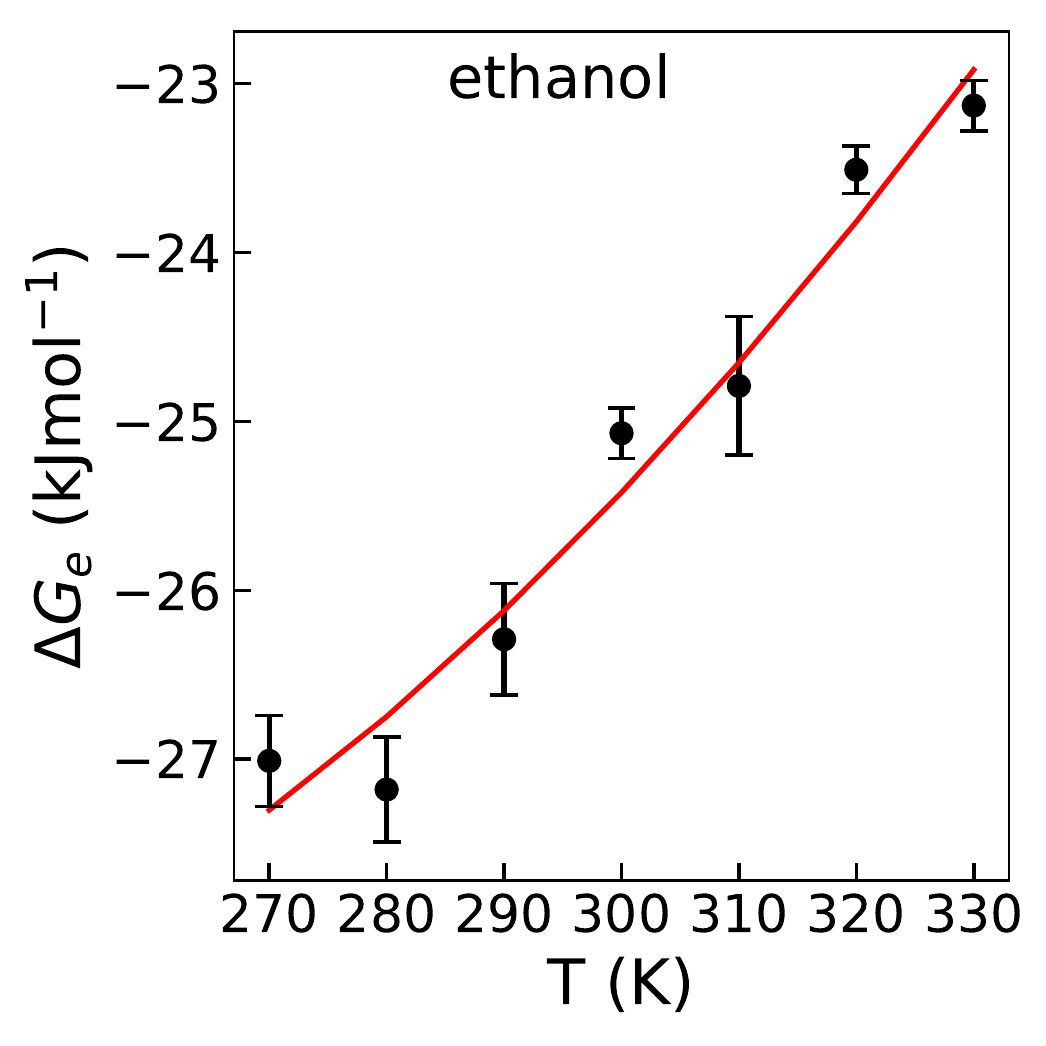}
\end{subfigure}&
\begin{subfigure}[c]{0.22\textwidth}
\includegraphics[height=4.5cm, width=\textwidth]{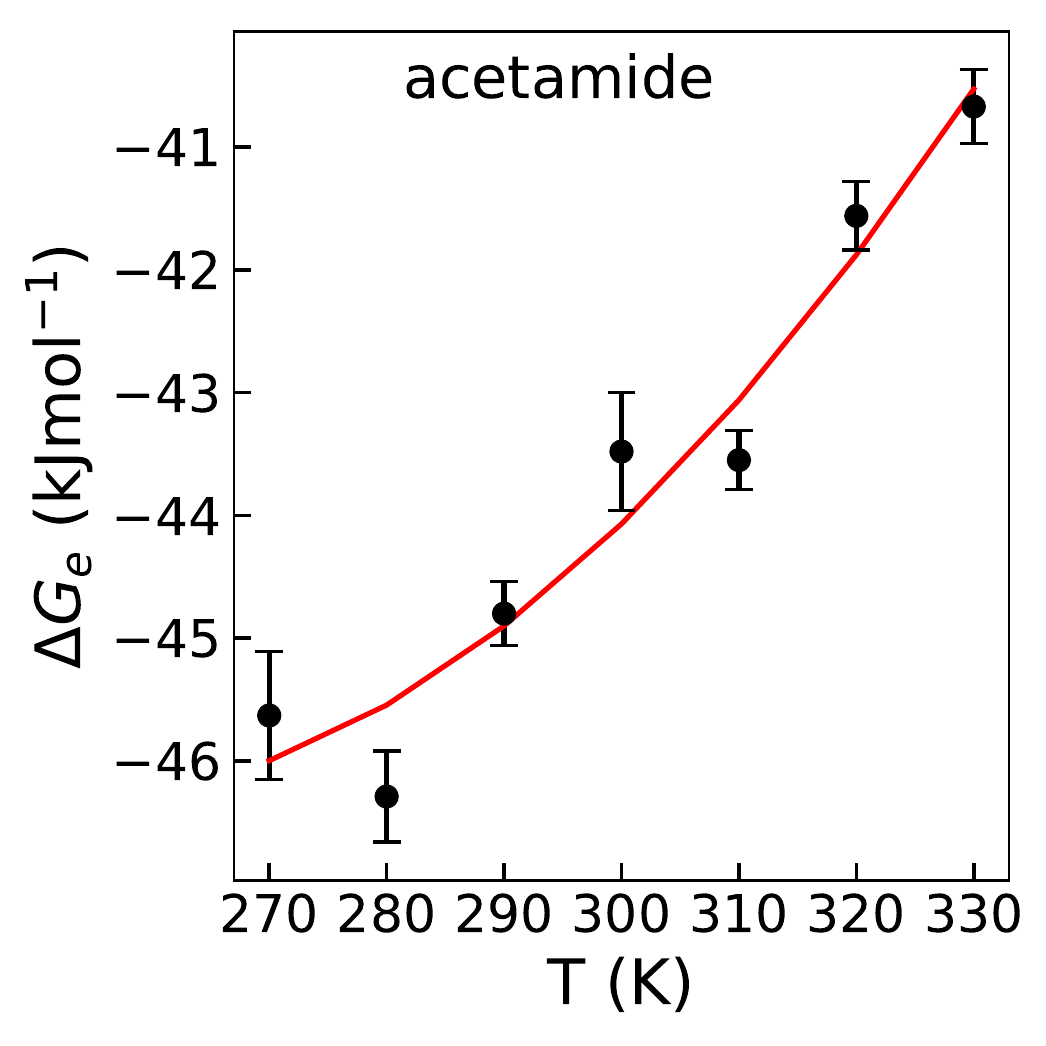}
\end{subfigure}&
\begin{subfigure}[c]{0.22\textwidth}
\includegraphics[height=4.5cm, width=\textwidth]{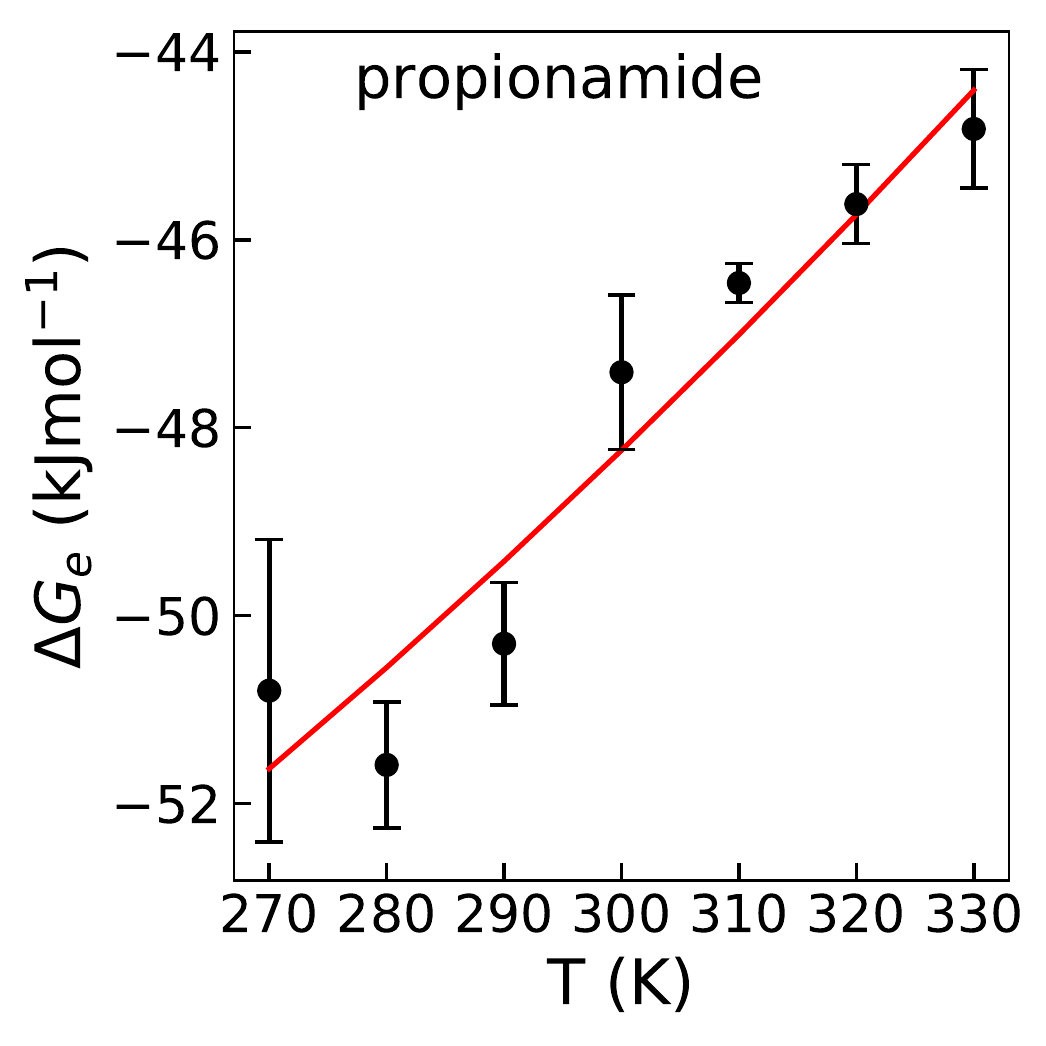}
\end{subfigure}\\
\begin{subfigure}[c]{0.22\textwidth}
\includegraphics[height=4.5cm, width=\textwidth]{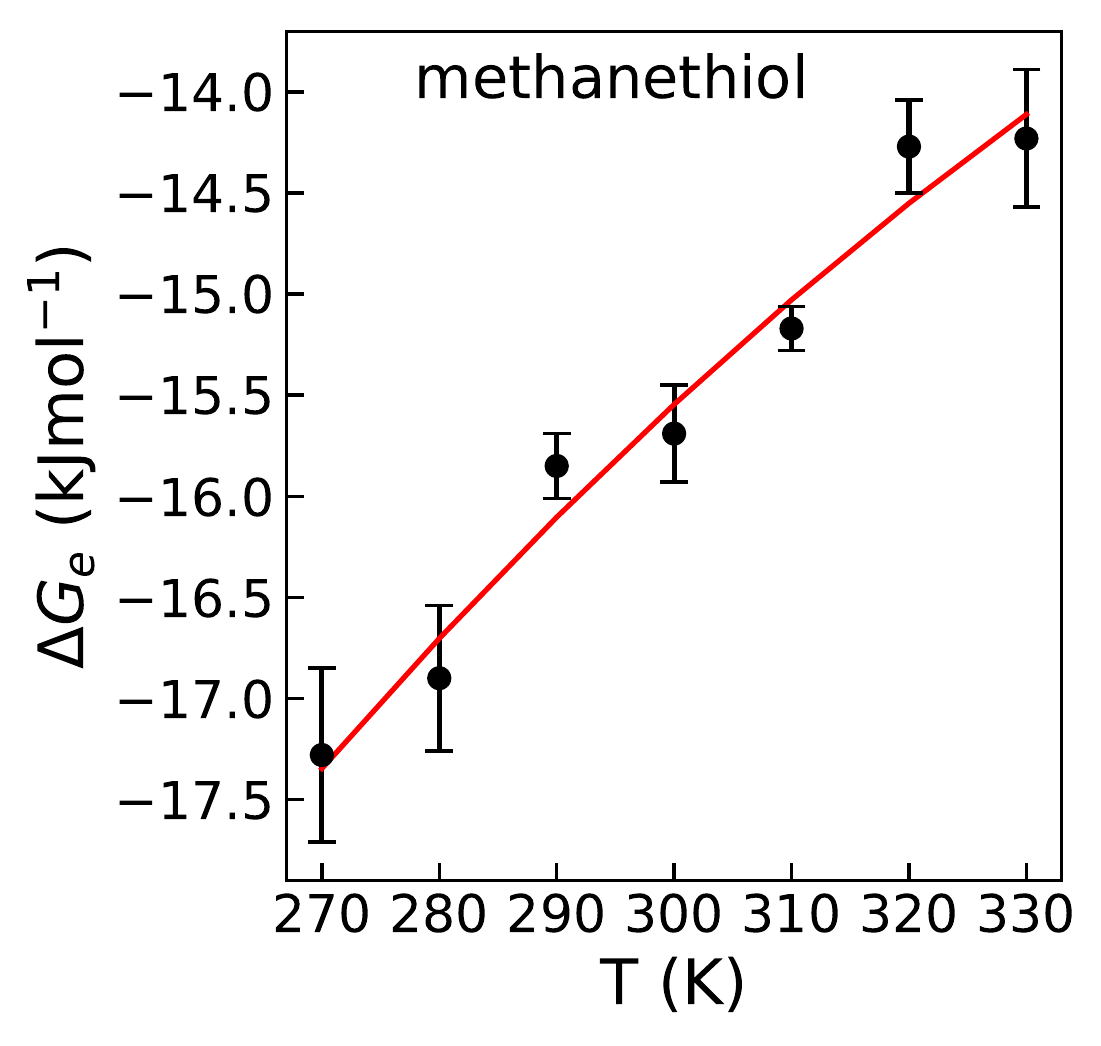}
\end{subfigure}&
\begin{subfigure}[c]{0.22\textwidth}
\includegraphics[height=4.5cm, width=\textwidth]{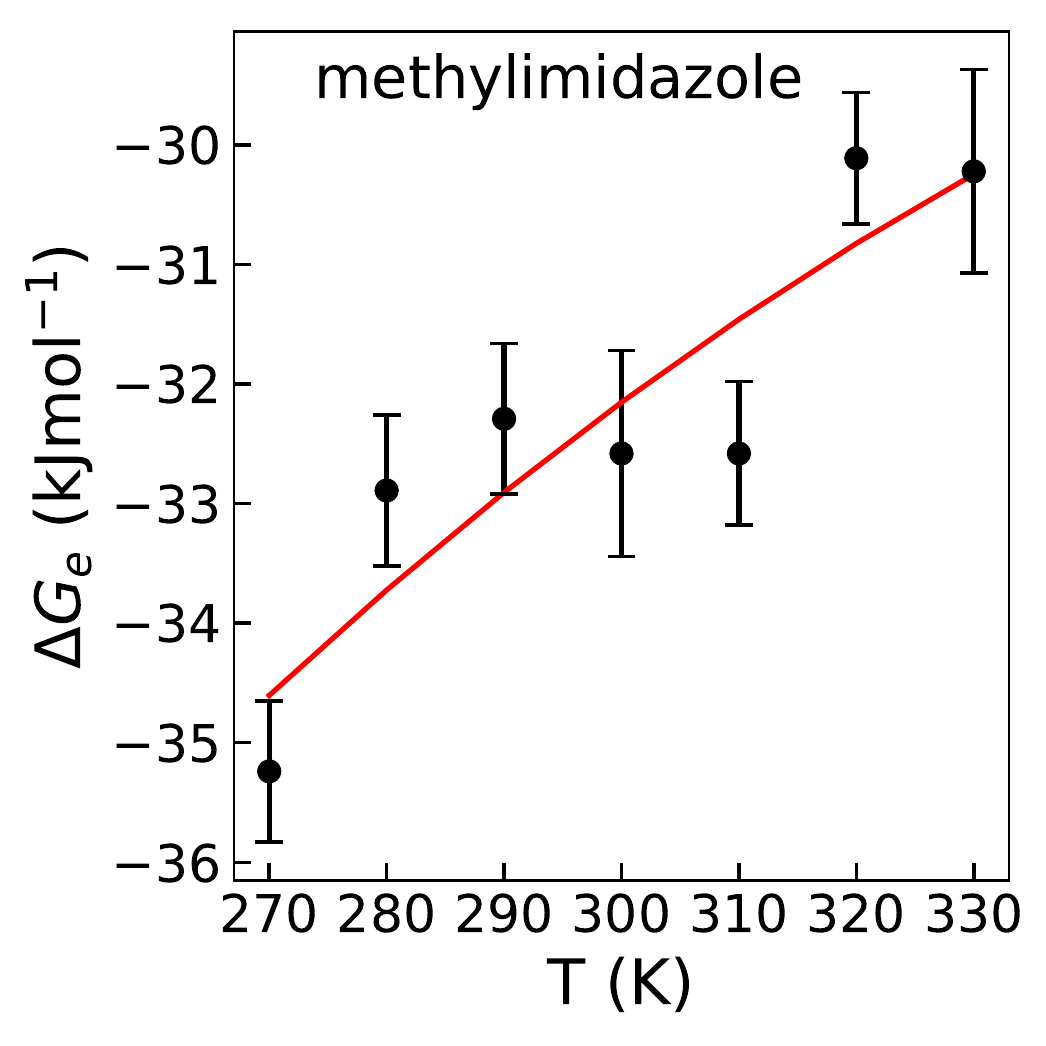}
\end{subfigure}&
\begin{subfigure}[c]{0.22\textwidth}
\includegraphics[height=4.5cm, width=\textwidth]{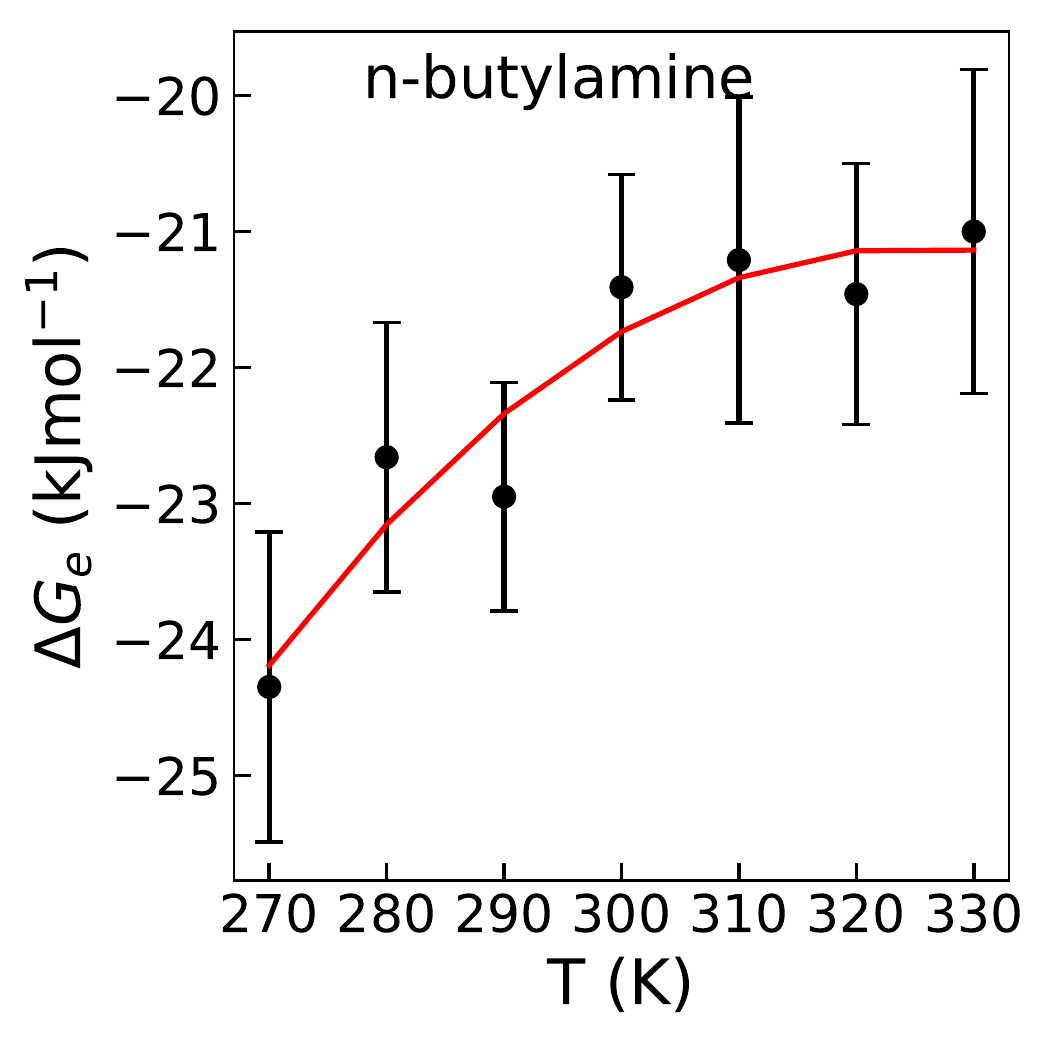}
\end{subfigure}&
\begin{subfigure}[c]{0.22\textwidth}
\includegraphics[height=4.5cm, width=\textwidth]{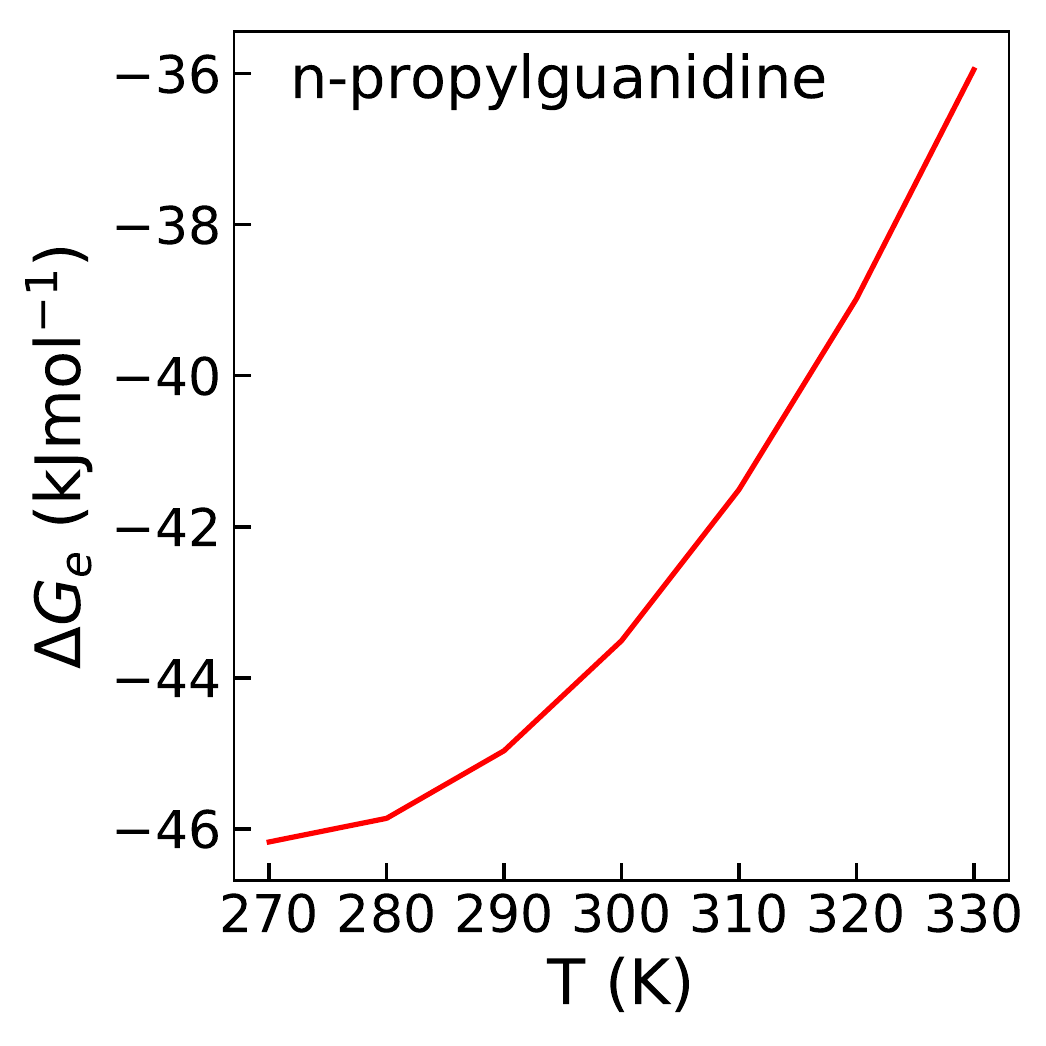}
\end{subfigure}&
\begin{subfigure}[c]{0.22\textwidth}
\includegraphics[height=4.5cm, width=\textwidth]{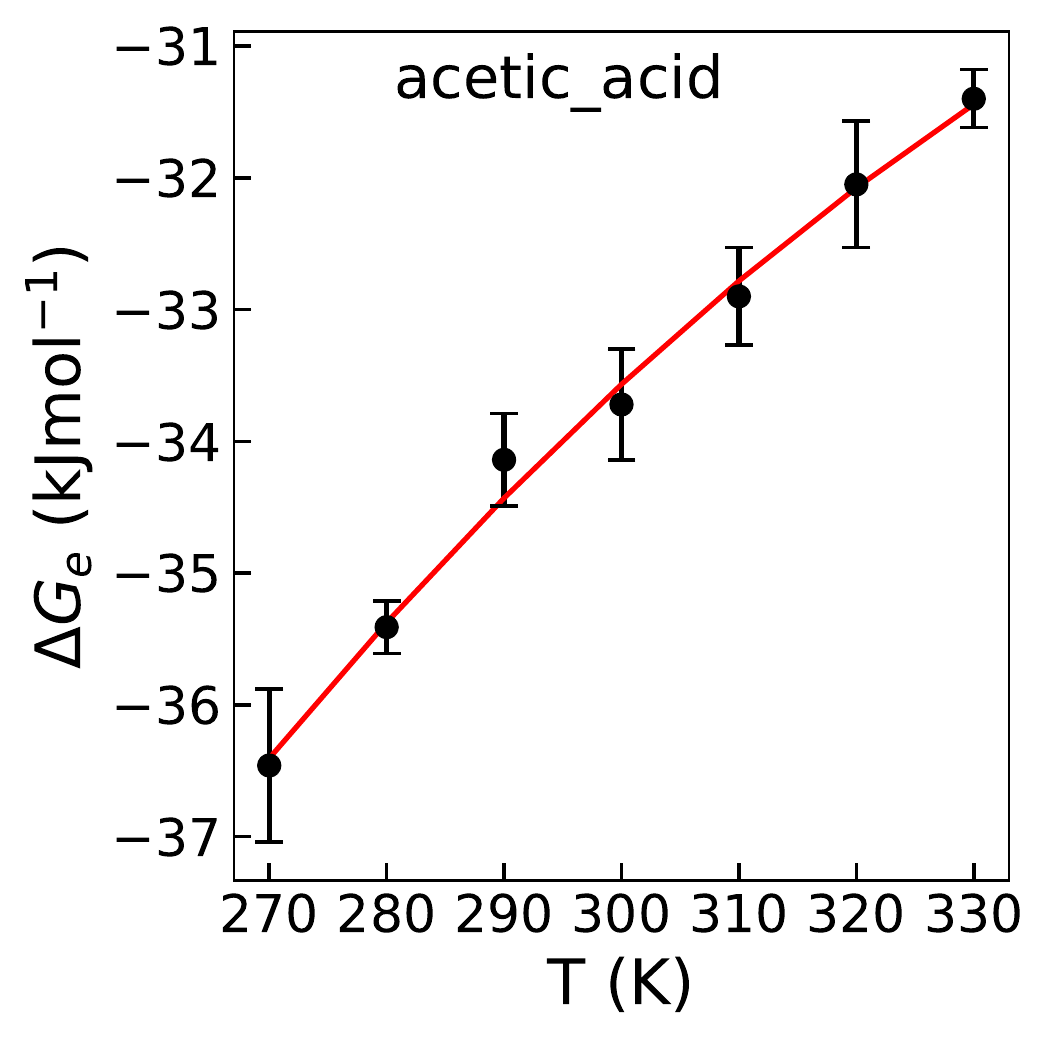}
\end{subfigure}&
\begin{subfigure}[c]{0.22\textwidth}
\includegraphics[height=4.5cm, width=\textwidth]{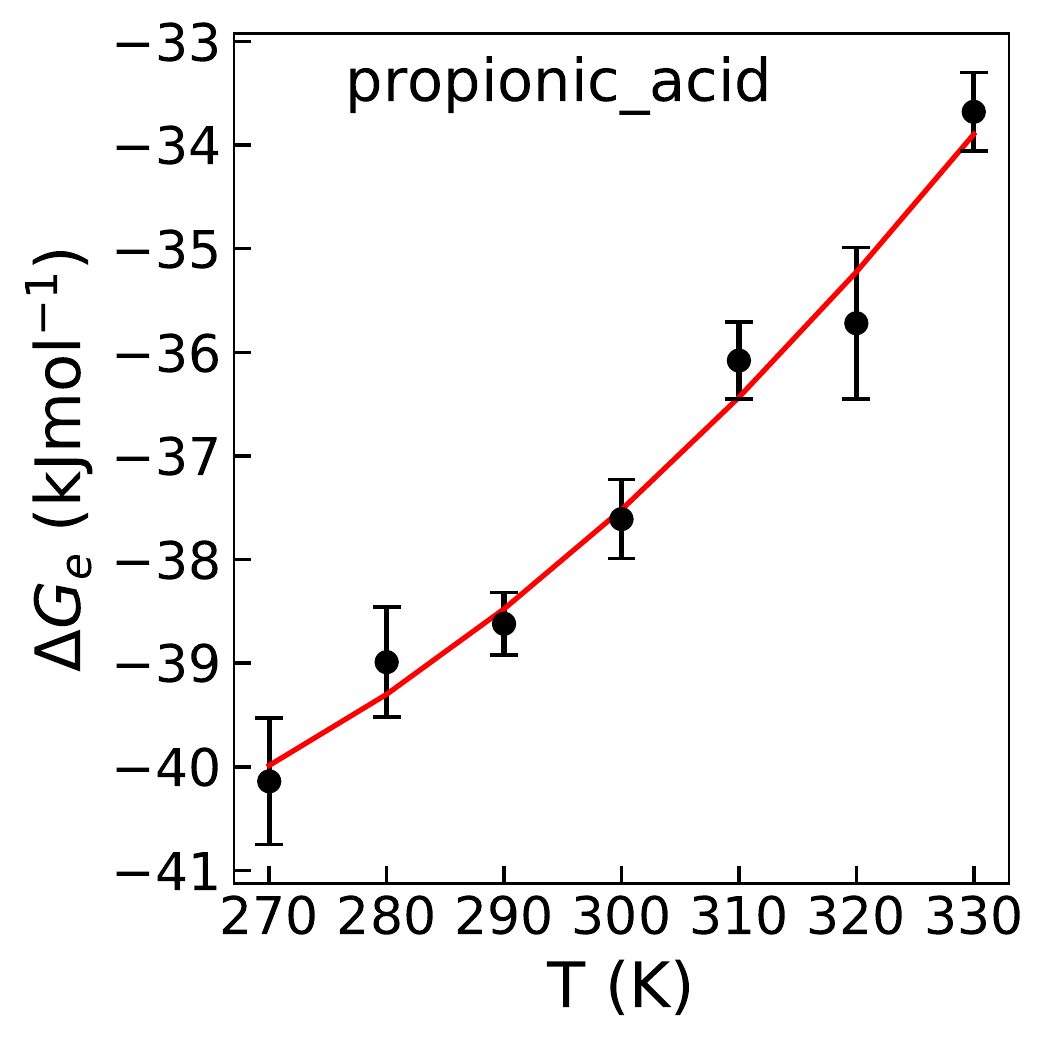}
\end{subfigure}\\        
\end{tabular}    
\caption{Temperature dependence of the solvation free energy from gas to ethanol \ce{EtOH}, $\Delta G_{e}$. It should be noted that large numerical fluctuations prevent the data collection in the case of n-propylguanidine (ARG). Therefore the corresponding plot should be viewed as an \textit{empirical}-like fitting which doesn't enable reliable extraction of thermodynamics constants.}
\label{fig:etoh}
\end{figure}
\end{turnpage}

\end{document}